\documentclass[preprint]{imsart}
\RequirePackage[OT1]{fontenc}
\RequirePackage{amsthm,amsmath}
\RequirePackage{natbib}
\RequirePackage[colorlinks,citecolor=blue,urlcolor=blue]{hyperref}

\usepackage{mathtools}
\usepackage{enumerate}
\usepackage{gensymb}
\usepackage{graphicx}
\usepackage{afterpage}
\usepackage{subfig}
\usepackage{microtype}
\usepackage{floatpag}
\usepackage{needspace}
\usepackage[utf8]{inputenc}

\DeclareMathOperator{\err}{err}
\DeclareMathOperator{\var}{var}
\DeclareMathOperator{\invW}{Inverse-Wishart}

\DeclareMathOperator{\I}{I}

\DeclareMathOperator{\E}{E}
\DeclareMathOperator{\argmax}{argmax}

\newcommand{\vect}[1]{\boldsymbol{#1}}
\newcommand{\matr}[1]{\boldsymbol{#1}}

\startlocaldefs
\numberwithin{equation}{section}
\theoremstyle{plain}

\endlocaldefs

\begin{document}

\begin{frontmatter}

\title{Functional Time Series Models for Ultrafine Particle Distributions\thanksref{T1}}
\thankstext{T1}{The study is supported by the Health Effects Institute’s Walter A. Rosenblith New Investigator Award under contract 4764-FRA06-3107-5.}
\begin{aug}
\author{\fnms{Heidi J.} \snm{Fischer}},
\author{\fnms{Qunfang} \snm{Zhang}},
\author{\fnms{Yifang} \snm{Zhu}}
\and
\author{\fnms{Robert E.} \snm{Weiss}}

\runauthor{H. Fischer et al.}

\affiliation{UCLA Fielding School of Public Health\thanksmark{m1}}

\address{Address of Heidi J. Fischer and Robert E. Weiss\\
Department of Biostatistics \\     
UCLA Fielding School of Public Health \\
Los Angeles, CA 90095-1772 USA \\
email: hfischer@ucla.edu; robweiss@ucla.edu}

\address{Address of Qunfang Zhang and Yifang Zhu\\
Department of Environmental Health Sciences \\
UCLA Fielding School of Public Health \\
Los Angeles, CA 90095-1772 USA \\
email: zhangqunfang@gmail.com; yifang@ucla.edu}
\end{aug}

\begin{abstract}
We propose Bayesian random effect functional time series models to model the impact of engine idling on ultrafine particle (UFP) counts inside school buses. UFPs are toxic to humans with health effects strongly linked to particle size. School engines emit particles primarily in the UFP size range and as school buses idle at bus stops, UFPs penetrate into cabins through cracks, doors, and windows. How UFP counts inside buses vary by particle size over time and under different idling conditions is not yet well understood.  We model UFP counts at a given time with a cubic B-spline basis as a function of size and allow counts to increase over time at a size dependent rate once the engine turns on. We explore alternate parametric models for the engine-on increase which also vary smoothly over size. The log residual variance over size is modeled using a quadratic B-spline basis to account for heterogeneity and an autoregressive model is used for the residual. Model predictions are communicated graphically.  These methods provide information needed for regulating vehicle emissions to minimize UFP exposure in the future.	
\end{abstract}

\begin{keyword}
\kwd{Bayesian Statistics}
\kwd{Hierarchical Models}
\kwd{Varying Coefficient Models}
\kwd{Heteroskedasticity}
\end{keyword}

\end{frontmatter}

\section{Introduction}
Ultrafine particles (UFPs) are particulate matter with diameters less than 100 nm. UFPs' small size and large surface area allow them to penetrate the lung, enter the circulatory system, and deposit in the brain~\citep{ober,samet} and it has been suggested that they are more toxic to humans than larger particles~\citep{aless,delfino,ferin,frampton}. The health effects of UFPs are linked to particle size which determines the region in the lung the particles deposit~\citep{mor08}.  Children are more sensitive than adults to UFPs because their physiological and immunological systems are still developing~\citep{bennett}. 

In the U.S., roughly 25 million children ride school buses daily. About 90 percent of buses are diesel powered, emitting particles primarily in the UFP size range~\citep{epa02,epa12}.  As school buses idle at bus stops, UFPs from diesel emissions penetrate into cabins through cracks, doors, and windows. This so-called “self-pollution” increases the exposure to UFPs of children on board~\citep{zhang12}.  How UFP counts vary by particle size as school buses idle over time and under different idling conditions is not yet well understood. 

Researchers collected particle counts inside buses first with the engine-off and then with the engine turned on and idling. A Scanning Mobility Particle Sizer (SMPS) counted particles per cubic centimeter in 102 \emph{size bins} that group particles by diameter ranging from the first size bin containing particles of the smallest diameters, 7.37--7.64 nm, to the last size bin containing sizes of 269.0--278.8 nm.  The ordered collection of counts in these 102 size bins at a single point in time is called a \emph{UFP size distribution}, even though (i) technically particles with diameters greater than 100 nm are too big to be UFPs and (ii) the counts are not a distribution in the statistical sense.  Size bin widths are approximately equally spaced on a log scale, so UFP size distributions have more bins for the smaller UFP particles of interest. UFP distributions were collected over time during multiple experiments, or \emph{runs}, making the data multivariate longitudinal. 

Numerous mathematical representations have been used to describe size distributions over size bin and time, typically via \emph{modal methods}~\citep{whitby78,whitby91}.  Modal methods model the particle size distribution as a mixture of densities~\citep{hussein05,whitby91}, ignoring the total number of particles. More recently, \citet{wraith09,wraith13} model time series of UFP size distributions using time-varying non-parameteric Bayesian mixture models.  Modal methods standardize particle counts: only information about the relative composition of particle size bins is retained. In contrast, in modeling vehicle emissions and in setting vehicle emissions policy, understanding actual particle counts is crucial; thus modal methods are insufficient for this application.  Modal methods have also not accounted for sampling variances in observed UFP counts. UFP counts in smaller size bins have larger variances than those in larger size bins, as they tend to be more unstable than larger particles, following dynamic processes which have them desorb, deposit, or combine to form larger particles at fairly rapid rates~\citep{kulmala04, kittleson06}.  In urban environments UFP counts vary for certain particle size bins more than others because they are created by anthropogenic processes which fluctuate over time. Statistical methods which model mean UFP counts by particle size bin and over time must account for increased residual variance for smaller particle size bins.

Modeling UFP size distributions using functional time-series methods allows for inference on particle counts while accounting for differences in residual variance across particle size bin. It is also an improvement on the methods in \citet{zhang12}, who modeled UFP counts inside idling school buses over time using separate univariate longitudinal models for each particle size bin. As we expect neighboring bins to have similar counts, the univariate longitudinal approach does not fully utilize the information in the data.

We propose Bayesian longitudinal functional time series models to model the impact of engine idling on UFP counts inside school buses. Our approach is a varying coefficient model as in~\citet{hastie93} or \citet{Lang04}.  We model UFP size distributions at a given time with a cubic B-spline basis~\citep{deboor} and allow counts to increase over time at a size bin dependent rate once the engine is turned on. We explore alternate models for the engine-on increase: a possible jump in counts at engine-on followed by either a quadratic or bent line time trend.

Steady meteorological and background traffic conditions during individual runs implies that UFP size distributions before the engine is on do not vary greatly with time, but variation in \emph{baseline} UFP size distributions is observed between runs.  Spline random effect models have been used numerous times in the literature to describe non-parametric processes with longitudinal study designs \citep[e.g.][]{ruppert03,shi96}, however our model is different in that B-splines model UFP size distributions, not time trends. Residuals are modeled with an autoregressive model over time to accommodate correlation over time.  To account for larger count variance for smaller particles relative to larger particle size bins, the log residual variance over size bin is modeled using a quadratic B-spline basis.

Interest centers on how mean particle counts change after the engine turns on as a function of size bin. Researchers are also interested in the mode of particle size counts, the mode height, and how both evolve after the engine turns on.  We provide summaries of how the mode and mode height evolve as the engine idles.  Plots are presented to aid in the interpretation of model inferences and make model output interpretable to non-statisticians.  Graphs also aid in diagnosis of lack of fit and can help suggest model improvements.  

In Section 2 we describe the dataset, Section 3 presents our model and Section 4 gives results.  Finally, Section 5 is discussion. 

\section{UFP Size Distributions Inside Buses}

UFP size distribution measurements were collected inside the bus every 2 minutes and the current analysis considers measurements taken during the time period between 15 minutes before the engine was turned on and 20 minutes afterwards.  A set of UFP size distributions collected over this time period defines one run, though a few runs are shorter than the defined time.  For certain runs, measurements occurred at odd numbered minutes while for other runs measurements occurred at even numbers.  Runs took place under one of two window positions: (1) all windows-closed, although some windows could not be closed tightly; and (2) eight rear windows, four on each side, open 20 cm.  There are 21 runs in this dataset: 12 for windows-open and 9 for windows-closed. The study was conducted in an open space under stable meteorological conditions without nearby UFP emission sources in Los Angeles, CA~\citep{zhang12}.

\begin{figure}[!ht]
\centering
\subfloat[]{\includegraphics[width=7.5cm]{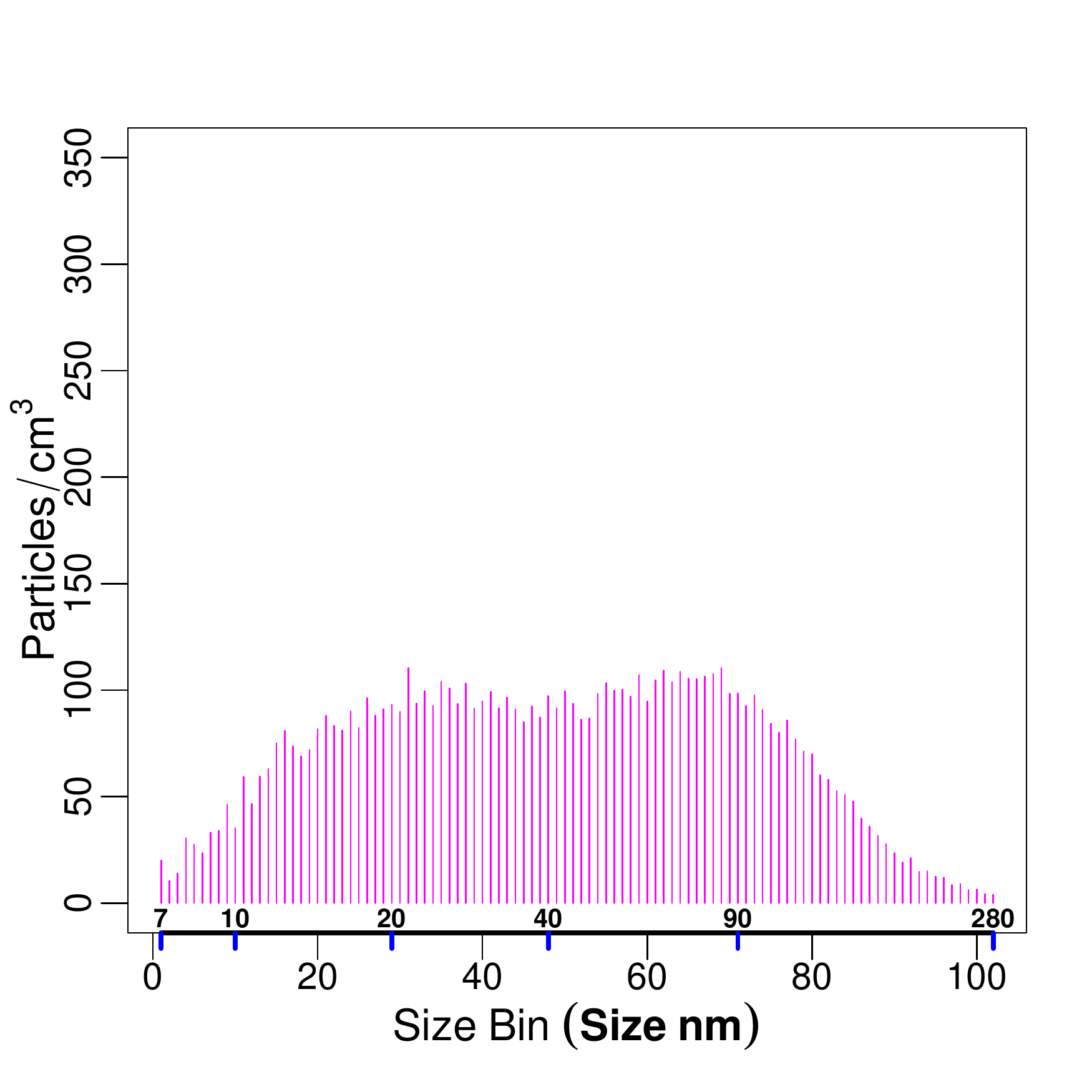}}
\subfloat[]{\includegraphics[width=7.5cm]{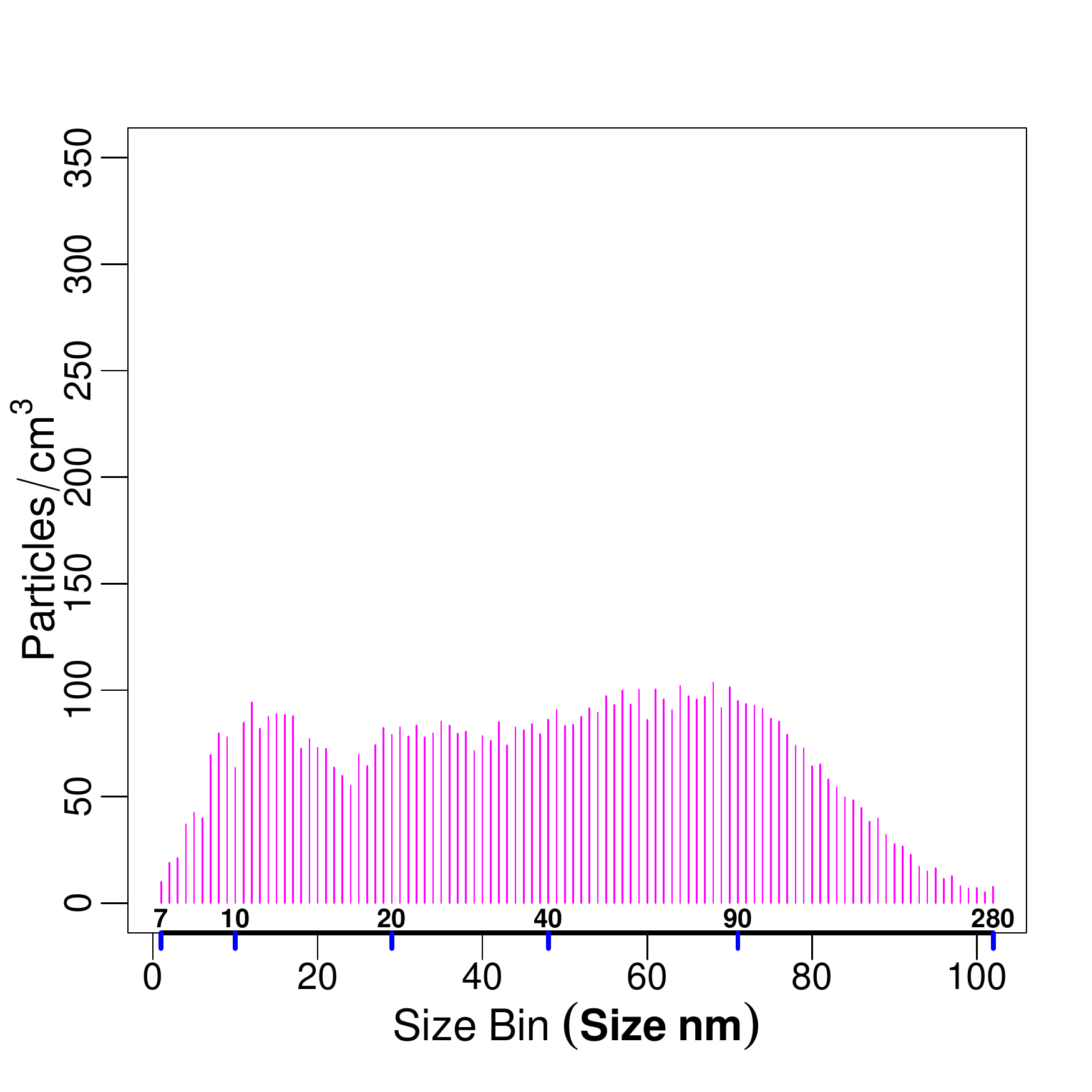}} \\
\subfloat[]{\includegraphics[width=7.5cm]{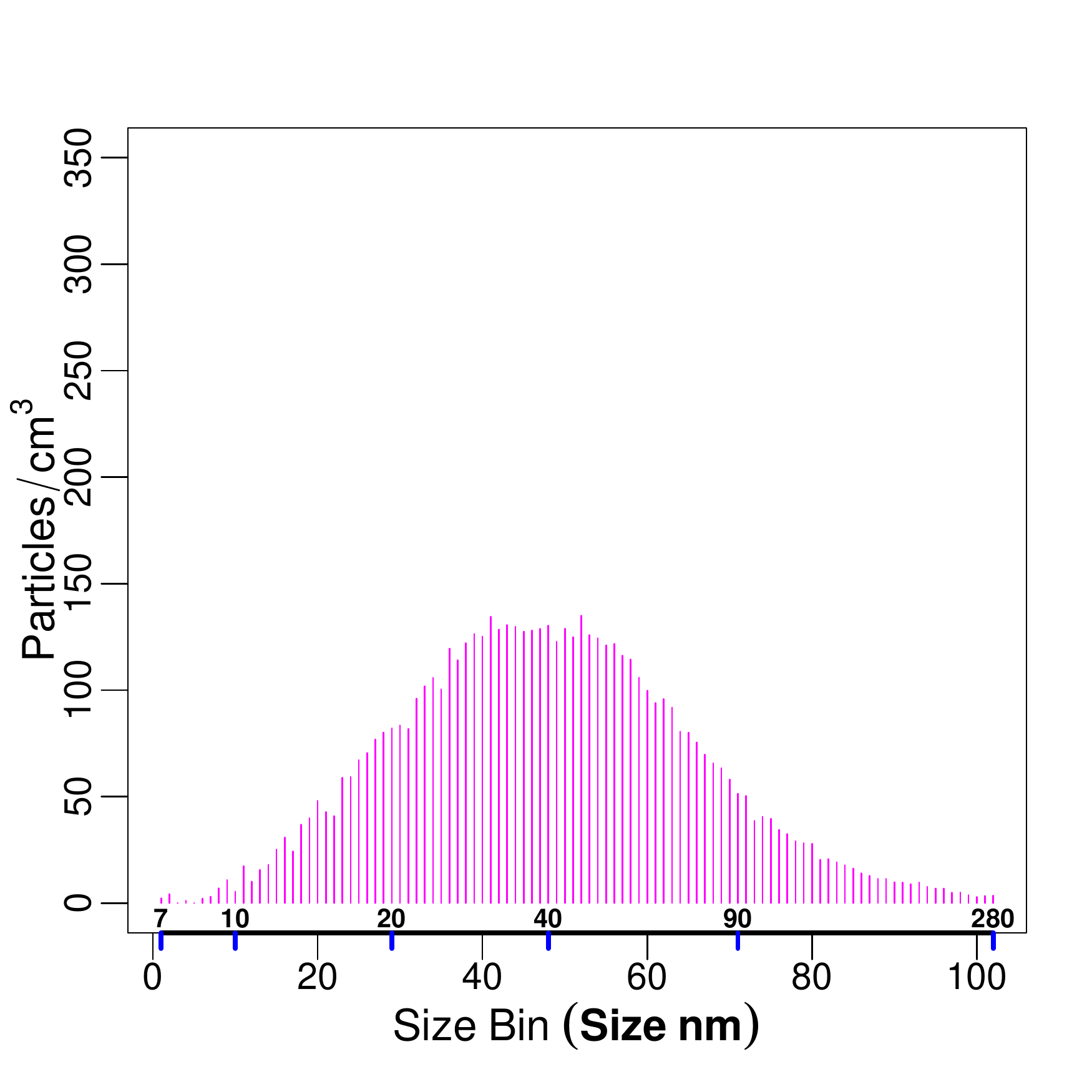}}
\subfloat[]{\includegraphics[width=7.5cm]{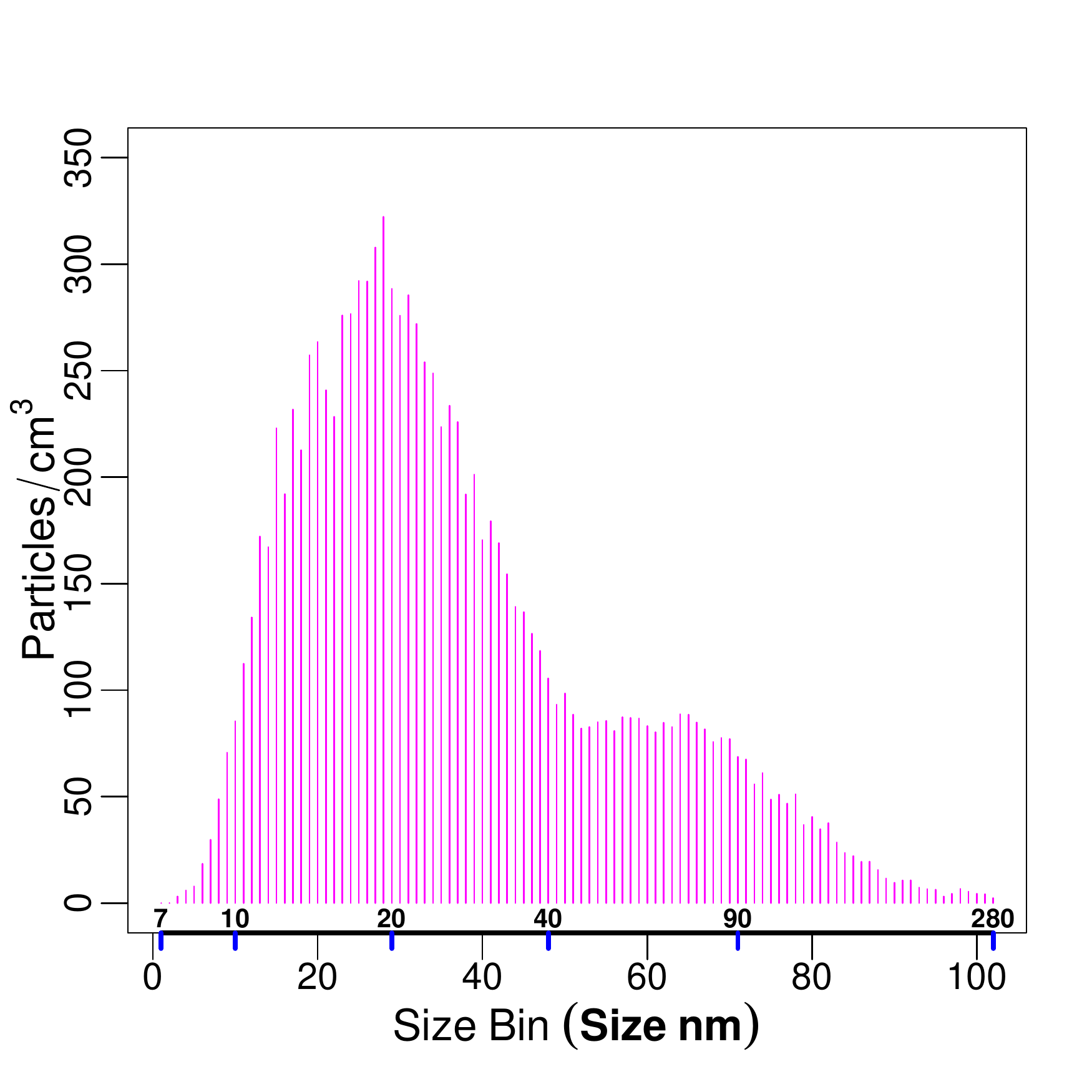}}
 \caption{Figures a) through d) show examples of engine-off UFP size distributions from different runs showing variation in particle counts by run.  Particle counts are shown as a bar chart with labels below the x-axis indicating size bin number and above the x-axis indicating approximate particle diameter (nm).}
\label{fig:sampleufp}
\end{figure}

Figure~\ref{fig:sampleufp} shows examples of engine-off UFP size distributions from different runs showing variation in particle counts by run. Particle counts are shown as a bar chart with labels below the x-axis indicating size bin number and labels just above the x-axis indicating particle diameter (nm).  

\begin{figure}[!ht]
  \centering
  \includegraphics[width=0.9\textwidth]{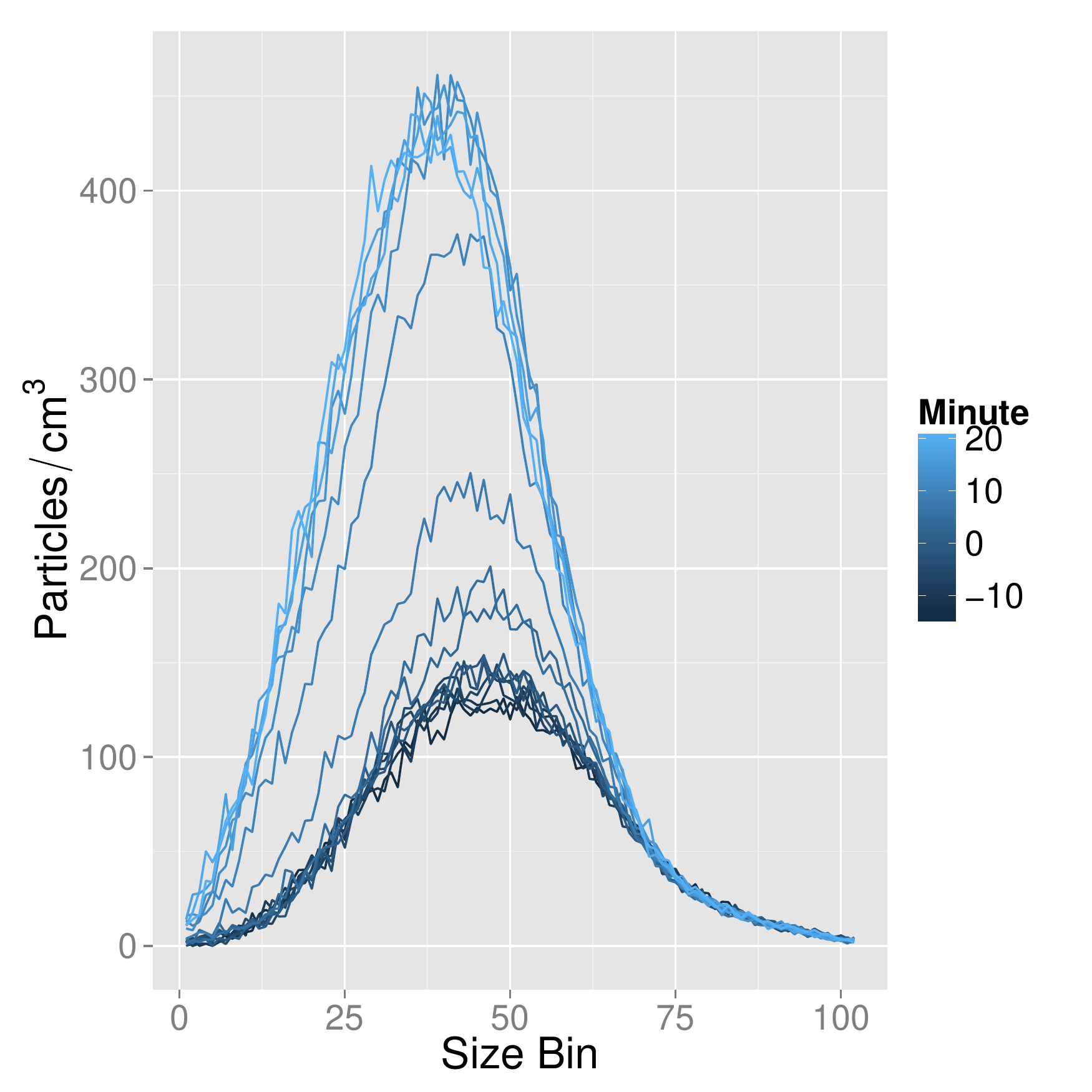}
  \caption{Plot of all UFP size distributions for one sample run. Time is measured in minutes from when the engine is turned on meaning measurements range from approximately -15 to 20 minutes.  The 7 UFP size distributions collected before the engine is turned on are plotted as darker curves and the 9 UFP distributions collected after the engine is turned on are plotted as progressively lighter curves.  UFP size distributions become increasingly more peaked the longer the engine runs. The rate of increase in particle counts after the engine turns on varies greatly by size bin, with little to no increase seen above bin 70, and much larger increases in the 10-60 size bin range.  This particular UFP size distribution has only one mode which occurs at smaller size bins as the engine continues to run while nearly tripling in magnitude.}
  \label{fig:ufpdist_overtime}
\end{figure}

Figure~\ref{fig:ufpdist_overtime} shows a plot of size distributions for a single run. Time is measured in minutes from when the engine is turned on and ranges from approximately -15 to 20 minutes.  The 7 UFP size distributions collected before the engine is turned on are plotted as darker curves and the 9 UFP distributions collected after the engine is turned on are plotted as progressively lighter curves.  UFP size distributions become increasingly more peaked the longer the engine runs. The rate of increase in particle counts after engine-on varies greatly by size bin, with little to no increase seen above bin 70, and much larger increases in the 10-60 size bin range.  This particular set of UFP size distributions has only one mode, and that mode occurs at smaller size bins as the engine runs while the height nearly triples in magnitude.  

\begin{figure}[!ht]
\centering
\subfloat[]{\includegraphics[width=7.5cm]{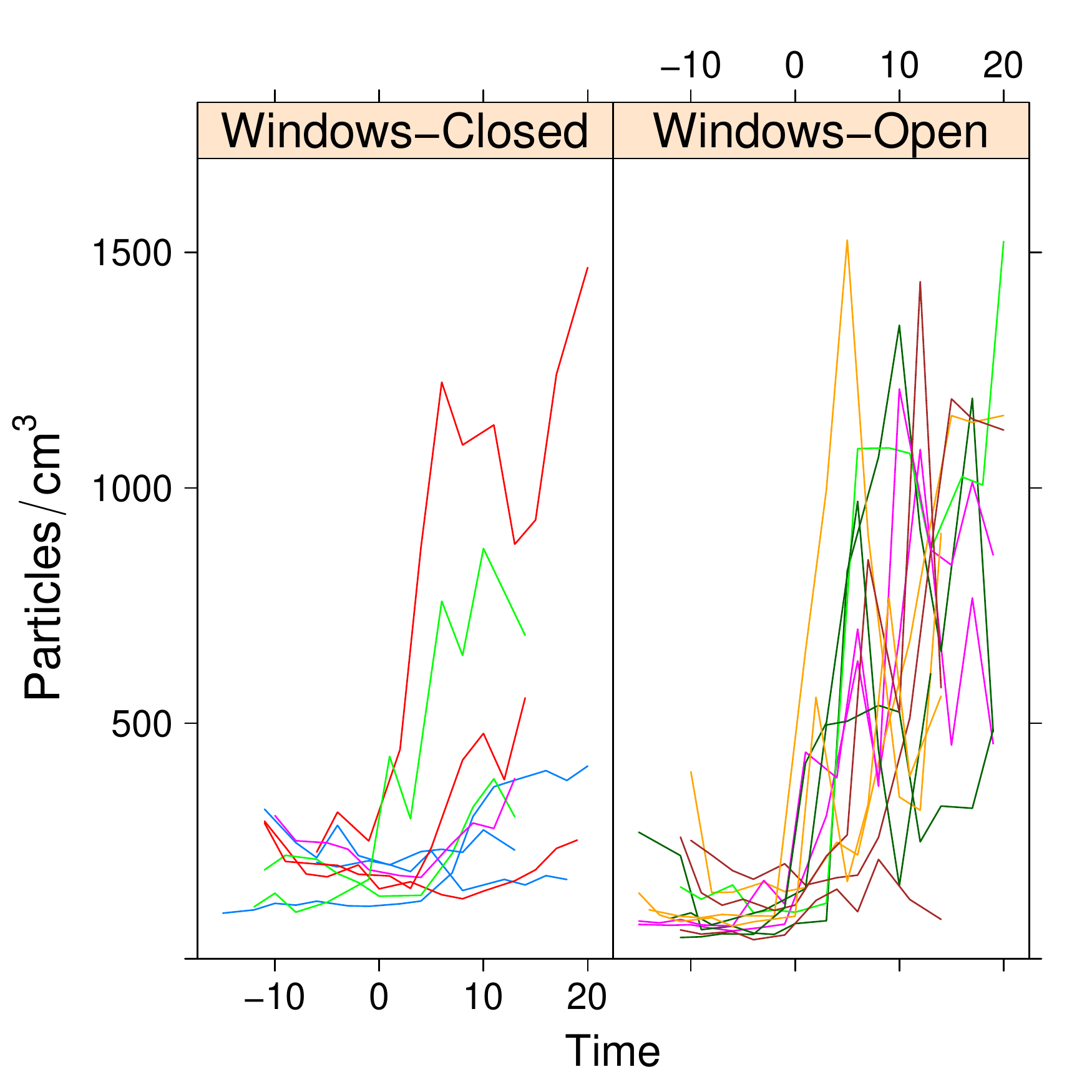}}\\
\subfloat[]{\includegraphics[width=7.5cm]{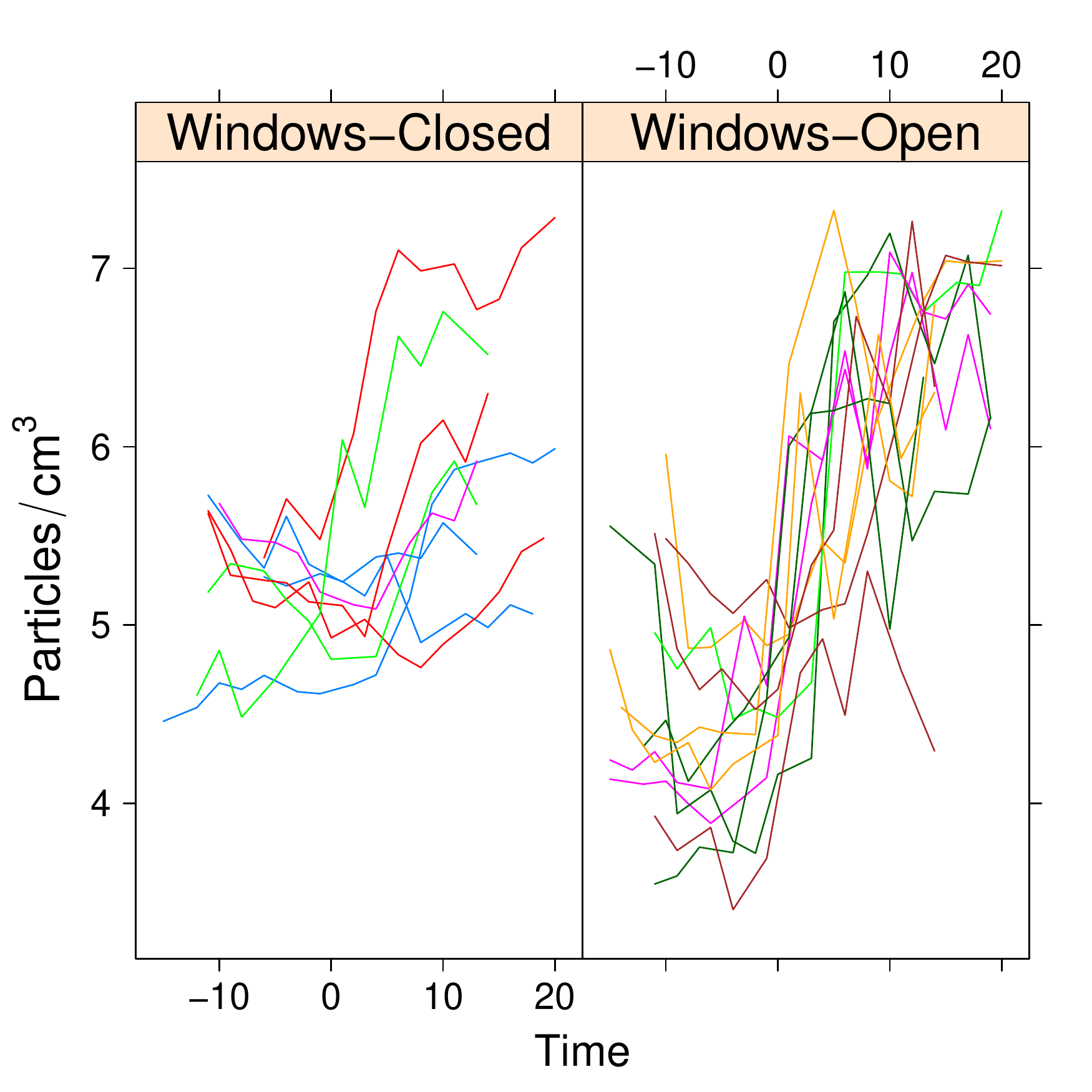}}
 \caption{Figures (a) and (b) show particle counts and log particle counts over time, respectively, for size bin 30 (20.9-21.7 nm) for all runs by window position.  Each line is a separate run. Counts for size bin 30 generally increase sharply when the engine first turns on then continue to increase at slower rates thereafter, though in some cases increases are not seen, particularly for the window closed position. A log transformation allows for easier temporal modeling of counts.}
\label{fig:size30overtime}
\end{figure}

Figure~\ref{fig:size30overtime}(a) and Figure~\ref{fig:size30overtime}(b) plot particle counts and log particle counts over time for size bin 30 (20.9-21.7 nm) for all runs by window position. Each line is a separate run. Counts for size bin 30 generally increase sharply when the engine first turns on then continue to increase at slower rates thereafter, though in some cases increases are not seen, particularly for the window closed position.  A log transformation allows for easier temporal modeling of counts~\citep{whitby91,wraith09,wraith13}.

\section{A Time Series Semi-Parametric Model for UFP Size Distributions}

Let $i$ index run, where $i=1 \ldots R$ and for our data $R=21$. Let $s$ index particle size bin, with $s=1 \ldots S$ and for our data $S=102$. Time, $t$, has a run dependent range of $t_{\min,i}$ to $t_{\max,i}$.  Time is measured in minutes and defined so that usually $t_{\min_,i}=-14$ or $-15$, always the engine is turned on at $t=0$, and usually $t_{\max,i}=19$ or $20$; there is modest variation by run for $t_{\min_,i}$  and $t_{\max_,i}$. Baseline refers to time before engine-on, when $t<0$. Let $z(i)$ be an indicator of window position where $z(i)=1$ corresponds to windows-open and $z(i)=0$ to corresponds to windows-closed. We write $z\equiv z(i)$ to simplify notation. The window position should only affect measurements after $t=0$, not before. Outcome $y_{ist}$ is the natural log of particle count plus 10 for run $i$, size bin $s$ at time $t$.

\subsection{Model}
Before engine-on, baseline mean log counts are expected to be constant over time and are modeled by a hierarchical model with random run intercepts which vary as a function of size bin $s$. At baseline, we expect $y_{ist}$ to have a size bin specific population mean, $\alpha_{s}$, and size bin specific random intercept, $\gamma_{is}$.  For $t<0,~y_{ist}$ are modeled as
	\begin{align} y_{ist} &= \alpha_{s} + \gamma_{is} +u_{ist}. \end{align}
\noindent
Residuals $u_{ist}$ are discussed shortly.

After engine-on, the $y_{ist}$ increase additively from baseline levels. Let $\vect{f(t)}$ be a $J \times 1$ vector of functions of $t$, with $j$-th element
\begin{align}
f_j(t) = \left\{ \begin{array}{rl}
 0 &\mbox{ $t<0$} \\
  m_j(t) &\mbox{ $t \geq 0.$} 
       \end{array} \right. \label{timedef}
\end{align}

\noindent
For a model with quadratic trend, $J=2$, $m_1(t)=t$ and $m_2(t)=t^2$.  Other choices for $\vect{f(t)}$ are discussed in Section \ref{sec:timetrend}.  

Engine-on increases in mean vary by window position $z$.  Let $\vect{\delta_{zs}}$ be a $J \times 1$ vector of size bin and window position specific regression coefficients for $\vect{f(t)}$.  Then, for given $s$ and $z$, the increase in mean log particle count from baseline is $\vect{\delta_{zs}^Tf(t)}$.

Residuals $u_{ist}$ are modeled with an autoregressive process over time, conditioning on the first observation. The model for any time $t$ is  
	\begin{align} y_{ist} &= \alpha_{s} + \gamma_{is} + \vect{\delta_{zs}^Tf(t)}+u_{ist}\\ u_{ist} &= \theta u_{i,s,t-1} + \epsilon_{ist}, ~~ \label{resideq} \\ \epsilon_{ist} &\sim N(0, \sigma^2_{s})  \label{erroreq} \end{align}
\noindent
where lag one correlation is $\theta$, and $\sigma^2_{s}$ is the variance of $\epsilon_{ist}$ given the previous residual $u_{i,s,t-1}$.

We want the baseline engine-off population mean $\alpha_s$, baseline random intercept $\gamma_{is}$, and the coefficients for the time trends $\delta_{zs}$ to vary smoothly by size bin $s$ and we model $\alpha_s$, $\gamma_{is}$, and $\delta_{zs}$ as cubic B-spline functions of $s$. Let $\vect{B(s)}$ be a cubic $(K \times 1)$ B-spline basis over size bin $s$ with $(K-4)$ knots.  Let $\vect{\alpha}$ be the coefficients of $\vect{B(s)}$ for the population mean, then 
	\begin{gather*} \alpha_s=\vect{\alpha^TB(s)},  \end{gather*}
\noindent
and let $\vect{\gamma_i}$ be the coefficients of $\vect{B(s)}$ for random intercept for run $i$, then 
	\begin{gather*} \gamma_{is}=\vect{\gamma_i^TB(s)}.  \end{gather*}
\noindent
The vectors of random coefficients describing baseline UFP size distribution variation by run, $\vect{\gamma_i}$, have distribution
\begin{gather} \vect{\gamma_i} \mid \matr{D} \sim N_K(0,\matr{D}), \label{randomeq}\end{gather}
	
\noindent
for, $i=1, \ldots, R$. Together $\vect{\alpha^TB(s)}+\vect{\gamma_i^TB(s)}$  model the baseline UFP size distribution for run $i$ and size bin $s$. 

Let $\delta_{zsj}$ be the $j$-th element of $\vect{\delta_{zs}}$.  Set $\delta_{zsj}=\matr{\Delta_{zj}^TB(s)}$.  We define  $\matr{\Delta_z}$ so that $\vect{\Delta_{zj}}$ is the $j$-th row of $\matr{\Delta_z}$, $j=1 \ldots J$.  Then
\begin{gather}
	\vect{f(t)}^T\matr{\Delta_{z}}\vect{B(s)}= \vect{\delta_{zs}^Tf(t)} \label{engoneq}
\end{gather}
\noindent
There is one $\matr{\Delta_z}$, $J \times K$, for each window position, closed, $\matr{\Delta_0}$, and open, $\matr{\Delta_1}$.  

The error variance $\sigma^2_{s}$,  is not constant and needs to vary smoothly over size bin.  The log variance is modeled with a separate quadratic $(L \times 1)$ B-spline basis over size bin, $\vect{B_{\err}(s)}$, with $(L-3)$ knots. Log variance is modeled
	\begin{align} \log(\sigma^2_{s})&=\vect{\eta}^T\vect{B_{\err}(s)}+ w_{s} \label{normreseq1} \\ w_{s} &\sim N(0,\tau^2_\eta) \label{normreseq2}\end{align}
\noindent
where the $w_{s}$ are small normal residual terms with known variance $\tau^2_\eta$ added to $\vect{\eta}^T\vect{B_{\err}(s)}$ to enable efficient MCMC sampling~\citep{ciprian07, bal05,mcmcglmm}.  We set $\tau^2_\eta$ to a small number to not add too much variation.

Now we discuss time trend models after engine-on.			
\subsection{Time Trend Models} \label{sec:timetrend}

Because the parametric form of count increases after the engine is on has not been explored in current research, we consider several models for the time trend after the engine-on. 

\subsubsection{Quadratic Time Trend} \label{sec:quadsec}

For the model with quadratic trend, $J=2$, $m_1(t)=t$, $m_2(t)=t^2$ and $\matr{\Delta_{z}}$ is a $2 \times K$ matrix with rows of coefficients of $\vect{B(s)}$ for the linear and quadratic time trend terms.

\subsubsection{Bent Line Time Trend} \label{sec:knotsec}

The change in counts after the engine turns on follow a linear spline with single knot, allowing the slope to change at time $t_{\Omega}$.  Here $J=2$, $m_1(t)=t$, and $m_2(t)=\max[0,t-t_{\Omega}]$.  We consider knots at $t_{\Omega}=8,10$ or $12$.

\subsubsection{Jump Models with a Quadratic or Linear Spline Time Trend} \label{sec:jumpsec}

We consider an immediate jump in counts when the engine first turns on followed by counts that change following either a quadratic or bent line time trend.  The same knot times $t_{\Omega}=8,10,12$ are considered.  For example, the jump model with bent line time trend and knot at $t_\Omega=8$ has $J=3$, $m_1(t)=1$ and $m_2(t)=t$ and $m_3(t)=\max[0,t-8]$.  

\subsection{Random Jumps for the Time Trend}

Initial log count increases modeled by the jump trend varied across run in part because the engine was turned on at different times relative to time $t$ across runs. For the jump models, we considered random jumps as part of the time trend to determine if further improvements in model fit were possible. The random jump time trend uses the same B-spline basis $\vect{B(s)}$ to vary time trend coefficients by size bin. Let $\vect{g(t)}$ be a $G \times 1$ vector of functions of $t$, defined analogously to $\vect{f(t)}$ in Equation (\ref{timedef}). Let $\matr{\Upsilon_{i}}$ be defined as $\matr{\Delta_z}$ in Equation (\ref{engoneq}), except $\matr{\Upsilon_{i}}$ varies by run, not window position, and $\matr{\Upsilon_{i}} \mid \matr{W} \sim N(0,\matr{W})$.  A model with a random jump would have $\vect{g(t)}=1$, $G=1$, and $\matr{\Upsilon_i}$ be a $K \times 1$ random vector. This notation allows for more complex random effects after engine-on, however residual analysis did not indicate a need for additional random effects, see web appendix figures A.4, A.5, and A.6.

The model now takes the form
			\begin{equation} y_{ist} = \vect{\alpha^T B(s)}+\vect{\gamma_i^T B(s)} + \vect{f(t)}^T\matr{\Delta_{z}}\vect{B(s)}+ \vect{g(t)}^T\matr{\Upsilon_{i}}\vect{B(s)} +u_{ist}, \label{maineq_final}\end{equation}
\noindent
along with Equations (\ref{resideq}), (\ref{erroreq}), (\ref{randomeq}), (\ref{normreseq1}), and (\ref{normreseq2}). 

Models with only quadratic or bent line time trends are referred to as \emph{non-jump} models, while models with an engine-on jump are referred to as \emph{jump} models. These models are described by Equation \ref{maineq_final} without the term $\vect{g(t)}^T\matr{\Upsilon_{i}}\vect{B(s)}$.  Models with the engine-on random jump are referred to as \emph{random jump} models and are described by Equation \ref{maineq_final}.

\subsection{Priors}

Proper priors are used on all parameters.  

Priors for the fixed effects $\vect{\alpha}$ and $\vect{\Delta_{zj}}$, $j=1 \dots J$, are
			\begin{align} \vect{\alpha} \mid \tau_{\alpha} &\sim N_{K}(0,\tau_{\alpha}^2\matr{\I_{K}}),  \\ \vect{\Delta_{zj}} \mid \tau_{\delta} &\sim N_{K}(0,\tau_{\delta}^2\matr{\I_{k}} ),\end{align}
\noindent
where $\matr{I_K}$ is the $K \times K$ identity matrix.   Both $\tau_{\alpha}^2$ and $\tau_{\delta}^2$ are set to large values so that the priors are non-informative.

The prior for the correlation parameter $\theta$ is taken to be truncated normal as conjugacy improves computational speed
\begin{gather} \theta \mid \mu_{\theta},\sigma^2_{\theta} \sim N(\mu_{\theta},\sigma^2_{\theta}) \I[\theta \in (-1,.9)]\end{gather}
\noindent
with the indicator function giving the lower $-1$ and upper $.9$ truncation points. The upper limit of $0.9$ was chosen to avoid singularity in the likelihood~\citep{palmer96}.  Covariance matrices $\matr{D}$ and $\matr{W}$ are given Inverse-Wishart priors
\begin{align} \matr{D} &\sim \invW(n_D,(n_D-K-1)*\matr{M}), \\
  \matr{W} &\sim \invW(G*K+1,\matr{I_{G*K}}),  \end{align}
\noindent
where the known $K \times K$ scale matrix $\matr{M}$ uses information from 65 other engine-off experiments reported in~\citet{zhang12} but not used in the current analysis: a linear mixed effects model using $\vect{B(s)}$ for the fixed and random UFP size distribution effects was run using log counts from this additional data.  The estimated covariance matrix for the random effects from this fit was used as the scale matrix $\matr{M}$ in the Inverse-Wishart prior and we set $n_D=65$.  This allowed information available about typical correlations in UFP size distribution shapes from urban environments to be used, supplementing the sparse information available in the current dataset with only 21 runs and a $K=7$ dimensional B-spline basis. Unfortunately there was no prior information as easily available for $\matr{W}$, we set the scale matrix to give marginal uniform distributions to the correlation coefficients with low degrees of freedom and a prior mean of one for the variances.

The prior for the coefficients of the log variance is taken to be
			\begin{equation} \vect{\eta} \sim N_L(\log(0.4),g^2\matr{I_L}) \end{equation}
\noindent
where $0.4$ was the mean residual error reported in the experiments of~\citet{zhang12} not examined here,  $\matr{I_L}$ is the $L \times L$ identity matrix, the hyperparameter $g$ is set to put a 95 probability around residual variances between $0.2$ and $0.7$. The prior is centered at a constant residual across size bins.
\subsection{Computing Overview} \label{sec:posgam}

Posterior estimates are obtained via Markov Chain Monte Carlo (MCMC) simulations~\citep{hastings70,gelfand90} using JAGS 3.3.0 \citep{Plummer03jags:a}. The MCMC convergence and mixing properties were assessed by visual inspection of the autocorrelation and chain histories of individual parameters. Convergence appeared to be immediate for all parameters. 

\subsection{Estimates of Interest}

Let $\vect{Y}$ be the vector of all $y_{ist}$.  To investigate model fit, we plotted subsets of Equation \ref{maineq_final}: baseline UFP size distributions, $\vect{\alpha^T B(s)}+\vect{\gamma_i^T B(s)}$ and engine-on posterior time trend changes, $\vect{f(t)}^T\matr{\Delta_{z}}\vect{B(s)}$. Integrating out random engine-off and engine-on effects, predicted mean log counts at size bin $s$ and time $t$ are $\psi_{st} = \vect{\alpha^T B(s)}+ \vect{f(t)}^T\matr{\Delta_{z}}\vect{B(s)}$

For emission monitoring purposes, posterior predictions must be transformed back to their original scale. 
Conditional on run $i$ and random effects, predicted counts have marginal variance, $\lambda^2_s \equiv \var(y_{ist} \mid \vect{\gamma_i},\matr{\Upsilon_{i}})$ that is the same for all times $t$.  Then, $\var(u_{ist}\mid \vect{\gamma_i},\matr{\Upsilon_{i}})= \var(\theta u_{i,s,t-1}+\epsilon_{ist} \mid \vect{\gamma_i},\matr{\Upsilon_{i}})$ so that $\lambda^2_s=\theta^2\lambda^2_s+\sigma^2_s$ to give $\lambda_s^2=\sigma^2_s/(1-\theta^2)$. Defining $\rho_s$ as $\vect{B(s)^t}(\matr{D}+\matr{W})\vect{B(s)}$, this implies a marginal predictive variance $\zeta_s^2 \equiv \rho_s + \lambda^2_s$ for size $s$ at new run $i$ for random jump models. No-jump, jump and random jump models define marginal predictive variance equivalently, but no-jump and jump models remove $\matr{W}$ from the definition of $\rho_s$.  The back transformed mean of predicted counts for size bin $s$ at time $t$ is then
 \begin{equation}\mu_{st}=\exp(\psi_{st}+\zeta_{s}^2/2) \label{predmean}\end{equation}

We define $s_t$ as the modal size bin and the modal particle count as $h_t \equiv \argmax(\mu_{st})$, where $\mu_{st}$ is defined in (\ref{predmean}). 

Residuals can be calculated as 
\begin{equation} e_{ist} = y_{ist}-(\vect{\alpha^T B(s)}+\vect{\gamma_i^T B(s)} + \vect{f(t)}^T\matr{\Delta_{z}}\vect{B(s)}+ \vect{g(t)}^T\matr{\Upsilon_{i}}\vect{B(s)}), \label{expres}\end{equation}
\noindent
removing both fixed and random effects \citep{chaloner88,chaloner94}.  Autoregressive error was not removed from Equation (\ref{expres}) to better visualize lack of fit in the models that may be picked up by the autoregressive term $\theta$. Residuals for no-jump and jump models are calculated removing $\vect{g(t)}^T\matr{\Upsilon_{i}}\vect{B(s)}$.

\section{Results}

We present results graphically to convey important model inferences as meaningful representations of conclusions are difficult to convey in table form.

\begin{figure}[!t]
\centering
\subfloat[5 minutes, Non-jump]{\includegraphics[width=5cm]{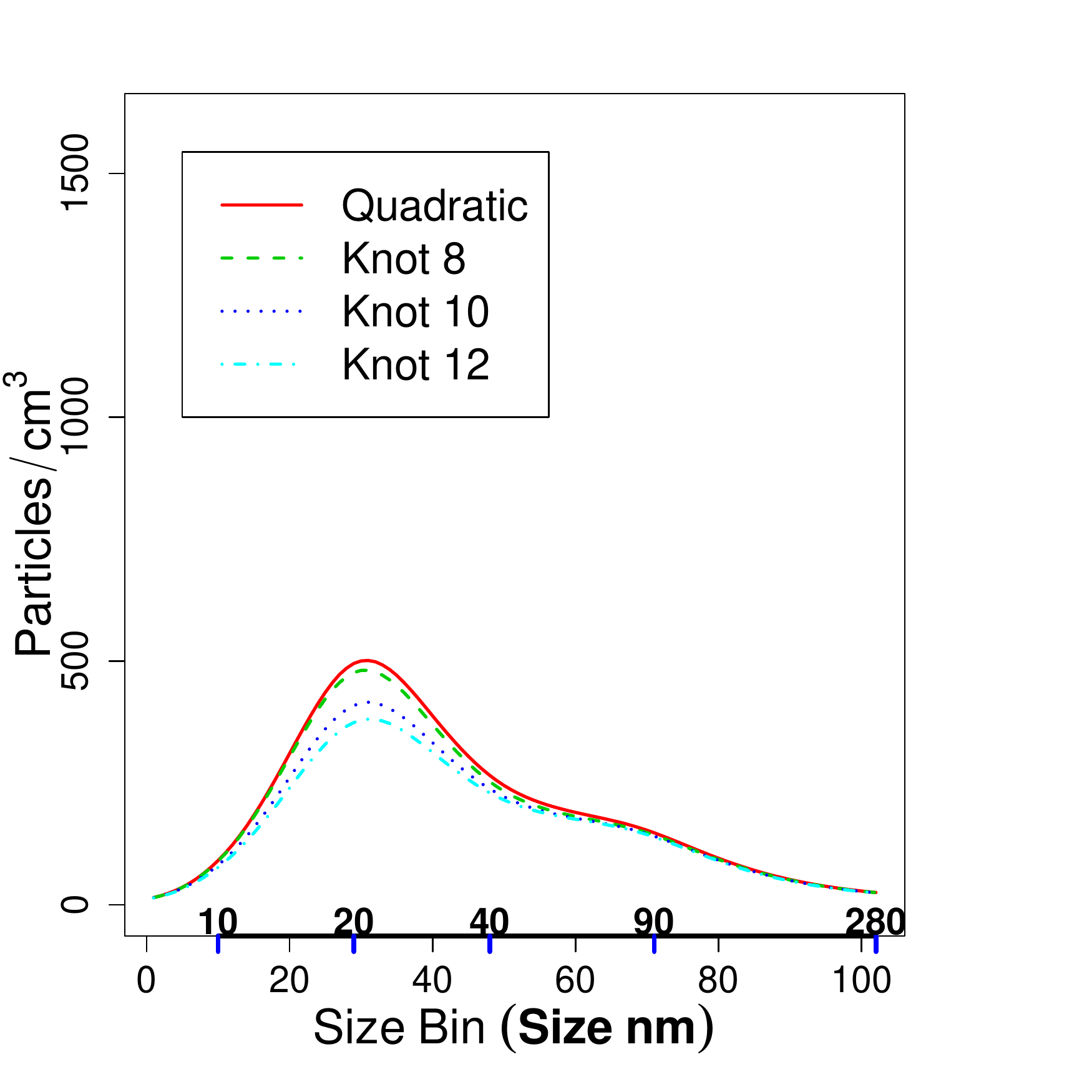}}
\subfloat[5 minutes, Jump]{\includegraphics[width=5cm]{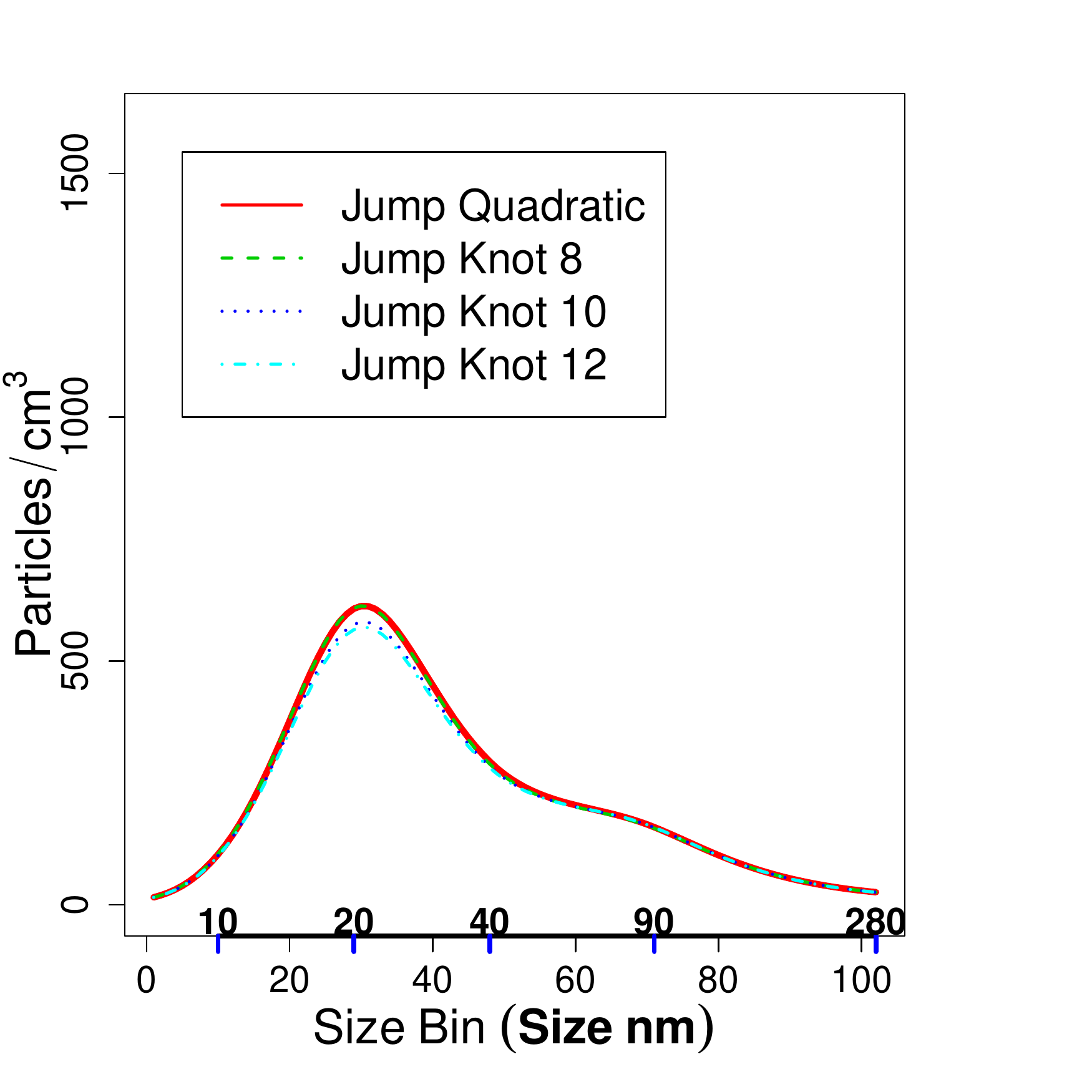}}
\subfloat[5 minutes, Random Jump]{\includegraphics[width=5cm]{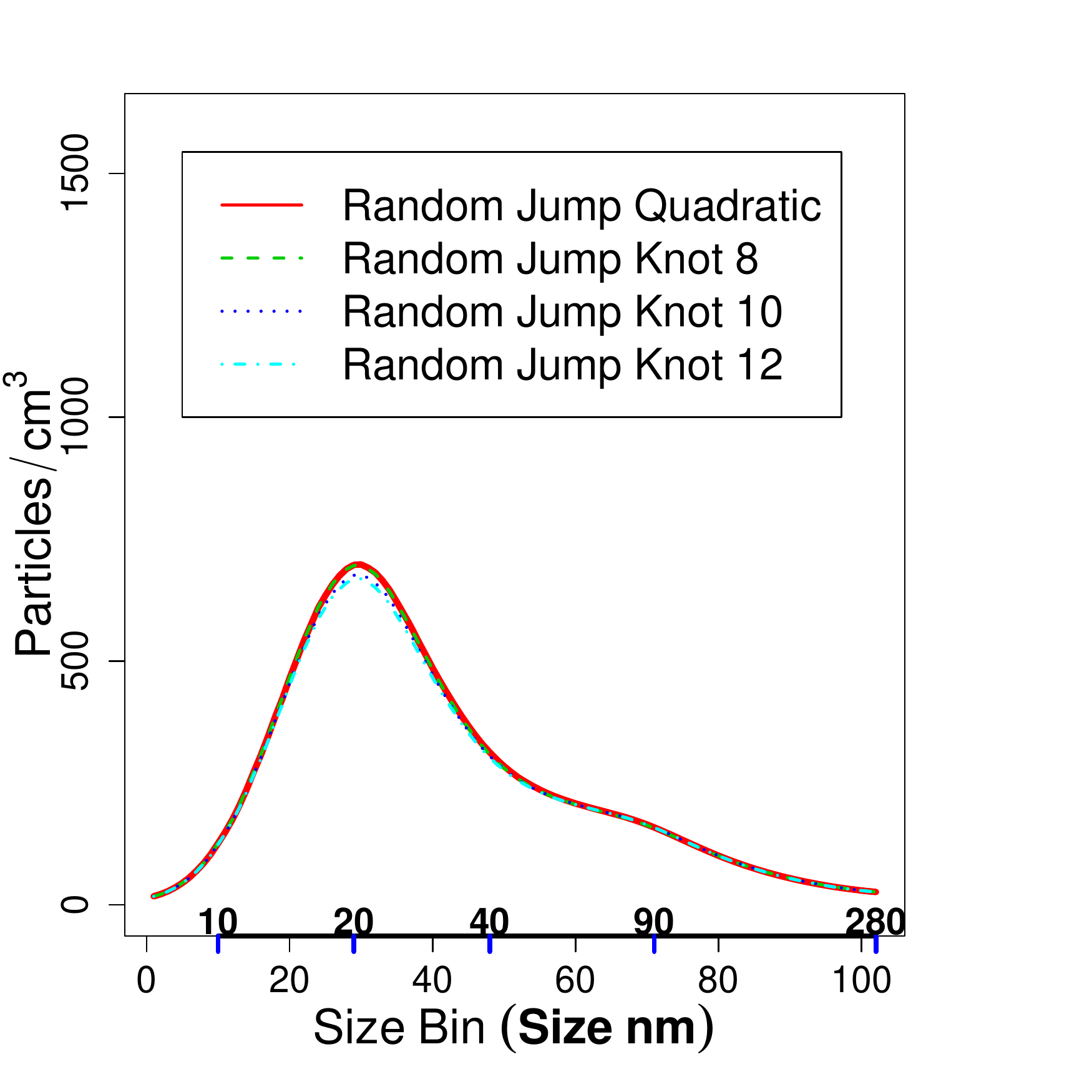}} \\
\vspace{-2\baselineskip}
\subfloat[15 minutes, Non-jump]{\includegraphics[width=5cm]{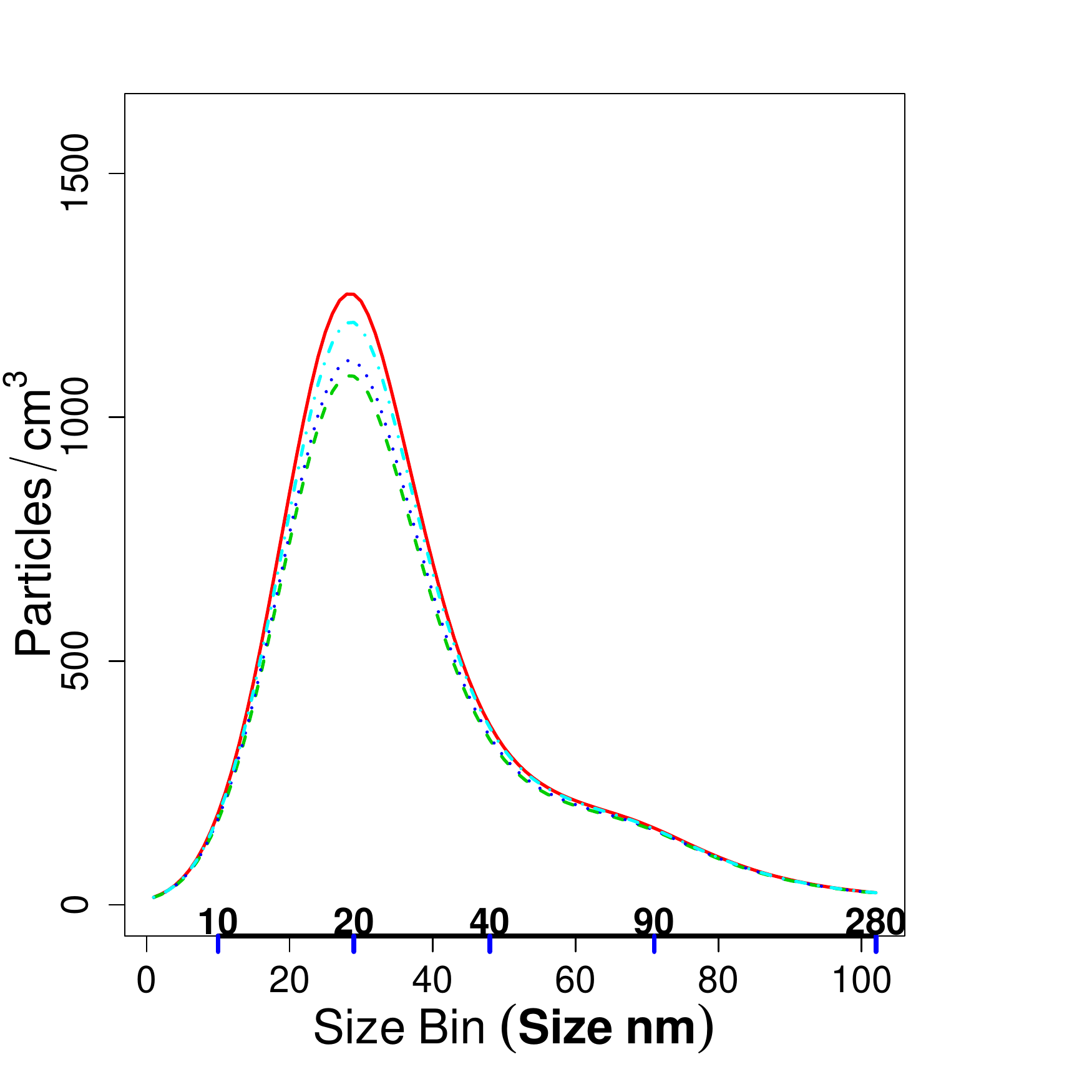}}
\subfloat[15 minutes, Jump]{\includegraphics[width=5cm]{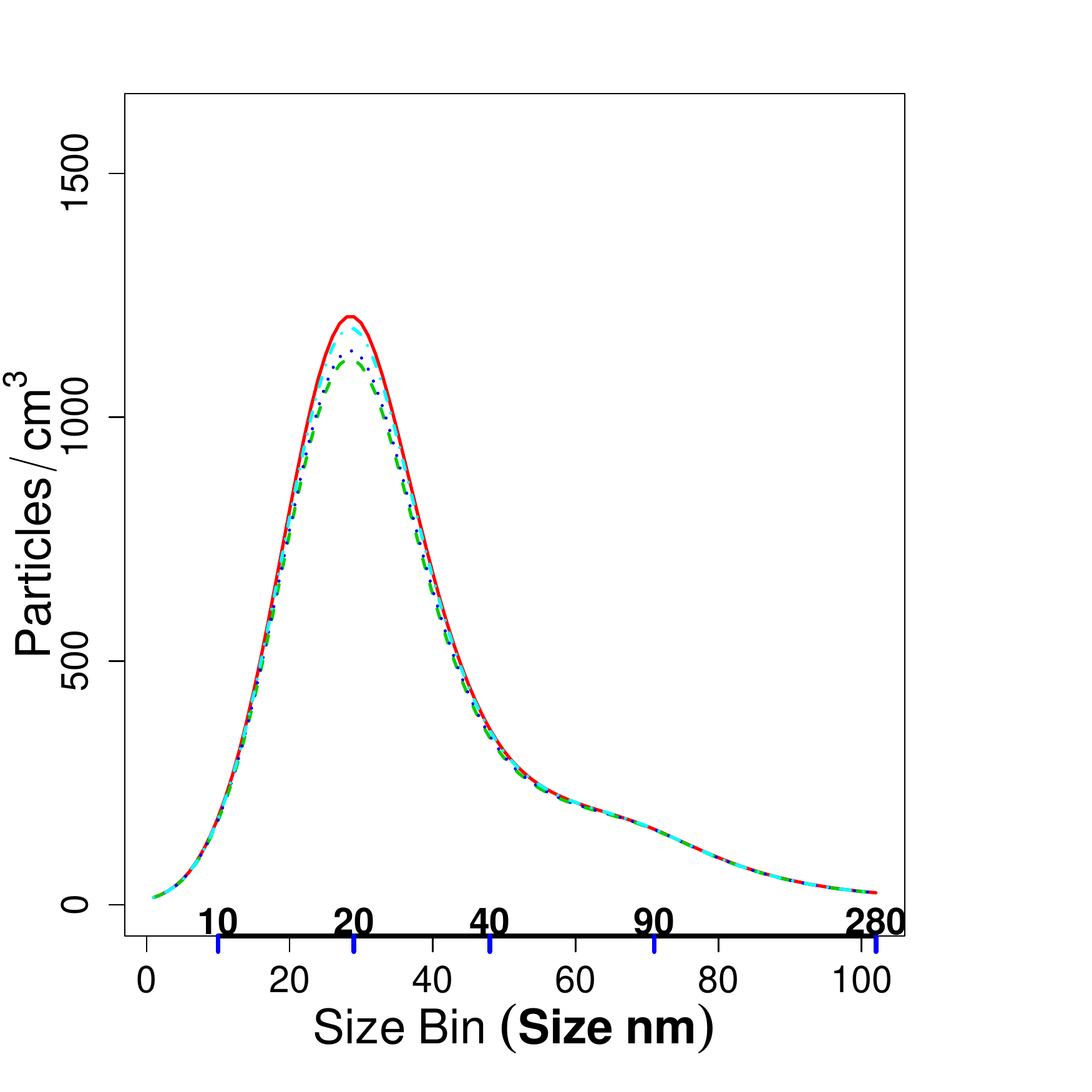}}
\subfloat[15 minutes, Random Jump]{\includegraphics[width=5cm]{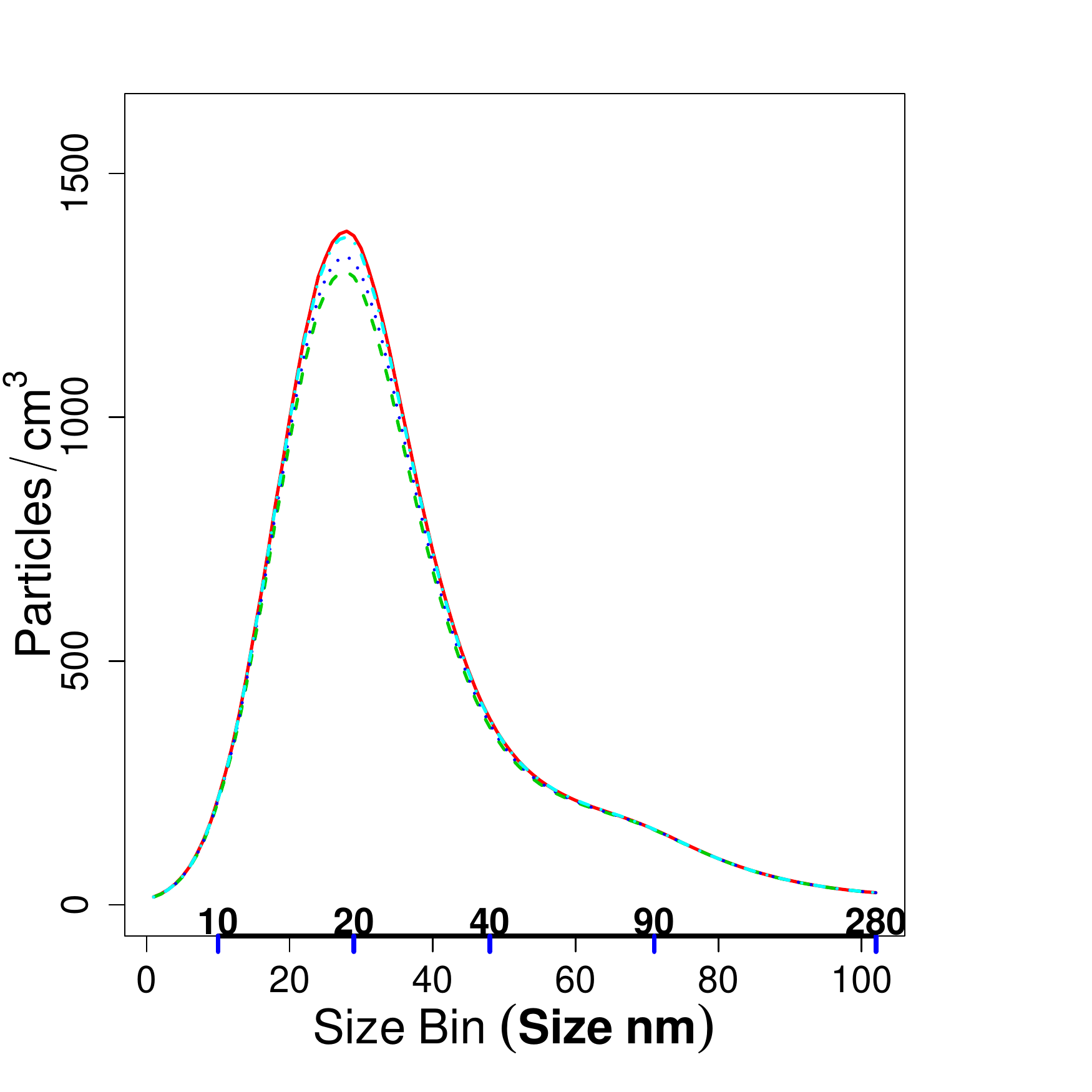}} \\
\vspace{-2\baselineskip}
\subfloat[20 minutes, Non-jump]{\includegraphics[width=5cm]{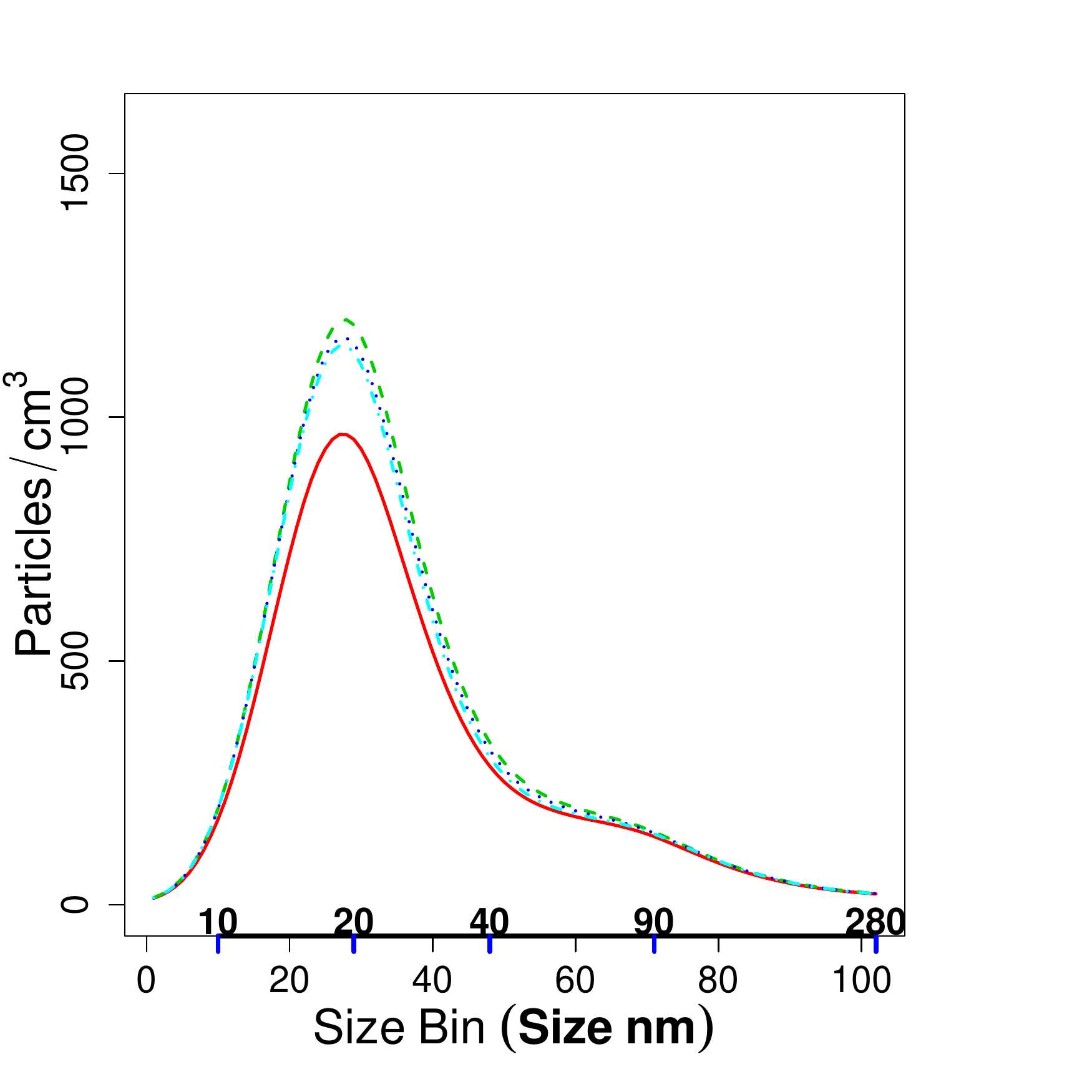}}
\subfloat[20 minutes, Jump]{\includegraphics[width=5cm]{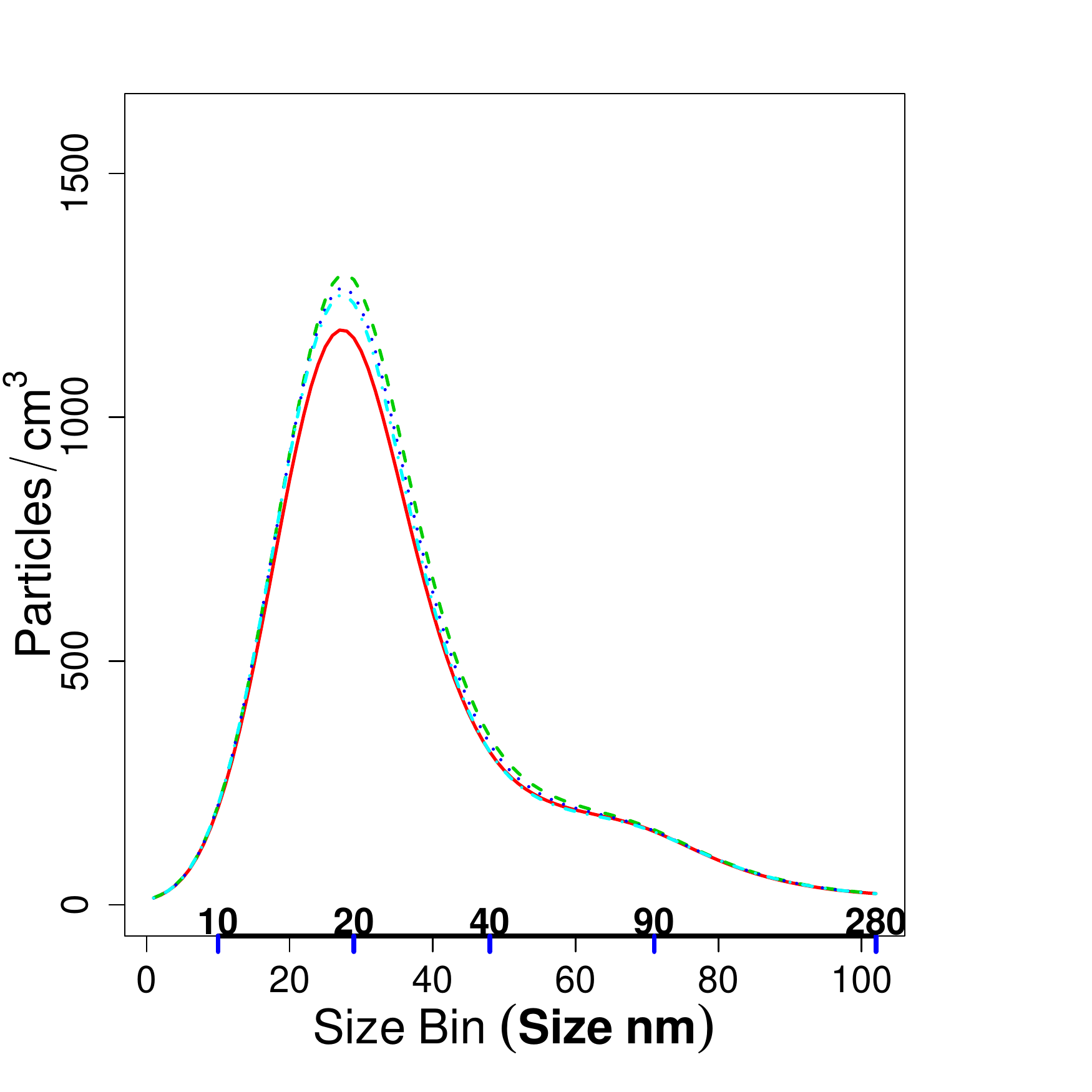}}
\subfloat[20 minutes, Random Jump]{\includegraphics[width=5cm]{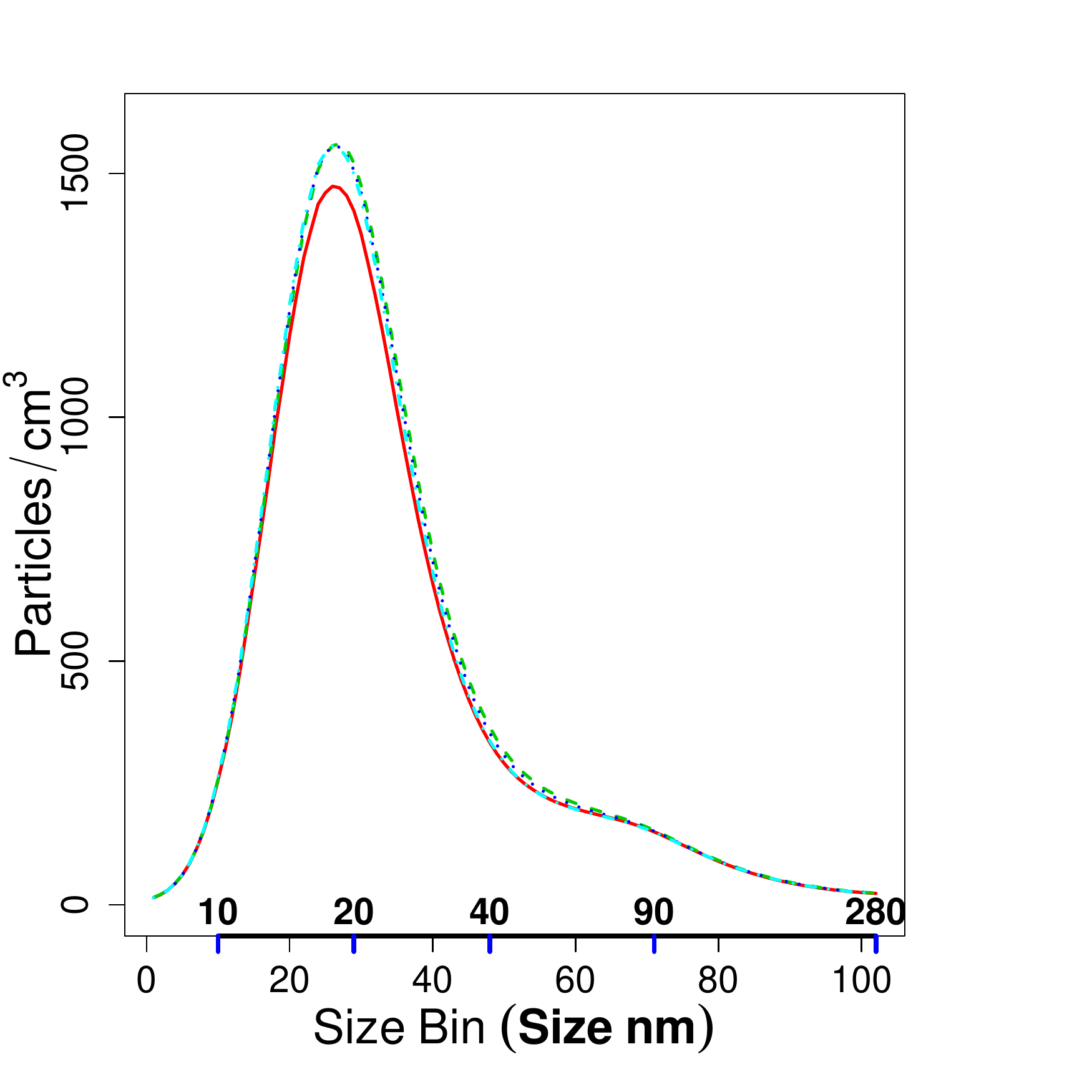}}
 \caption{Mean predictive posterior counts for the windows-open position for all models after the engine has been idling for $t=5$, row 1(a)(b)(c); $t=15$, row 2(d)(e)(f); and $t=20$, row 3(g)(h)(i). Non-jump models are in the left column (a)(c)(c), jump models are in the middle column (d)(e)(f), and random jump models are in the right column (g)(h)(i). The random jump models predict the highest counts at all time periods. The quadratic model differs from knot models in that it predicts counts to eventually decrease over time by $t=20$, however this decrease is diminished with the addition of the jump, and removed with the random jump model. Credible intervals are not plotted to ease visual interpretation, but there is overlap in these intervals for all models.}
\label{fig:windowopenallcurves}
\end{figure}

Figure~\ref{fig:windowopenallcurves} presents window-open posterior predictive means for $\mu_{st}$ as functions of $s$ for no-jump, jump, and random jump models at $t=5,15$ and $20$ minutes after the engine is turned on: each sub-figure has estimates for the quadratic and the three knot models. Predictive mean counts differ between no-jump models, but are very similar within the jump and random jump models. The non-jump quadratic model differs from knot models in that it predicts counts to eventually decrease by time $t=20$, however this decrease is diminished with the addition of the jump, and removed with the random jump. The random jump models give the highest predicted means across size bin for all time points. Credible intervals are not plotted to ease visual interpretation, but there is overlap in these intervals for all models.  The $\mu_{st}$ for the window-closed position are very similar across all models over time and are presented in the web appendix.

Model choice was through a combination of DIC \citep{dic} and graphical examination of residuals. DIC and components for each model are presented in Table~\ref{table:dic}.

\begin{table}[ht] 
\caption{DIC Calculations for Models} 
\centering 
\begin{tabular}{c c c c} 
\hline\hline 
Model & DIC & DIC bar & pD \\ [0.5ex] 
\hline 
Quadratic & ~3,082.3 & ~2,858.7 & 223.6 \\ 
Knot 8 & ~3,509.7 & ~3,285.8 & 223.9 \\ 
Knot 10 & ~3,715.9 & ~3,492.4 & 223.4 \\ 
Knot 12 & ~3,913.6 & ~3,690.9 & 222.7 \\
\hline
Jump Quadratic & ~1,587.2 & ~1,348.6 & 238.6 \\  
Jump Knot 8 & ~1,640.8 & ~1,402.3 & 238.6 \\ 
Jump Knot 10 & ~1,624.4 & ~1,386.0 & 238.4 \\ 
Jump Knot 12 & ~1,614.5 & ~1,376.1 & 238.4 \\ 
\hline
Random Jump Quadratic & -3,928.3 & -4,291.0 & 362.7 \\  
Random Jump Knot 8 & -3,887.6 & -4,250.2 & 362.6 \\ 
Random Jump Knot 10 & -3,903.6 & -4,265.6 & 362.1 \\ 
Random Jump Knot 12 & -3,896.3 & -4,258.8 & 362.4 \\ [1ex] 
\hline 
\end{tabular} 
\label{table:dic} 
\end{table}

The model with the lowest DIC is the preferred model.  The random jump models have much lower DICs than jump models, which have much lower DICs than the non-jump models.  

\begin{figure}[!t]
\centering
\subfloat[Non-jump Quadratic $\bar{e}_{ist}$ Over Time]{\includegraphics[width=6cm]{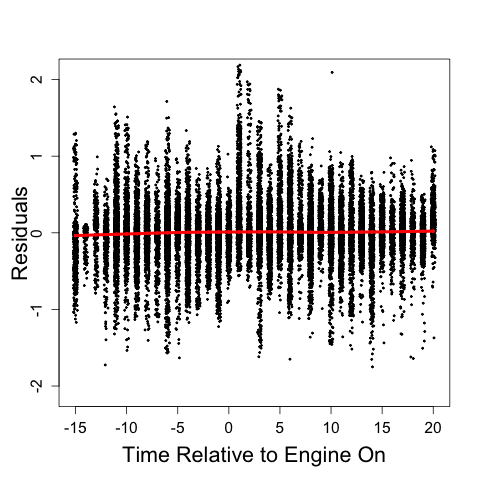}}
\subfloat[Non-jump Quadratic $\bar{e}_{ist}$ Residuals Over Size]{\includegraphics[width=6cm]{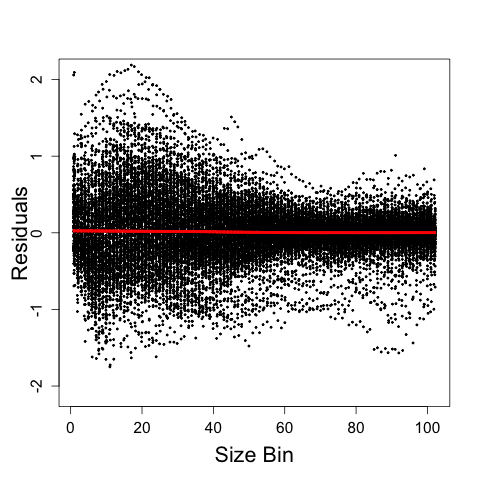}} \\
\subfloat[Jump Quadratic $\bar{e}_{ist}$ Over Time]{\includegraphics[width=6cm]{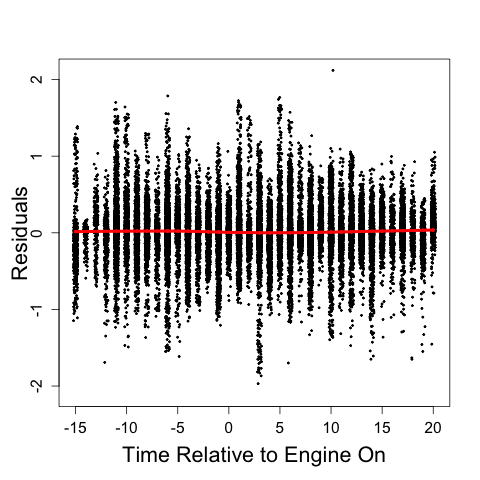}} 
\subfloat[Jump Quadratic $\bar{e}_{ist}$ Over Size]{\includegraphics[width=6cm]{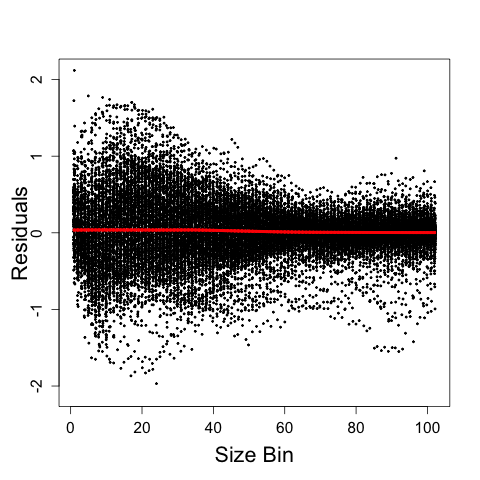}} \\
\subfloat[Random Jump Quadratic Over Time]{\includegraphics[width=6cm]{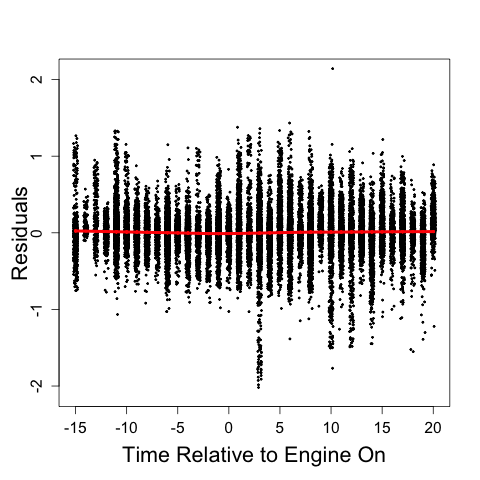}} 
\subfloat[Random Jump Quadratic Over Size]{\includegraphics[width=6cm]{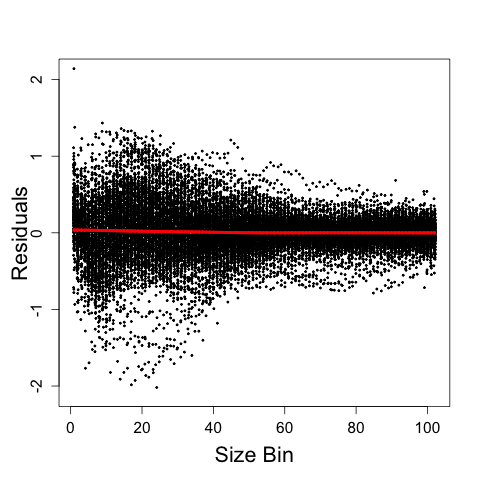}} \\
 \caption{Residuals $\bar{e}_{ist}$ for the quadratic, jump quadratic, and random jump models plotted against time and size bin.  Adding the jump component to the model narrowed the range of mean posterior residuals and reduced outliers.  Adding the random jump component reduced these to an even greater extent.}
\label{fig:res20}
\end{figure}

Models were further evaluated by plotting mean posterior residuals, $\bar{e}_{ist} \equiv E[e_{ist} \mid \vect{Y}]$, by time and by size. Figure~\ref{fig:res20} presents residuals $\bar{e}_{ist}$ for the quadratic, jump quadratic, and random jump models plotted against size and time.  Adding the jump component to the model narrowed the range of mean posterior residuals and reduced outliers.  Adding the random jump component reduced these to an even greater extent. 

The posterior median of the AR parameter $\hat{\theta}$ for the jump quadratic model without any random effect was $.90$ with a $95$ percent credible interval$~(.89,.90)$, recall the prior was truncated at .9.  Adding engine off random effects reduced this to $0.55$ with credible interval $(.54,.56)$, while adding the engine on random jump effect further reduced the estimate to $.37$ with credible interval $(.35,.38)$. Residual estimates and AR estimates for other models show similar patterns. 

The random jump quadratic model had the best (lowest) DIC score and, along with all random jump models, superior fit when looking at the residuals, and the lower estimates for the AR parameter. Additionally, the fit of engine-off random effects was vastly improved. Plots showing improvement in fit are presented in the web appendix.  This model was selected to present further inferences.

\begin{figure}[!t]
\centering

\includegraphics[width=1\textwidth]{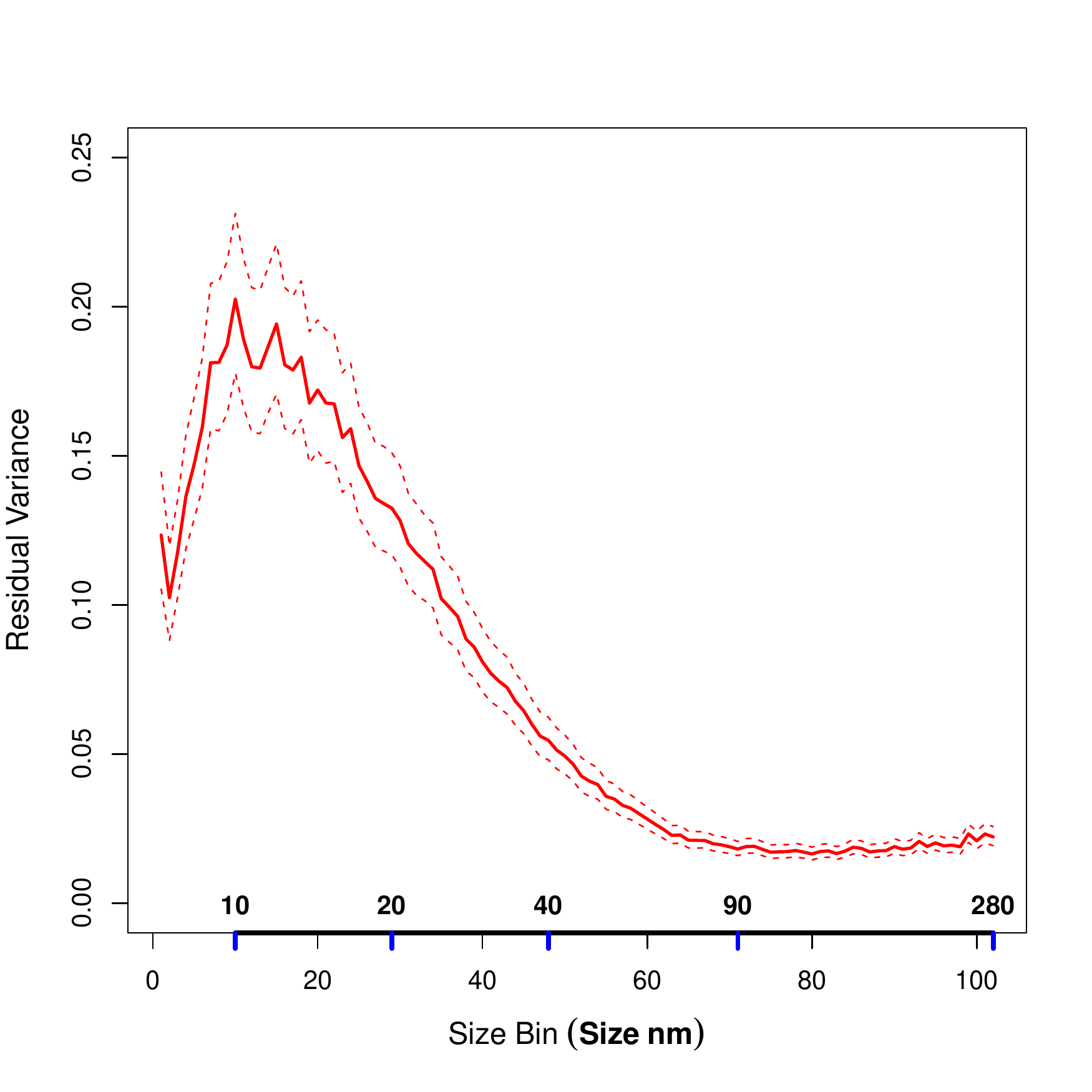}

 \caption{Posterior medians and 95 percent credible intervals for ${\sigma}_s^2$ as a function of size bin.}
\label{fig:ncv}
\end{figure}

Residual variance shows heteroskedasticity over size bin, as shown in Figure~\ref{fig:res20}, and Figure \ref{fig:ncv} plots posterior median and 95 percent credible intervals for $\sigma_s^2$ by bin size.

\begin{figure}[!ht]
\centering
\includegraphics[width=1\textwidth]{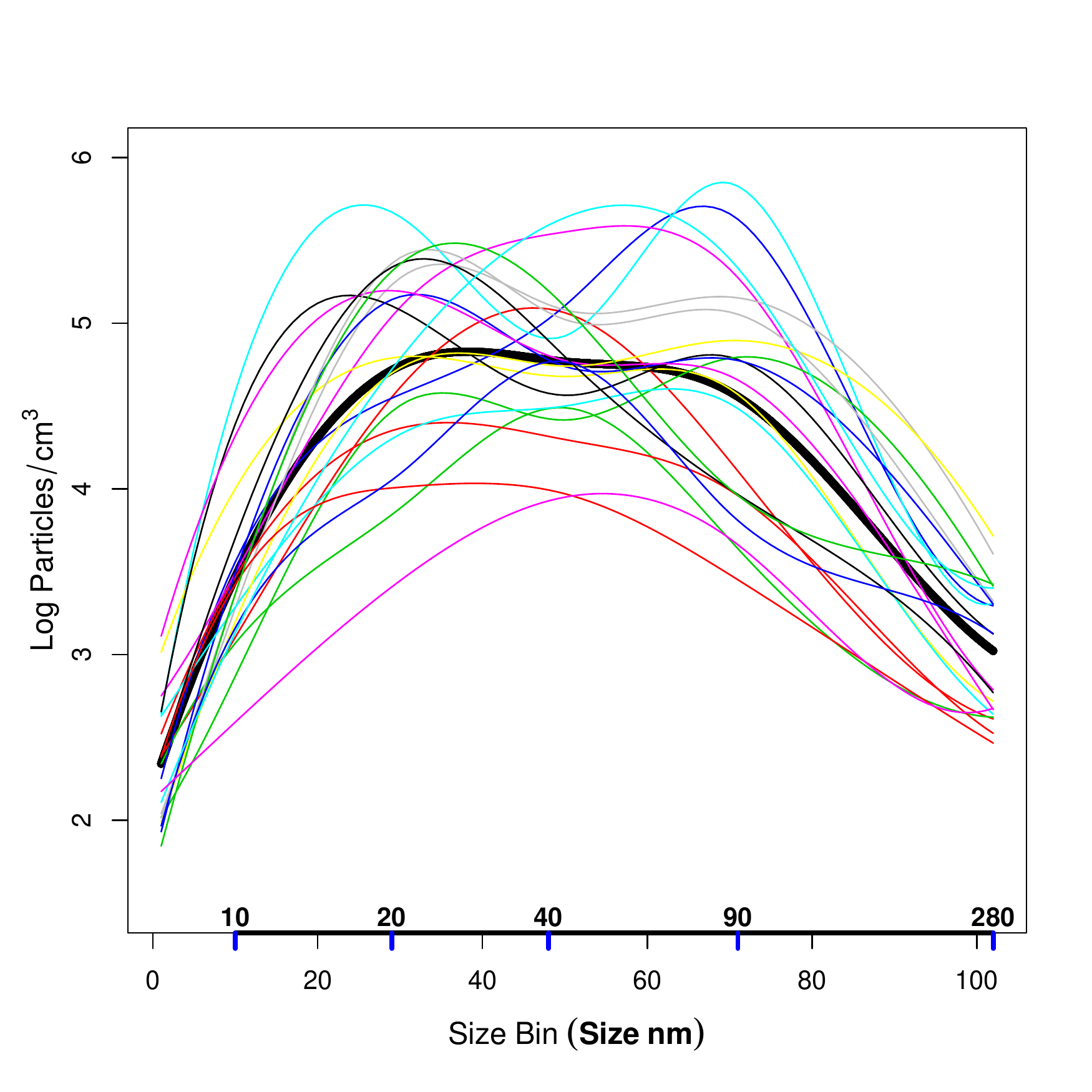}
\caption{Posterior mean log baseline counts by size bin for $t<0$.  The thick black curve is the posterior population mean and the lighter curves show posterior mean log counts by size bin for individual run $i$.}
\label{fig:pred_off_random}
\end{figure}

Figure~\ref{fig:pred_off_random} plots baseline posterior mean log counts by size bin for $t<0$ for all runs $i$. The plots show significant variation in baseline UFP size distributions across run. 

\begin{figure}[!t]
\centering
\subfloat[Initial Engine-On Jump]{\includegraphics[width=7cm]{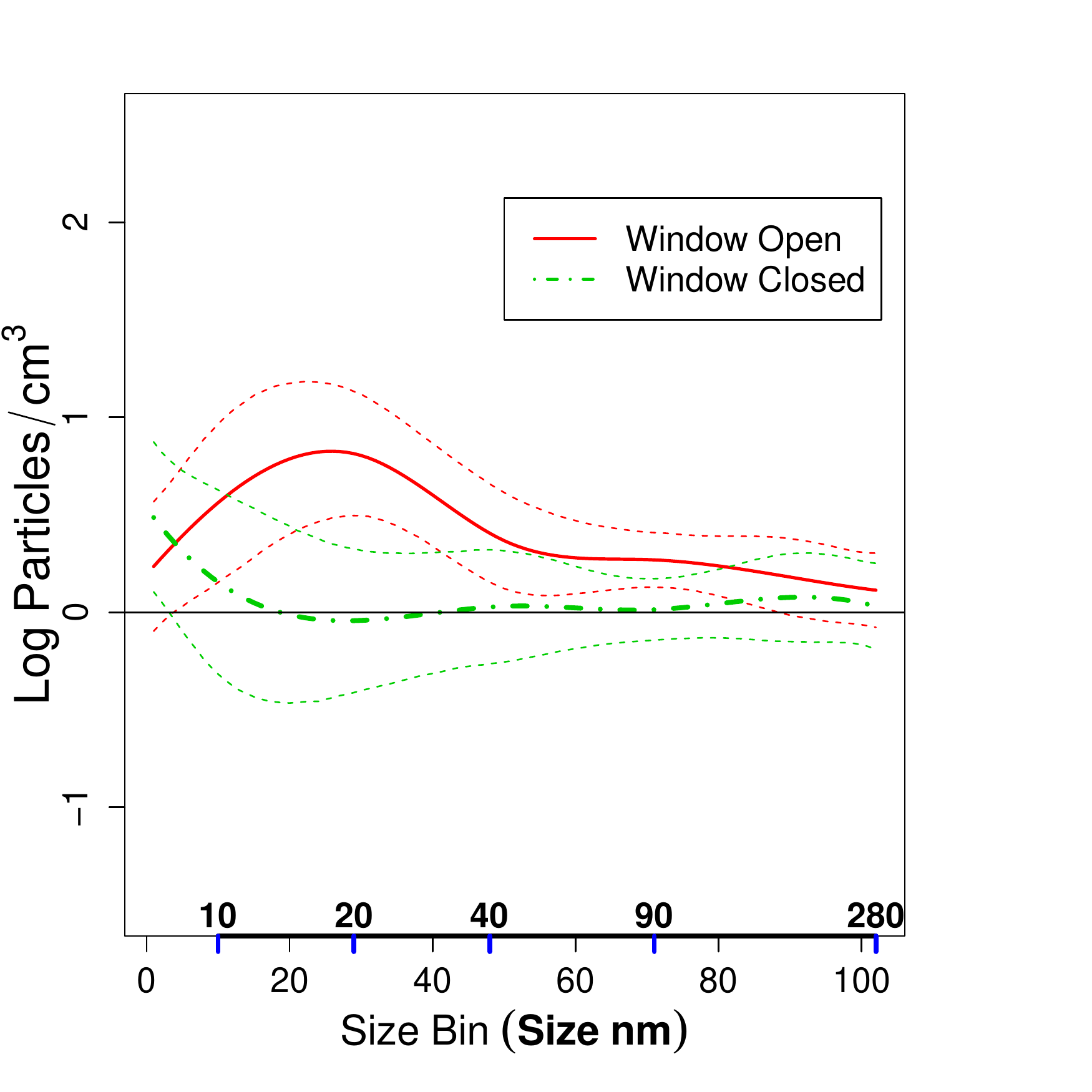}} \\
\subfloat[Linear Trend]{\includegraphics[width=7cm]{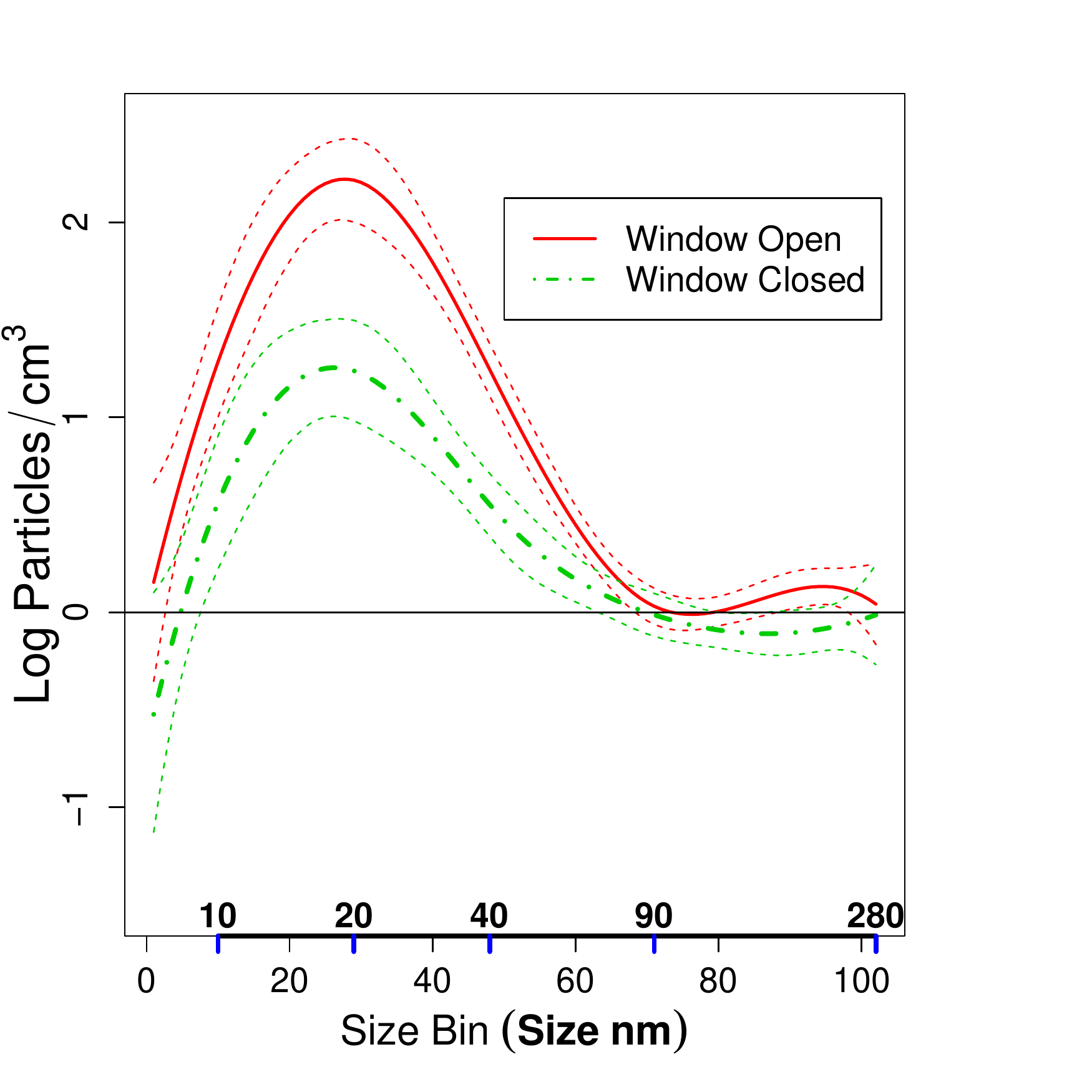}}
\subfloat[Quadratic Trend]{\includegraphics[width=7cm]{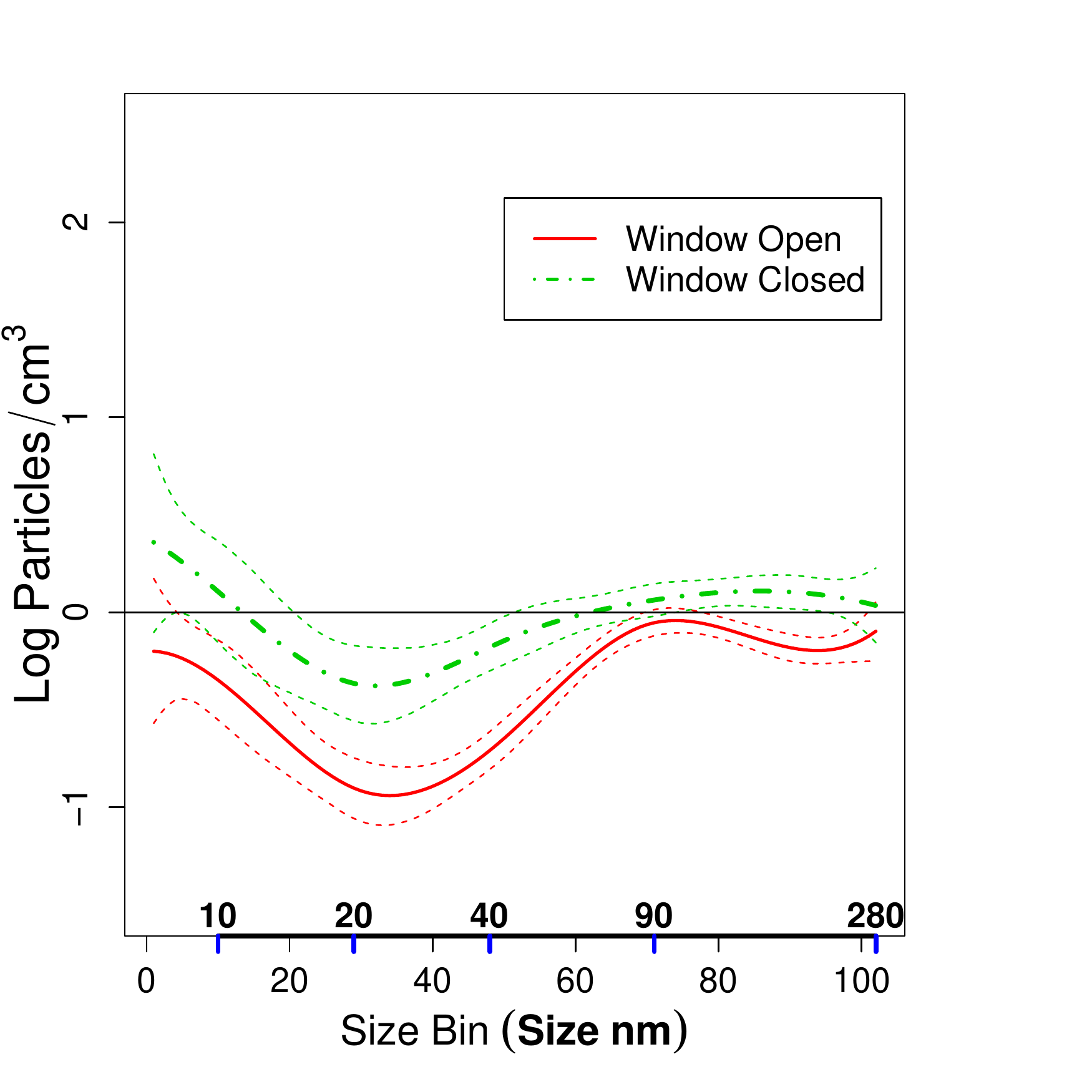}}

 \caption{Fixed time trend components on the log scale. Figure (a) plots the jump component, $\vect{\Delta_{z1}^TB(s)}$, (b) plots the linear component, $15*\vect{\Delta_{z2}^TB(s)}$, and (c) plots the quadratic component $15^2\vect{\Delta_{z3}^TB(s)}$ for both window positions. Adding these curves together gives the net log particle count increase after 15 minutes. Dotted lines display ninety-five percent posterior intervals.  The jump component is negligible for the window closed position, but there is an initial jump estimated for the window open position for all size bins. The linear component is larger for the window open position than the window closed position for most size bins.  Negative values across size bin for both window positions in the quadratic component implies that that particle counts increase at slower rates for all bins as the engine continues to run.}
\label{fig:jumptrends}
\end{figure}

Subsequent inferences are discussed in terms of particle sizes rather than size bins to facilitate interpretation of results. Time trend components are presented at $t=15$ in Figure \ref{fig:jumptrends}.  Figure \ref{fig:jumptrends}(a) plots the jump component, $\vect{\Delta_{z1}^TB(s)}$,  (b) plots the 15 times linear component, $15*\vect{\Delta_{z2}^TB(s)}$, and (c) plots the $15^2$ times quadratic component $15^2\vect{\Delta_{z3}^TB(s)}$ for both window positions. Adding these curves together gives the net log particle count increase after 15 minutes. The jump component is negligible for the window closed position, but there is an initial jump estimated for the window open position for all size bins, with the greatest increase for particles between 10 and 40nm. Credible intervals do not overlap between window positions for sizes between 15 and 30 nm. For particle sizes between 10 and 90 nm, the linear component is larger for the window open position than the window closed position as credible intervals do not overlap.  Negative values across size bin for both window positions in the quadratic component implies that that particle counts increase at slower rates for all bins as the engine continues to run, though these rate decreases are smaller across size for the window closed position.

\begin{figure}[!t]
\centering
\subfloat[Idling 5 minutes]{\includegraphics[width=7cm]{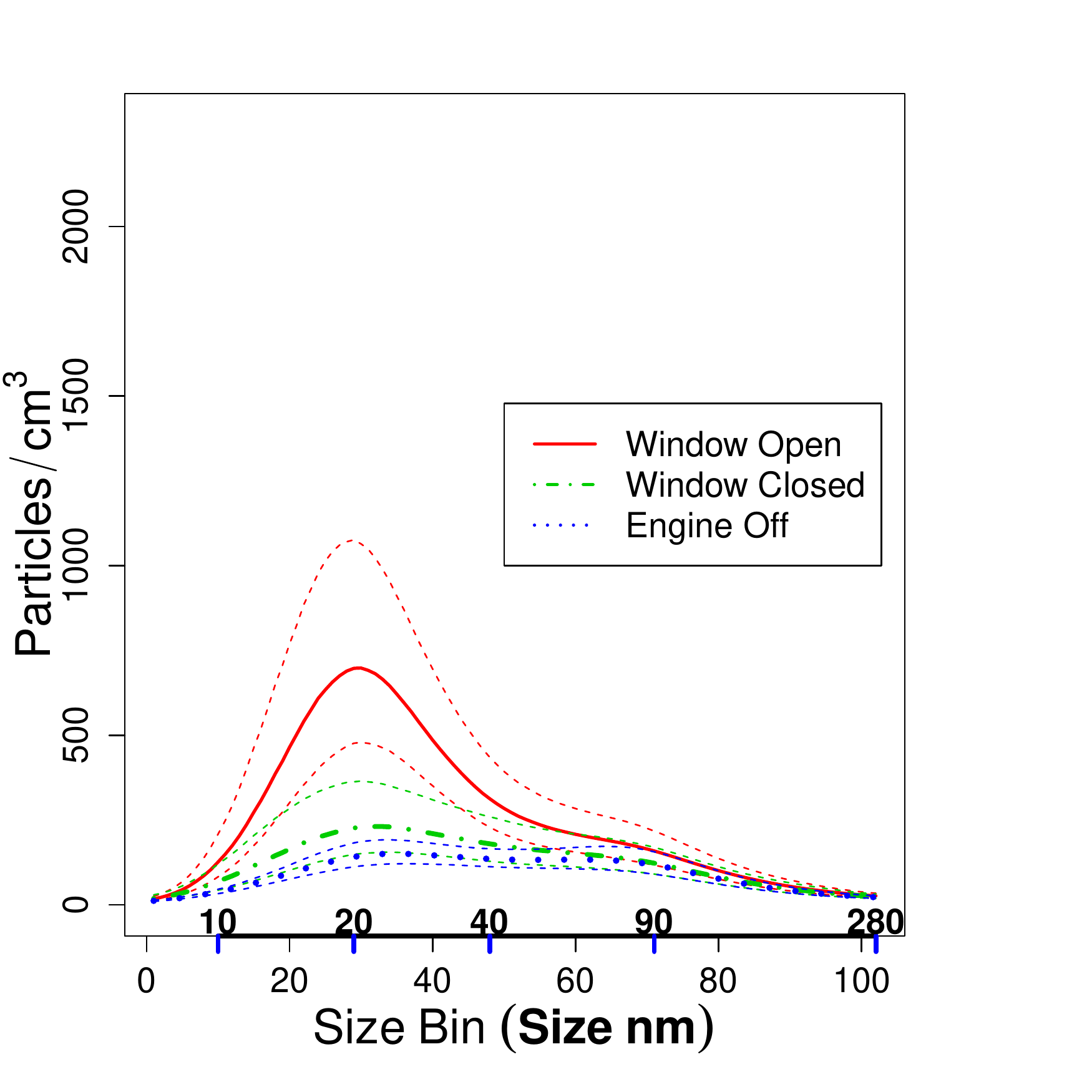}}
\subfloat[Idling 10 minutes]{\includegraphics[width=7cm]{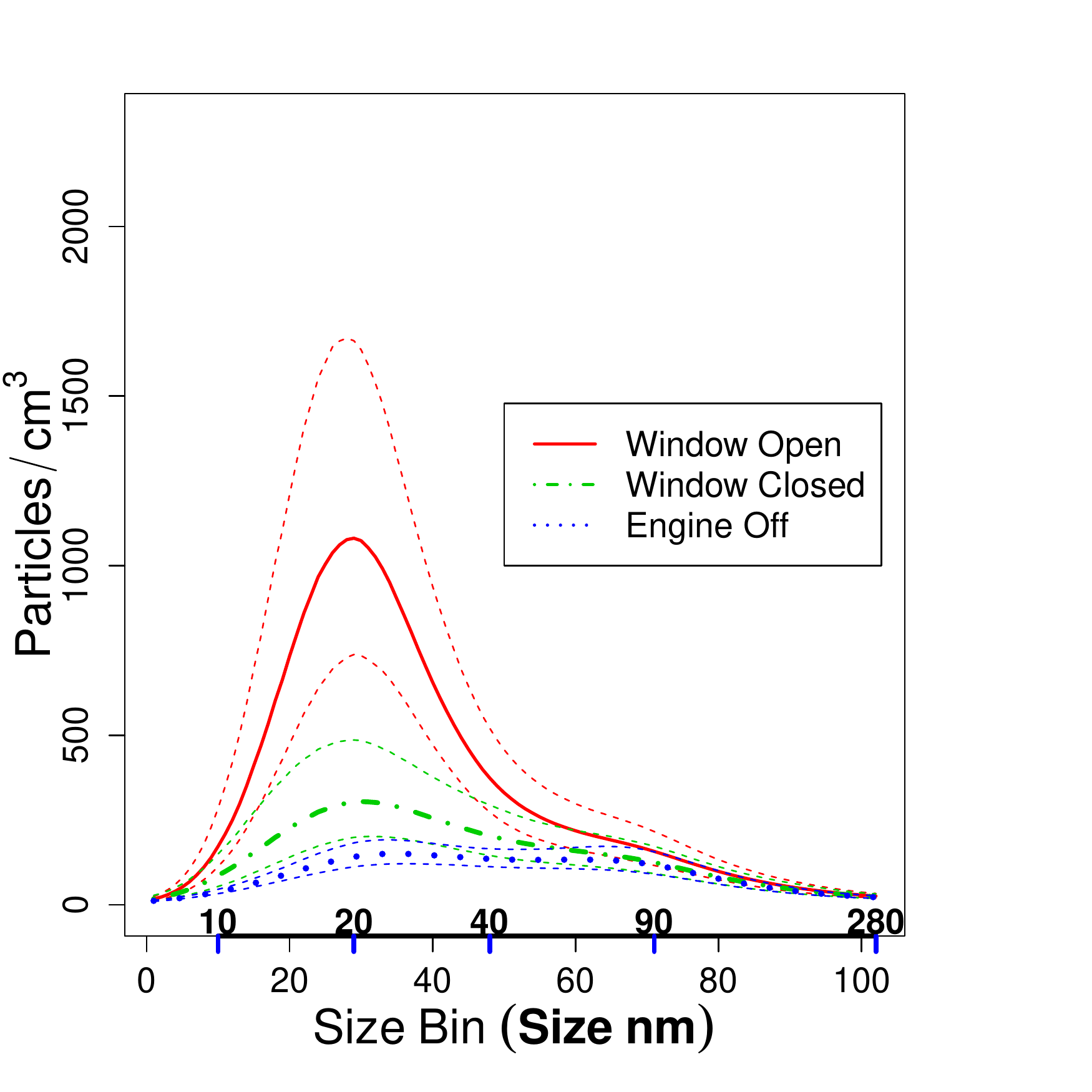}} \\
\subfloat[Idling 15 minutes]{\includegraphics[width=7cm]{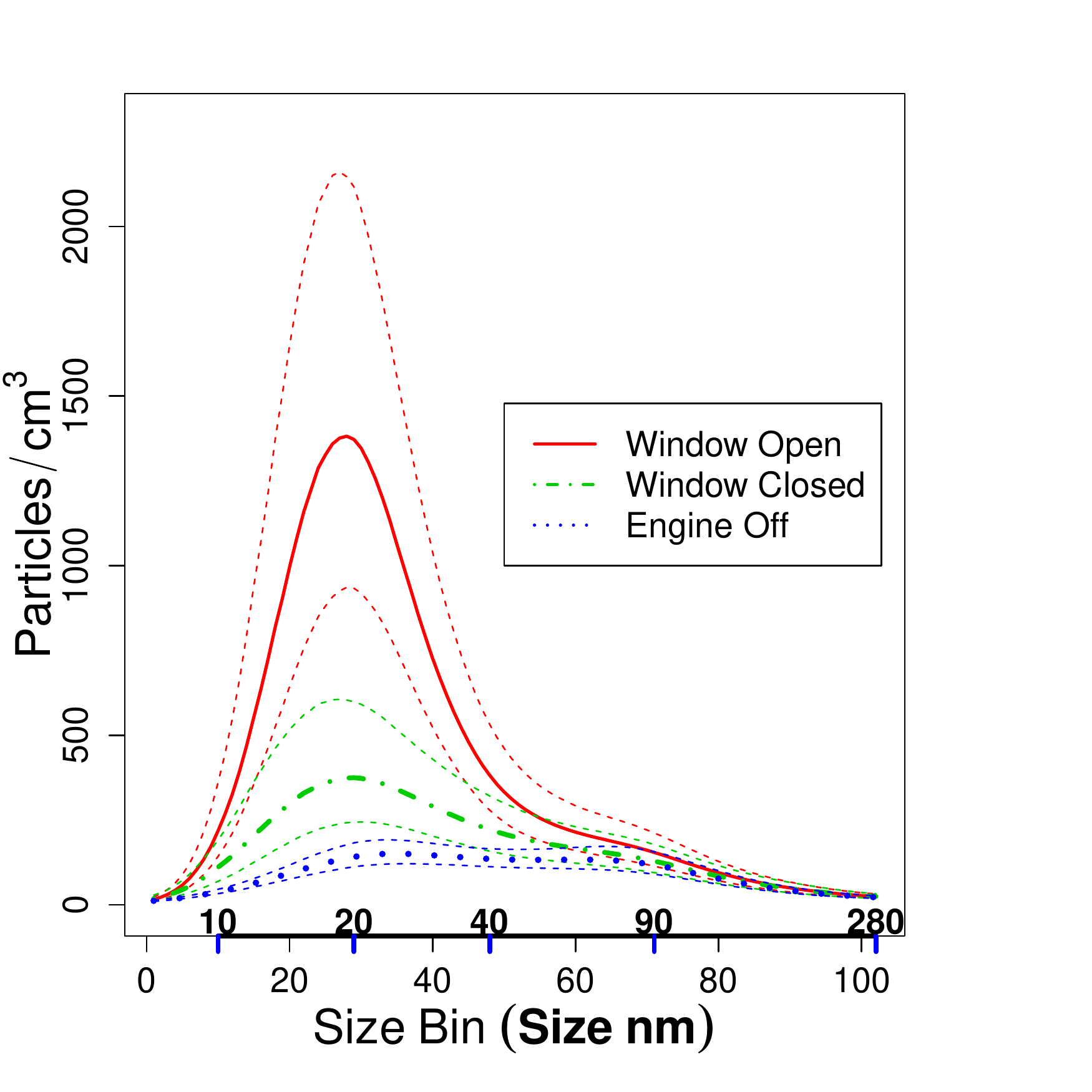}}
\subfloat[Idling 20 minutes]{\includegraphics[width=7cm]{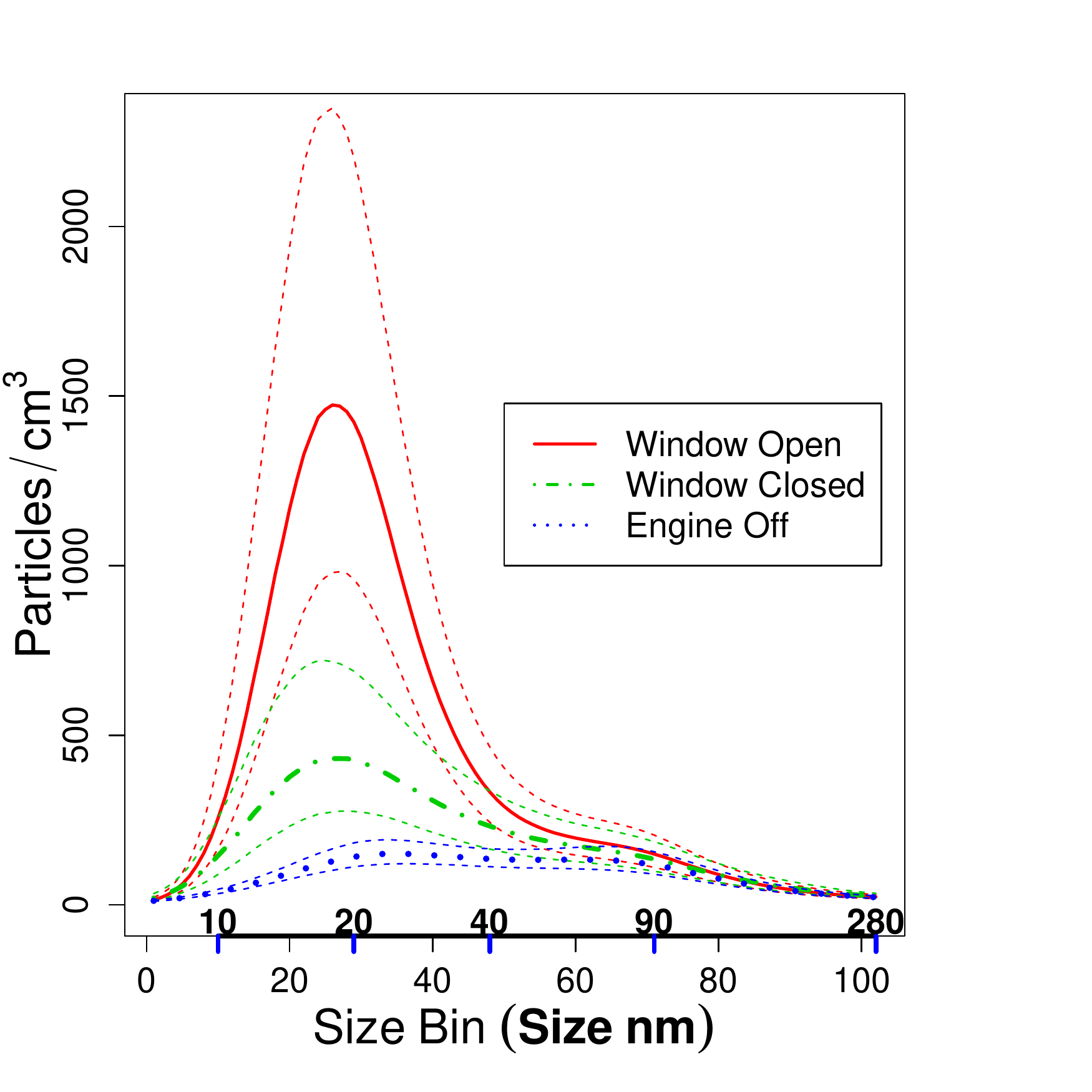}} 
 \caption{UFP size distributions, $\mu_{st}$ for engine-off is plotted in all figures; window open and closed after the engine has been idling for (a) 5, (b) 10, (c) 15, and (d) 20 minutes for the random jump quadratic model.  Solid lines are medians and dashed are 95 percent credible intervals. Adding the random jump both increased predicted means and widened credible intervals. The window open position has higher predicted counts for sizes between about 15 and 40 nm at all time points compared to the window closed position and also becomes much more peaked over time as the mode gets higher and higher.  The windows-closed position predictions increase over time at a slower, more constant rate.}
\label{fig:jumpallufp}
\end{figure}

In Figure \ref{fig:jumpallufp}, UFP size bin distributions, $\mu_{st}$, are plotted for engine-off and window open and closed after the engine has been idling for (a) 5, (b) 10, (c) 15, and (d) 20 minutes for the random jump quadratic model. Adding the random jump both increased predicted means and widened credible intervals compared to models without it. The window open position has higher predicted counts for sizes between about 15 and 40 nm at all time points compared to the window closed position and also becomes much more peaked over time as its mode gets higher and higher.  The windows-closed position predictions increase over time at a slower, more constant rate.  While for the first 10 minutes credible intervals between the engine-off and window closed curves overlap slightly, at 15 minutes particles of sizes less than 30nm show increases from the engine-off state which increase further at 20 minutes. Closing the windows hinders self pollution over time but does not prevent it.

\begin{figure}[!t]
\centering
\subfloat[]{\includegraphics[width=7cm]{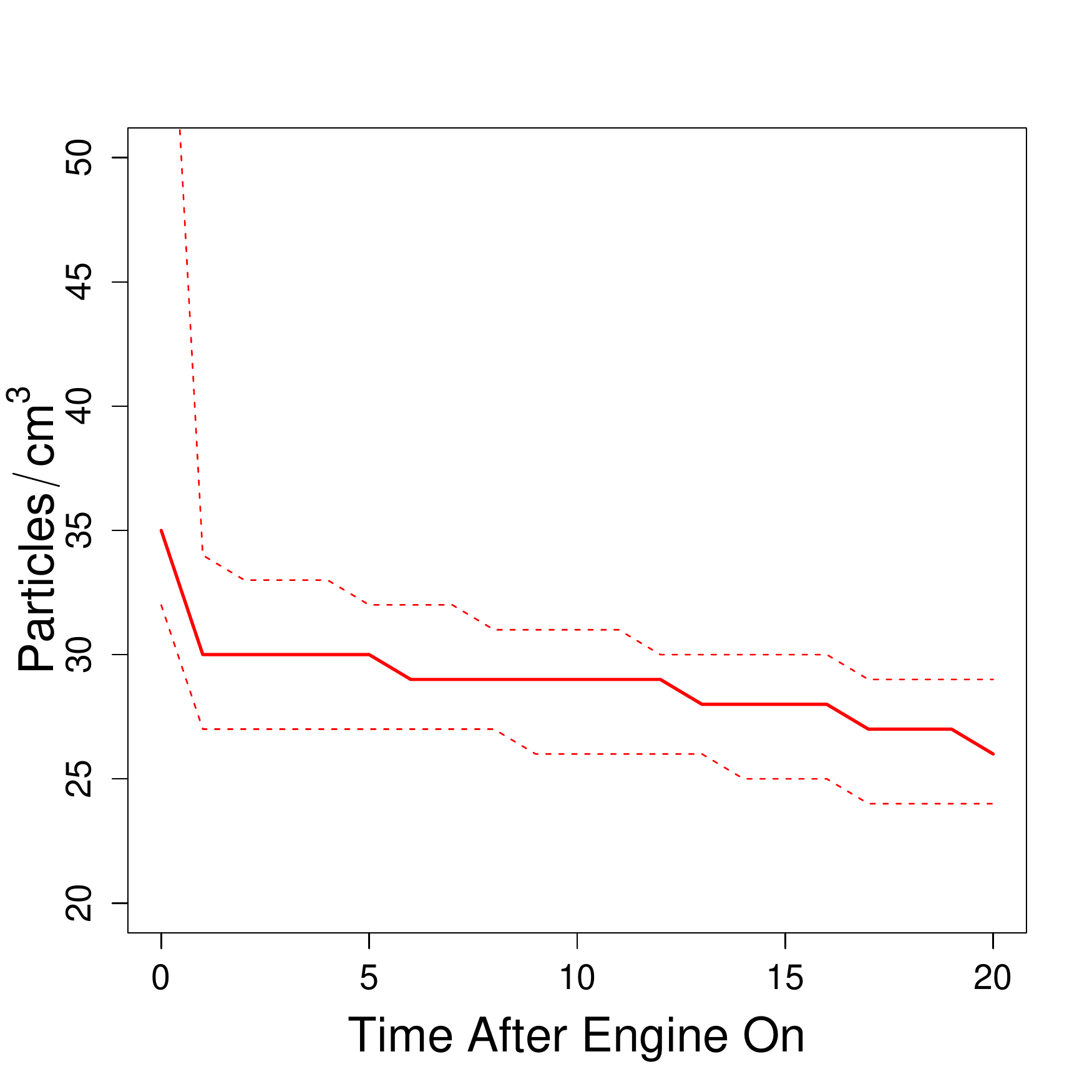}} 
\subfloat[]{\includegraphics[width=7cm]{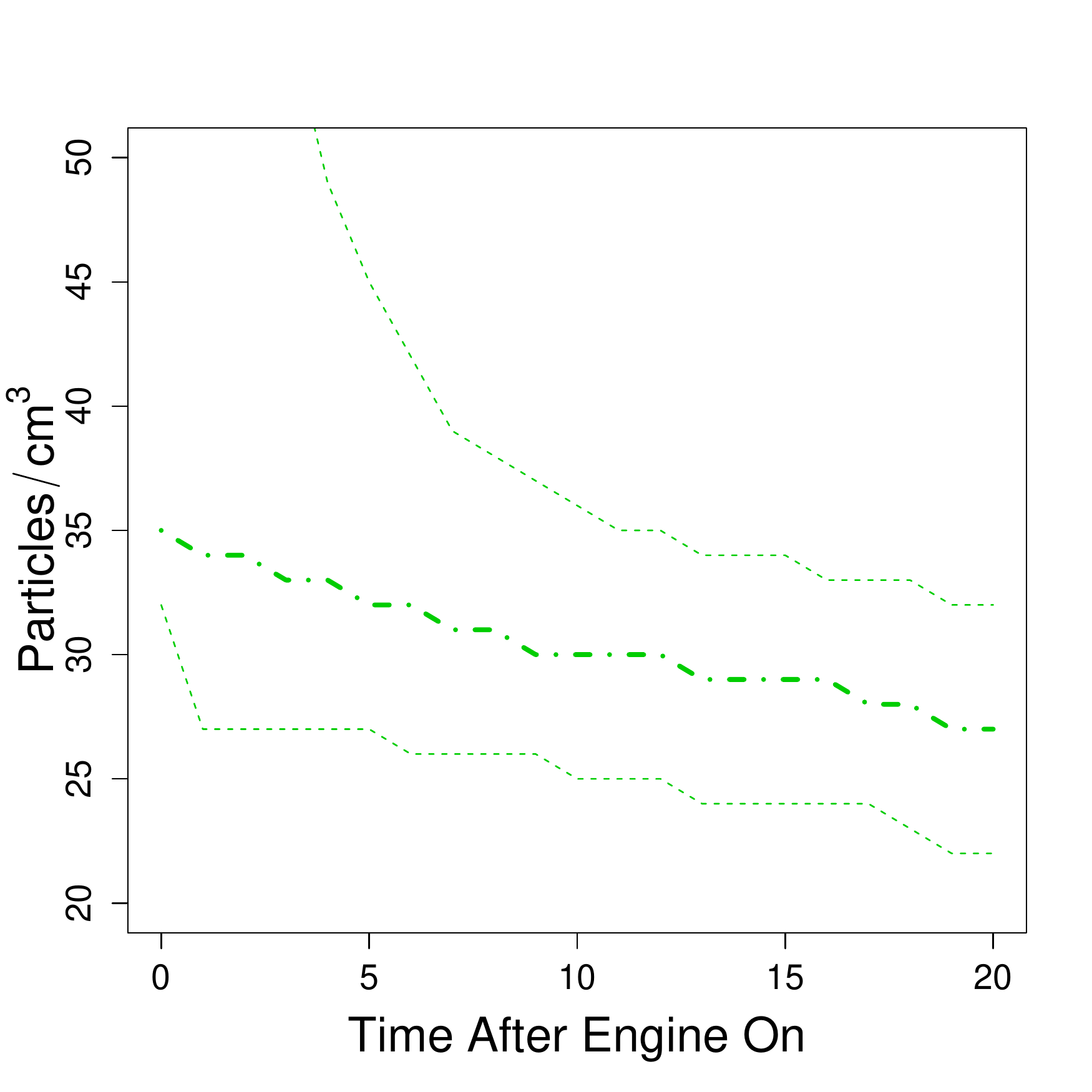}} \\
\subfloat[]{\includegraphics[width=7cm]{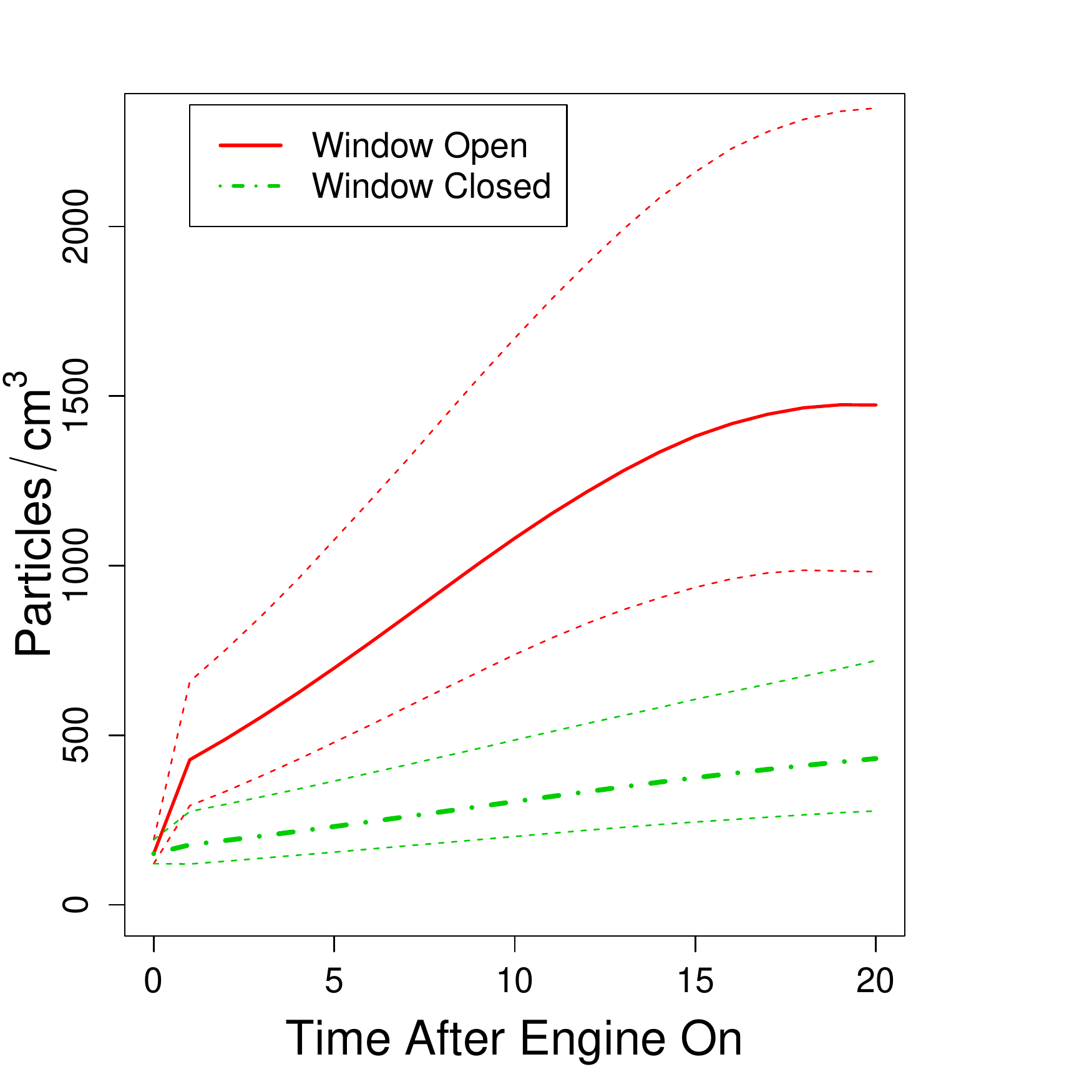}}

 \caption{Median posterior mode locations $s_t$ for (a) window-open and (b) window-closed and (c) heights over time $h_t$ for both window positions with 95 percent credible intervals. The location of the mode in the engine-off state is uncertain, with 95 percent credible intervals $(32,66)$, equivalent to a size range of $22-76$nm (full credible interval not shown on plot). In Figure (a), the window-open mode location rapidly narrows while in (b) the window-closed position mode location decreases at a slower rate and with more uncertainty.  Posterior mode heights $h_t$ in Figure (c) for the window-closed position increase at a slower almost constant linear rate over time.  Credible intervals do not overlap, but get wider over time for both window positions.}
\label{fig:modejump}
\end{figure}

Figure \ref{fig:modejump} plots medians of the posterior mode locations, $s_t$ and heights, ${h}_t$.  The location of the mode in the engine-off state is uncertain, with 95 percent credible intervals $(32,66)$, equivalent to a size range of $22-76$nm (full credible interval not shown on plot). In \ref{fig:modejump}(a), the window-open mode location decreases in size and the credible interval rapidly narrows to finally reach $(24,29)$ at 20 minutes.  The window-closed position mode location in \ref{fig:modejump}(b) decreases at a slower rate and with more uncertainty: at 20 minutes the 95 percent credible interval is $(22,32)$.  In Figure (c), posterior medians of the heights $h_t$ reach counts of $1,475$ for window open at $t=20$ compared to about $390$ for window-closed.  The window-closed position has posterior modeheights that increase at an almost linear rate over time that is slower than for the windows-open position.  Credible intervals do not overlap, but get wider over time for both window positions.

\section{Discussion}

Our use of functional time-series methods greatly reduce the size of confidence intervals compared to the non-functional univariate longitudinal methods used in \citet{zhang12} which show overlap for both window positions at all size bins.  Our methods permit inference about actual particle counts, not available using modal methods. 

At all time points the windows-open position had higher predictions for particles between about 15 and 40 nm than windows-closed, but predictions for the windows-closed position continued to increase over time.  Having the window closed does not prevent UFPs from entering the bus cabin, particularly for particles less than 40 nm.  Though predictions varied between no-jump, jump, and random jump models, predictions within jump or random jump models were similar, indicating conclusions are not extremely sensitive to parametric choice after accounting for engine on jump.   

A major benefit of these methods are the ease in which model fit can be evaluated, both through the use of DIC and through residual checks allowing visual inspection of model fit across size bin and time.  Graphical techniques also allow for the evaluation of the fit of random effects.  Plots demonstrating a subset of these residual checks for random effects are included in the supplemental materials.  Our methods are easily adapted to different model specifications, such as a different form for the time trends or engine-on random effects and other variance structures.  

Our methods allow uncertainty in predictions to be quantified and communicated graphically in ways easily interpretable to a non-statistician.  We provide statistical summaries of useful quantities currently used by researchers such as mode locations and heights. 

Vehicle emission standards are currently based on particulate matter mass measurements which fail to capture exposure to UFPs.  There is growing concern about the suitability of a single metric based on particle mass as the standard for regulation~\citep{ning10}. Our methods provide a basis to monitor exposure to UFPs based on particle counts separated into particles of different groups of size bins.  Given UFPs' toxic nature with health effects strongly related to particle size bin, these methods provide information needed for regulating vehicle emissions to minimize UFP exposure in the future.  Our methods allow evaluation and comparison of emissions by particle size for UFPs not captured in current emission standards.

\Needspace{10\baselineskip}
\clearpage

	\bibliographystyle{imsart-nameyear}
	\bibliography{proposal}

\newcommand{\noop}[1]{}
\begin{thebibliography}{34}

\bibitem[\protect\citeauthoryear{Alessandrini et~al.}{2006}]{aless}
\begin{barticle}[author]
\bauthor{\bsnm{Alessandrini},~\bfnm{F.}\binits{F.}},
  \bauthor{\bsnm{Schulz},~\bfnm{H.}\binits{H.}},
  \bauthor{\bsnm{Takenaka},~\bfnm{S.}\binits{S.}},
  \bauthor{\bsnm{Lentner},~\bfnm{B.}\binits{B.}},
  \bauthor{\bsnm{Karg},~\bfnm{E.}\binits{E.}},
  \bauthor{\bsnm{Behrendt},~\bfnm{H.}\binits{H.}} \AND
  \bauthor{\bsnm{Jakob},~\bfnm{T.}\binits{T.}}
(\byear{2006}).
\btitle{Effects of ultrafine carbon particle inhalation on allergic
  inflammation of the lung.}
\bjournal{Journal of Allergy and Clinical Immunology}
\bvolume{117}
\bpages{824-830}.
\end{barticle}
\endbibitem

\bibitem[\protect\citeauthoryear{Baladandayuthapani, Mallick and
  Carroll}{2005}]{bal05}
\begin{barticle}[author]
\bauthor{\bsnm{Baladandayuthapani},~\bfnm{V.}\binits{V.}},
  \bauthor{\bsnm{Mallick},~\bfnm{B.~K.}\binits{B.~K.}} \AND
  \bauthor{\bsnm{Carroll},~\bfnm{R.~J.}\binits{R.~J.}}
(\byear{2005}).
\btitle{Spatially adaptive {Bayesian} penalized regression splines
  (P-splines)}.
\bjournal{Journal of Computational and Graphical Statistics}
\bvolume{14}
\bpages{378–394}.
\end{barticle}
\endbibitem

\bibitem[\protect\citeauthoryear{Bennett and Zeman}{1998}]{bennett}
\begin{barticle}[author]
\bauthor{\bsnm{Bennett},~\bfnm{W.~D.}\binits{W.~D.}} \AND
  \bauthor{\bsnm{Zeman},~\bfnm{K.~L.}\binits{K.~L.}}
(\byear{1998}).
\btitle{Deposition of fine particles in children spontaneously breathing at
  rest.}
\bjournal{Inhalation Toxicology}
\bvolume{10}
\bpages{831-842}.
\end{barticle}
\endbibitem

\bibitem[\protect\citeauthoryear{Chaloner}{1994}]{chaloner94}
\begin{binbook}[author]
\bauthor{\bsnm{Chaloner},~\bfnm{K.}\binits{K.}}
(\byear{1994}).
\btitle{In Aspects of Uncertainty: A Tribute to DV Lindley}
\bchapter{Residual analysis and outliers in Bayesian hierarchical models},
\bpages{149-157}.
\bpublisher{New York: Wiley}.
\end{binbook}
\endbibitem

\bibitem[\protect\citeauthoryear{Chaloner and Brant}{1988}]{chaloner88}
\begin{barticle}[author]
\bauthor{\bsnm{Chaloner},~\bfnm{K.}\binits{K.}} \AND
  \bauthor{\bsnm{Brant},~\bfnm{R.}\binits{R.}}
(\byear{1988}).
\btitle{A Bayesian approach to outlier detection and residual analysis}.
\bjournal{Biometrika}
\bvolume{75(4)}
\bpages{651-659}.
\end{barticle}
\endbibitem

\bibitem[\protect\citeauthoryear{Crainiceanu et~al.}{2007}]{ciprian07}
\begin{barticle}[author]
\bauthor{\bsnm{Crainiceanu},~\bfnm{C.}\binits{C.}},
  \bauthor{\bsnm{Ruppert},~\bfnm{D.}\binits{D.}},
  \bauthor{\bsnm{Carroll},~\bfnm{R.}\binits{R.}},
  \bauthor{\bsnm{Joshi},~\bfnm{A.}\binits{A.}} \AND
  \bauthor{\bsnm{Goodner},~\bfnm{B.}\binits{B.}}
(\byear{2007}).
\btitle{Spatially adaptive {Bayesian} penalized splines with heteroscedastic
  errors}.
\bjournal{Journal of Computational and Graphical Statistics}
\bvolume{16}
\bpages{265-288}.
\end{barticle}
\endbibitem

\bibitem[\protect\citeauthoryear{de~Boor}{1978}]{deboor}
\begin{bbook}[author]
\bauthor{\bparticle{de} \bsnm{Boor},~\bfnm{C.}\binits{C.}}
(\byear{1978}).
\btitle{A Practical Guide to Splines}.
\bpublisher{New York: Springer}.
\end{bbook}
\endbibitem

\bibitem[\protect\citeauthoryear{Delfino, Sioutas and Malik}{2005}]{delfino}
\begin{barticle}[author]
\bauthor{\bsnm{Delfino},~\bfnm{R.~J.}\binits{R.~J.}},
  \bauthor{\bsnm{Sioutas},~\bfnm{C.}\binits{C.}} \AND
  \bauthor{\bsnm{Malik},~\bfnm{S.}\binits{S.}}
(\byear{2005}).
\btitle{Potential role of ultrafine particles in associations between airborne
  particle mass and cardiovascular health.}
\bjournal{Environmental Health Perspectives}
\bvolume{113}
\bpages{934-946}.
\end{barticle}
\endbibitem

\bibitem[\protect\citeauthoryear{EPA}{2002}]{epa02}
\begin{bmanual}[author]
\bauthor{\bsnm{EPA}}
(\byear{2002}).
\btitle{Health assessment document for diesel engine exhaust.}
\bpublisher{The National Technical Information Service},
\baddress{Springfield, VA}.
\end{bmanual}
\endbibitem

\bibitem[\protect\citeauthoryear{EPA}{2014}]{epa12}
\begin{bmanual}[author]
\bauthor{\bsnm{EPA}}
(\byear{2014}).
\btitle{Clean school bus {USA}.}
\bnote{Available from http://www.epa.gov/cleanschoolbus/csb-overview.htm.
  Accessed 10/28/14}.
\end{bmanual}
\endbibitem

\bibitem[\protect\citeauthoryear{Ferin et~al.}{1990}]{ferin}
\begin{barticle}[author]
\bauthor{\bsnm{Ferin},~\bfnm{J.}\binits{J.}},
  \bauthor{\bsnm{Oberdorster},~\bfnm{G.}\binits{G.}},
  \bauthor{\bsnm{Penney},~\bfnm{D.~P.}\binits{D.~P.}},
  \bauthor{\bsnm{Soderholm},~\bfnm{S.~C.}\binits{S.~C.}},
  \bauthor{\bsnm{Gelein},~\bfnm{R.}\binits{R.}} \AND
  \bauthor{\bsnm{Piper},~\bfnm{H.~C.}\binits{H.~C.}}
(\byear{1990}).
\btitle{Increased pulmonary toxicity of ultrafine particles?}
\bjournal{Journal of Aerosol Science}
\bvolume{21}
\bpages{384-387}.
\end{barticle}
\endbibitem

\bibitem[\protect\citeauthoryear{Frampton et~al.}{2006}]{frampton}
\begin{barticle}[author]
\bauthor{\bsnm{Frampton},~\bfnm{M.~W.}\binits{M.~W.}},
  \bauthor{\bsnm{Stewart},~\bfnm{J.~C.}\binits{J.~C.}},
  \bauthor{\bsnm{Oberdorster},~\bfnm{G.}\binits{G.}},
  \bauthor{\bsnm{Morrow},~\bfnm{P.~E.}\binits{P.~E.}},
  \bauthor{\bsnm{Chalupa},~\bfnm{D.}\binits{D.}},
  \bauthor{\bsnm{Pietropaoli},~\bfnm{A.~P.}\binits{A.~P.}},
  \bauthor{\bsnm{Frasier},~\bfnm{L.~M.}\binits{L.~M.}},
  \bauthor{\bsnm{Speers},~\bfnm{D.~M.}\binits{D.~M.}},
  \bauthor{\bsnm{Cox},~\bfnm{C.}\binits{C.}},
  \bauthor{\bsnm{Huang},~\bfnm{L.~S.}\binits{L.~S.}} \AND
  \bauthor{\bsnm{Utell},~\bfnm{M.~J.}\binits{M.~J.}}
(\byear{2006}).
\btitle{Inhalation of ultrafine particles alters blood leukocyte expression of
  adhesion molecules in humans.}
\bjournal{Environmental Health Perspectives}
\bvolume{114}
\bpages{51-58}.
\end{barticle}
\endbibitem

\bibitem[\protect\citeauthoryear{Gelfand and Smith}{1990}]{gelfand90}
\begin{barticle}[author]
\bauthor{\bsnm{Gelfand},~\bfnm{A.}\binits{A.}} \AND
  \bauthor{\bsnm{Smith},~\bfnm{A.~F.~M.}\binits{A.~F.~M.}}
(\byear{1990}).
\btitle{Sampling-based approaches to calculating marginal densities.}
\bjournal{Journal of the American Statistical Association}
\bvolume{85}
\bpages{398-409}.
\end{barticle}
\endbibitem

\bibitem[\protect\citeauthoryear{Hadfield}{2010}]{mcmcglmm}
\begin{barticle}[author]
\bauthor{\bsnm{Hadfield},~\bfnm{Jarrod~D}\binits{J.~D.}}
(\byear{2010}).
\btitle{{MCMC} Methods for multi-response generalized linear mixed Models: The
  {MCMCglmm} {R} Package}.
\bjournal{Journal of Statistical Software}
\bvolume{33}
\bpages{1--22}.
\end{barticle}
\endbibitem

\bibitem[\protect\citeauthoryear{Hastie and Tibshirani}{1993}]{hastie93}
\begin{barticle}[author]
\bauthor{\bsnm{Hastie},~\bfnm{T.}\binits{T.}} \AND
  \bauthor{\bsnm{Tibshirani},~\bfnm{R.}\binits{R.}}
(\byear{1993}).
\btitle{Varying-coefficient models}.
\bjournal{Journal of the Royal Statistical Society}
\bvolume{55}
\bpages{757-796}.
\end{barticle}
\endbibitem

\bibitem[\protect\citeauthoryear{Hastings}{1970}]{hastings70}
\begin{barticle}[author]
\bauthor{\bsnm{Hastings},~\bfnm{W.~K.}\binits{W.~K.}}
(\byear{1970}).
\btitle{Monte {Carlo} sampling methods using {Markov} chains and their
  applications}.
\bjournal{Biometrika}
\bvolume{57}
\bpages{97-109}.
\end{barticle}
\endbibitem

\bibitem[\protect\citeauthoryear{Hussein et~al.}{\noop{3001}2005}]{hussein05}
\begin{barticle}[author]
\bauthor{\bsnm{Hussein},~\bfnm{T.}\binits{T.}},
  \bauthor{\bsnm{Maso},~\bfnm{M.~D.}\binits{M.~D.}},
  \bauthor{\bsnm{Petäjä},~\bfnm{T.}\binits{T.}},
  \bauthor{\bsnm{Koponen},~\bfnm{I.~K.}\binits{I.~K.}},
  \bauthor{\bsnm{Paatero},~\bfnm{P.}\binits{P.}},
  \bauthor{\bsnm{Aalto},~\bfnm{P.}\binits{P.}},
  \bauthor{\bsnm{Hämeri},~\bfnm{K.}\binits{K.}} \AND
  \bauthor{\bsnm{Kulmala},~\bfnm{M.}\binits{M.}}
(\byear{\noop{3001}2005}).
\btitle{Evaluation of an automatic algorithm for fitting the particle number
  size distributions}.
\bjournal{Boreal Environment Research}
\bvolume{10}
\bpages{337-355}.
\end{barticle}
\endbibitem

\bibitem[\protect\citeauthoryear{Kittelson, Watts and
  Johnson}{2006}]{kittleson06}
\begin{barticle}[author]
\bauthor{\bsnm{Kittelson},~\bfnm{D.~B.}\binits{D.~B.}},
  \bauthor{\bsnm{Watts},~\bfnm{W.~F.}\binits{W.~F.}} \AND
  \bauthor{\bsnm{Johnson},~\bfnm{J.~P.}\binits{J.~P.}}
(\byear{2006}).
\btitle{On-road and laboratory evaluation of combustion aerosols part 1:
  summary of diesel engine results}.
\bjournal{Journal of Aerosol Science}
\bvolume{37}
\bpages{913-930}.
\end{barticle}
\endbibitem

\bibitem[\protect\citeauthoryear{Kulmala et~al.}{2004}]{kulmala04}
\begin{barticle}[author]
\bauthor{\bsnm{Kulmala},~\bfnm{M.}\binits{M.}},
  \bauthor{\bsnm{Vehkamäki},~\bfnm{H.}\binits{H.}},
  \bauthor{\bsnm{Petäjä},~\bfnm{T.}\binits{T.}},
  \bauthor{\bsnm{Maso},~\bfnm{M.~Dal}\binits{M.~D.}},
  \bauthor{\bsnm{Lauri},~\bfnm{A.}\binits{A.}},
  \bauthor{\bsnm{Kerminen},~\bfnm{V.~M.}\binits{V.~M.}},
  \bauthor{\bsnm{Birmili},~\bfnm{W.}\binits{W.}} \AND
  \bauthor{\bsnm{McMurry},~\bfnm{P.~H.}\binits{P.~H.}}
(\byear{2004}).
\btitle{Formation and growth rates of ultrafine atmospheric particles: a review
  of observations.}
\bjournal{Journal of Aerosol Science}
\bvolume{35}
\bpages{143–176}.
\end{barticle}
\endbibitem

\bibitem[\protect\citeauthoryear{Lang and Brezger}{2004}]{Lang04}
\begin{barticle}[author]
\bauthor{\bsnm{Lang},~\bfnm{Stefan}\binits{S.}} \AND
  \bauthor{\bsnm{Brezger},~\bfnm{Andreas}\binits{A.}}
(\byear{2004}).
\btitle{Bayesian {P-Splines}}.
\bjournal{Journal of Computational and Graphical Statistics}
\bvolume{13 No. 1}
\bpages{183-212}.
\end{barticle}
\endbibitem

\bibitem[\protect\citeauthoryear{Morawska et~al.}{2008}]{mor08}
\begin{barticle}[author]
\bauthor{\bsnm{Morawska},~\bfnm{L.}\binits{L.}},
  \bauthor{\bsnm{Ristovski},~\bfnm{Z.}\binits{Z.}},
  \bauthor{\bsnm{Jayaratne},~\bfnm{E.~R.}\binits{E.~R.}},
  \bauthor{\bsnm{Keogh},~\bfnm{D.~U.}\binits{D.~U.}} \AND
  \bauthor{\bsnm{Ling},~\bfnm{X.}\binits{X.}}
(\byear{2008}).
\btitle{Ambient nano and ultrafine particles from motor vehicle emissions:
  characteristics, ambient processing and implications on human exposure}.
\bjournal{Atmospheric Environment}
\bvolume{42}
\bpages{8113-8138}.
\end{barticle}
\endbibitem

\bibitem[\protect\citeauthoryear{Ning and Sioutas}{2010}]{ning10}
\begin{barticle}[author]
\bauthor{\bsnm{Ning},~\bfnm{Z.}\binits{Z.}} \AND
  \bauthor{\bsnm{Sioutas},~\bfnm{C.}\binits{C.}}
(\byear{2010}).
\btitle{Atmospheric processes influencing aerosols generated by combustion and
  the inference of their impact on public exposure: a review}.
\bjournal{Aerosol and Air Quality Research}
\bvolume{10}
\bpages{43-58}.
\end{barticle}
\endbibitem

\bibitem[\protect\citeauthoryear{Oberdorster et~al.}{2004}]{ober}
\begin{barticle}[author]
\bauthor{\bsnm{Oberdorster},~\bfnm{G.}\binits{G.}},
  \bauthor{\bsnm{Sharp},~\bfnm{Z.}\binits{Z.}},
  \bauthor{\bsnm{Atudorei},~\bfnm{V.}\binits{V.}},
  \bauthor{\bsnm{Elder},~\bfnm{A.}\binits{A.}},
  \bauthor{\bsnm{Gelein},~\bfnm{R.}\binits{R.}},
  \bauthor{\bsnm{Kreyling},~\bfnm{W.}\binits{W.}} \AND
  \bauthor{\bsnm{Cox},~\bfnm{C.}\binits{C.}}
(\byear{2004}).
\btitle{Translocation of inhaled ultrafine particles to the brain.}
\bjournal{Inhalation Toxicology}
\bvolume{16}
\bpages{437-445}.
\end{barticle}
\endbibitem

\bibitem[\protect\citeauthoryear{Palmer and Pettit}{1996}]{palmer96}
\begin{barticle}[author]
\bauthor{\bsnm{Palmer},~\bfnm{J.}\binits{J.}} \AND
  \bauthor{\bsnm{Pettit},~\bfnm{Lawrence}\binits{L.}}
(\byear{1996}).
\btitle{Risks of using improper priors with Gibbs sampling and autocorrelated
  errors}.
\bjournal{Journal of Computational and Graphical Statistics}
\bvolume{5}
\bpages{245-249}.
\end{barticle}
\endbibitem

\bibitem[\protect\citeauthoryear{Plummer}{2003}]{Plummer03jags:a}
\begin{bmisc}[author]
\bauthor{\bsnm{Plummer},~\bfnm{Martyn}\binits{M.}}
(\byear{2003}).
\btitle{JAGS: A program for analysis of Bayesian graphical models using Gibbs
  sampling}.
\end{bmisc}
\endbibitem

\bibitem[\protect\citeauthoryear{Ruppert, Wand and Carroll}{2003}]{ruppert03}
\begin{bbook}[author]
\bauthor{\bsnm{Ruppert},~\bfnm{D.}\binits{D.}},
  \bauthor{\bsnm{Wand},~\bfnm{M.~P.}\binits{M.~P.}} \AND
  \bauthor{\bsnm{Carroll},~\bfnm{R.~J}\binits{R.~J.}}
(\byear{2003}).
\btitle{Semiparametric Regression}.
\bpublisher{Cambridge: Cambridge Press}.
\end{bbook}
\endbibitem

\bibitem[\protect\citeauthoryear{Samet et~al.}{2009}]{samet}
\begin{barticle}[author]
\bauthor{\bsnm{Samet},~\bfnm{J.~M.}\binits{J.~M.}},
  \bauthor{\bsnm{Rappold},~\bfnm{A.}\binits{A.}},
  \bauthor{\bsnm{Graff},~\bfnm{D.}\binits{D.}},
  \bauthor{\bsnm{Cascio},~\bfnm{W.~E.}\binits{W.~E.}},
  \bauthor{\bsnm{Berntsen},~\bfnm{J.~H.}\binits{J.~H.}},
  \bauthor{\bsnm{Huang},~\bfnm{Y.~C.~T.}\binits{Y.~C.~T.}},
  \bauthor{\bsnm{Herbst},~\bfnm{M.}\binits{M.}},
  \bauthor{\bsnm{Bassett},~\bfnm{M.}\binits{M.}},
  \bauthor{\bsnm{Montilla},~\bfnm{T.}\binits{T.}},
  \bauthor{\bsnm{Hazucha},~\bfnm{M.~J.}\binits{M.~J.}},
  \bauthor{\bsnm{Bromberg},~\bfnm{P.~A.}\binits{P.~A.}} \AND
  \bauthor{\bsnm{Devlin},~\bfnm{R.~B.}\binits{R.~B.}}
(\byear{2009}).
\btitle{Concentrated ambient ultrafine particle exposure induces cardiac
  changes in young healthy volunteers.}
\bjournal{American Journal of Respiratory and Critical Care Medicine}
\bvolume{179}
\bpages{1034-1042}.
\end{barticle}
\endbibitem

\bibitem[\protect\citeauthoryear{Shi, Weiss and Taylor}{1996}]{shi96}
\begin{barticle}[author]
\bauthor{\bsnm{Shi},~\bfnm{M}\binits{M.}},
  \bauthor{\bsnm{Weiss},~\bfnm{R.~E.}\binits{R.~E.}} \AND
  \bauthor{\bsnm{Taylor},~\bfnm{J.~M.~G.}\binits{J.~M.~G.}}
(\byear{1996}).
\btitle{An analysis of paediatric {CD4} counts for acquired immune deficiency
  syndrome using flexible random curves}.
\bjournal{Journal of the Royal Statistical Society. Series C (Applied
  Statistics)}
\bvolume{45}
\bpages{151-163}.
\end{barticle}
\endbibitem

\bibitem[\protect\citeauthoryear{Spiegelhalter et~al.}{2002}]{dic}
\begin{barticle}[author]
\bauthor{\bsnm{Spiegelhalter},~\bfnm{D.~J.}\binits{D.~J.}},
  \bauthor{\bsnm{Best},~\bfnm{NG}\binits{N.}},
  \bauthor{\bsnm{Carlin},~\bfnm{B.~P.}\binits{B.~P.}} \AND
  \bauthor{\bparticle{der} \bsnm{Linde},~\bfnm{A.~Van}\binits{A.~V.}}
(\byear{2002}).
\btitle{Bayesian measures of model complexity and fit (with discussion)}.
\bjournal{Journal of the Royal Statistical Society - Series B}
\bvolume{64}
\bpages{583-616}.
\end{barticle}
\endbibitem

\bibitem[\protect\citeauthoryear{Whitby}{1978}]{whitby78}
\begin{bmanual}[author]
\bauthor{\bsnm{Whitby},~\bfnm{P.~H.}\binits{P.~H.}}
(\byear{1978}).
\btitle{Modal aerosol dynamics modelling. Technical report}
\bpublisher{U.S. Environment Protection Agency, Atmospheric Research and
  Exposure Assessment Laboratory}.
\end{bmanual}
\endbibitem

\bibitem[\protect\citeauthoryear{Whitby et~al.}{1991}]{whitby91}
\begin{bmanual}[author]
\bauthor{\bsnm{Whitby},~\bfnm{P.~H.}\binits{P.~H.}},
  \bauthor{\bsnm{McMurry},~\bfnm{P.~H.}\binits{P.~H.}},
  \bauthor{\bsnm{Shanker},~\bfnm{U.}\binits{U.}} \AND
  \bauthor{\bsnm{Binkowski},~\bfnm{F.~S.}\binits{F.~S.}}
(\byear{1991}).
\btitle{Modal aerosol dynamics modeling. Technical report}
\bpublisher{U.S. Environment Protection Agency, Atmospheric Research and
  Exposure Assessment Laboratory}.
\end{bmanual}
\endbibitem

\bibitem[\protect\citeauthoryear{Wraith et~al.}{2009}]{wraith09}
\begin{barticle}[author]
\bauthor{\bsnm{Wraith},~\bfnm{D.}\binits{D.}},
  \bauthor{\bsnm{Alston},~\bfnm{C.}\binits{C.}},
  \bauthor{\bsnm{Mengersen},~\bfnm{K.}\binits{K.}} \AND
  \bauthor{\bsnm{Hussein},~\bfnm{T.}\binits{T.}}
(\byear{2009}).
\btitle{Bayesian mixture model estimation of aerosol particle size
  distributions}.
\bjournal{Environmetrics}
\bvolume{22}
\bpages{23-34}.
\end{barticle}
\endbibitem

\bibitem[\protect\citeauthoryear{Wraith et~al.}{2014}]{wraith13}
\begin{barticle}[author]
\bauthor{\bsnm{Wraith},~\bfnm{D.}\binits{D.}},
  \bauthor{\bsnm{Mengersen},~\bfnm{K.}\binits{K.}},
  \bauthor{\bsnm{Alston},~\bfnm{C.}\binits{C.}},
  \bauthor{\bsnm{Rousseu},~\bfnm{J.}\binits{J.}} \AND
  \bauthor{\bsnm{Hussein},~\bfnm{T.}\binits{T.}}
(\byear{2014}).
\btitle{Using informative priors in the estimation of mixtures over time with
  application to aerosol particle size distributions}.
\bjournal{Annals of Applied Statistics}
\bvolume{8}
\bpages{232-258}.
\end{barticle}
\endbibitem

\bibitem[\protect\citeauthoryear{Zhang et~al.}{2012}]{zhang12}
\begin{barticle}[author]
\bauthor{\bsnm{Zhang},~\bfnm{Qunfang}\binits{Q.}},
  \bauthor{\bsnm{Fischer},~\bfnm{Heidi~J.}\binits{H.~J.}},
  \bauthor{\bsnm{Weiss},~\bfnm{Robert~E.}\binits{R.~E.}} \AND
  \bauthor{\bsnm{Zhu},~\bfnm{Yifang}\binits{Y.}}
(\byear{2012}).
\btitle{Ultrafine particle concentrations in and around idling school buses}.
\bjournal{Atmospheric Environment}
\bvolume{69}
\bpages{65-75}.
\end{barticle}
\endbibitem

\end{thebibliography}


\begin{thebibliography}{}

\bibitem[Alessandrini {\rm et~al.}, 2006]{aless}
Alessandrini, F., Schulz, H., Takenaka, S., Lentner, B., Karg, E., Behrendt, H.
   and Jakob, T. (2006{\rm{}}).
\newblock Effects of ultrafine carbon particle inhalation on allergic
  inflammation of the lung.
\newblock {\rm Journal of Allergy and Clinical Immunology } \emph{J 117},
  824--830.

\bibitem[Bennett and Zeman, 1998]{bennett}
Bennett, W. and Zeman, K. (1998{\rm{}}).
\newblock Deposition of fine particles in children spontaneously breathing at
  rest.
\newblock {\rm Inhalation Toxicology } \emph{10}, 831--842.

\bibitem[Crainiceanu {\rm et~al.}, 2007]{ciprian07}
Crainiceanu, C., Ruppert, D., Carroll, R., Joshi, A.  and Goodner, B.
  (2007{\rm{}}).
\newblock Spatially Adaptive Bayesian Penalized Splines With Heteroscedastic
  Errors.
\newblock {\rm Journal of Computational and Graphical Statistics } \emph{16 No.
  2}, 265--288.

\bibitem[de~Boor, 1978]{deboor}
de~Boor, C. (1978{\rm{}}).
\newblock {\rm A Practical Guide to Splines}.
\newblock New York: Springer.

\bibitem[Delfino {\rm et~al.}, 2005]{delfino}
Delfino, R., Sioutas, C.  and Malik, S. (2005{\rm{}}).
\newblock Potential role of ultrafine particles in associations between
  airborne particle mass and cardiovascular health.
\newblock {\rm Environmental Health Perspectives } \emph{113}, 934--946.

\bibitem[EPA, 2002]{epa02}
EPA, U. (2002{\rm{}}).
\newblock {\rm Health assessment document for diesel engine exhaust.}
\newblock The National Technical Information Service Springfield, VA
  epa/600/8-90/057f edition.

\bibitem[EPA, 2012]{epa12}
EPA, U. (2012{\rm{}}).
\newblock {\rm Clean school bus USA.}
\newblock Available from http://www.epa.gov/cleanschoolbus/basicinfo.htm.
  Accessed 4/1/12.

\bibitem[Ferin {\rm et~al.}, 1990]{ferin}
Ferin, J., Oberdorster, G., Penney, D., Soderholm, S., Gelein, R.  and Piper,
  H. (1990{\rm{}}).
\newblock Increased pulmonary toxicity of ultrafine particles? I. particle
  clearance, translocation, morphology.
\newblock {\rm Journal of Aerosol Science } \emph{21}, 384--387.

\bibitem[Frampton {\rm et~al.}, 2006]{frampton}
Frampton, M., Stewart, J., Oberdorster, G., Morrow, P., Chalupa, D.,
  Pietropaoli, A., Frasier, L., Speers, D., Cox, C., Huang, L.  and Utell, M.
  (2006{\rm{}}).
\newblock Inhalation of ultrafine particles alters blood leukocyte expression
  of adhesion molecules in humans.
\newblock {\rm Environmental Health Perspectives } \emph{114}, 51--58.

\bibitem[Hussein {\rm et~al.}, 2005]{hussein05}
Hussein, T., Maso, M., Petäjä, T., Koponen, I.~K., Paatero, P., Aalto, P.~P.,
  Hämeri, K.  and Kulmala, M. (2005{\rm{}}).
\newblock Evaluation of an automatic algorithm for fitting the particle number
  size distributions.
\newblock {\rm Boreal Environment Research } \emph{10}, 337--355.

\bibitem[Lang and Brezger, 2004]{Lang04}
Lang, S. and Brezger, A. (2004{\rm{}}).
\newblock Bayesian P-Splines.
\newblock {\rm Journal of Computational and Graphical Statistics } \emph{13 No.
  1}, 183--212.

\bibitem[Oberdorster {\rm et~al.}, 2004]{ober}
Oberdorster, G., Sharp, Z., Atudorei, V., Elder, A., Gelein, R., Kreyling, W.
  and Cox, C. (2004{\rm{}}).
\newblock Translocation of inhaled ultrafine particles to the brain.
\newblock {\rm Inhalation Toxicology } \emph{16}, 437--445.

\bibitem[Samet {\rm et~al.}, 2009]{samet}
Samet, J., Rappold, A., Graff, D., Cascio, W., Berntsen, J., Huang, Y.,
  M.~Herbst, M.~B., Montilla, T., Hazucha, M., Bromberg, P.  and Devlin, R.
  (2009{\rm{}}).
\newblock Concentrated ambient ultrafine particle exposure induces cardiac
  changes in young healthy volunteers.
\newblock {\rm American Journal of Respiratory and Critical Care Medicine }
  \emph{179}, 1034--1042.

\bibitem[Shi {\rm et~al.}, 1996]{shi96}
Shi, M., Weiss, R.~E.  and Taylor, J. (1996{\rm{}}).
\newblock An Analysis of Paediatric CD4 Counts for Acquired Immune Deficiency
  Syndrome using Flexible Random Curves.
\newblock {\rm Journal of the Royal Statistical Society. Series C (Applied
  Statistics) } \emph{45, No 2}, 151--163.

\bibitem[Whitby {\rm et~al.}, 1991]{whitby91}
Whitby, McMurry, P.  and Shanker, U. (1991{\rm{}}).
\newblock {\rm Modal aerosol dynamics modelling. Technical Report}.
\newblock U.S. Environment Protection Agency, Atmospheric Research and Exposure
  Assessment Laboratory.

\bibitem[Whitby and OTHERS, 1978]{whitby78}
Whitby and OTHERS (1978{\rm{}}).
\newblock {\rm Modal aerosol dynamics modelling. Technical Report}.
\newblock U.S. Environment Protection Agency, Atmospheric Research and Exposure
  Assessment Laboratory.

\bibitem[Wraith {\rm et~al.}, 2009]{wraith09}
Wraith, D., Alston, C., Mengersen, K.  and Hussein, T. (2009{\rm{}}).
\newblock Bayesian mixture model estimation of aerosol particle size
  distributions.
\newblock {\rm Environmetrics } \emph{22}, 23--34.

\bibitem[Yau and Kohn, 2003]{yau03}
Yau, P. and Kohn, R. (2003{\rm{}}).
\newblock Estimation and variable selection in nonparametric heteroscedastic
  regression.
\newblock {\rm Statistics and Computing } \emph{13}, 191--208.

\bibitem[Zhang and Zhu, 2010]{zhang10}
Zhang, Q. and Zhu, Y. (2010{\rm{}}).
\newblock {\rm Atmospheric Environment } \emph{44}, 253--261.

\end{thebibliography}

\appendix

\section{Supplementary Material}

\setcounter{figure}{0}

\makeatletter 
\renewcommand{\thefigure}{A\@arabic\c@figure}
\makeatother

\subsection{Mean Predictive Posterior Counts for Windows-closed Position}

\begin{figure}[!t]
	\centering
	\subfloat[5 minutes, Non-jump]{\includegraphics[width=5cm]{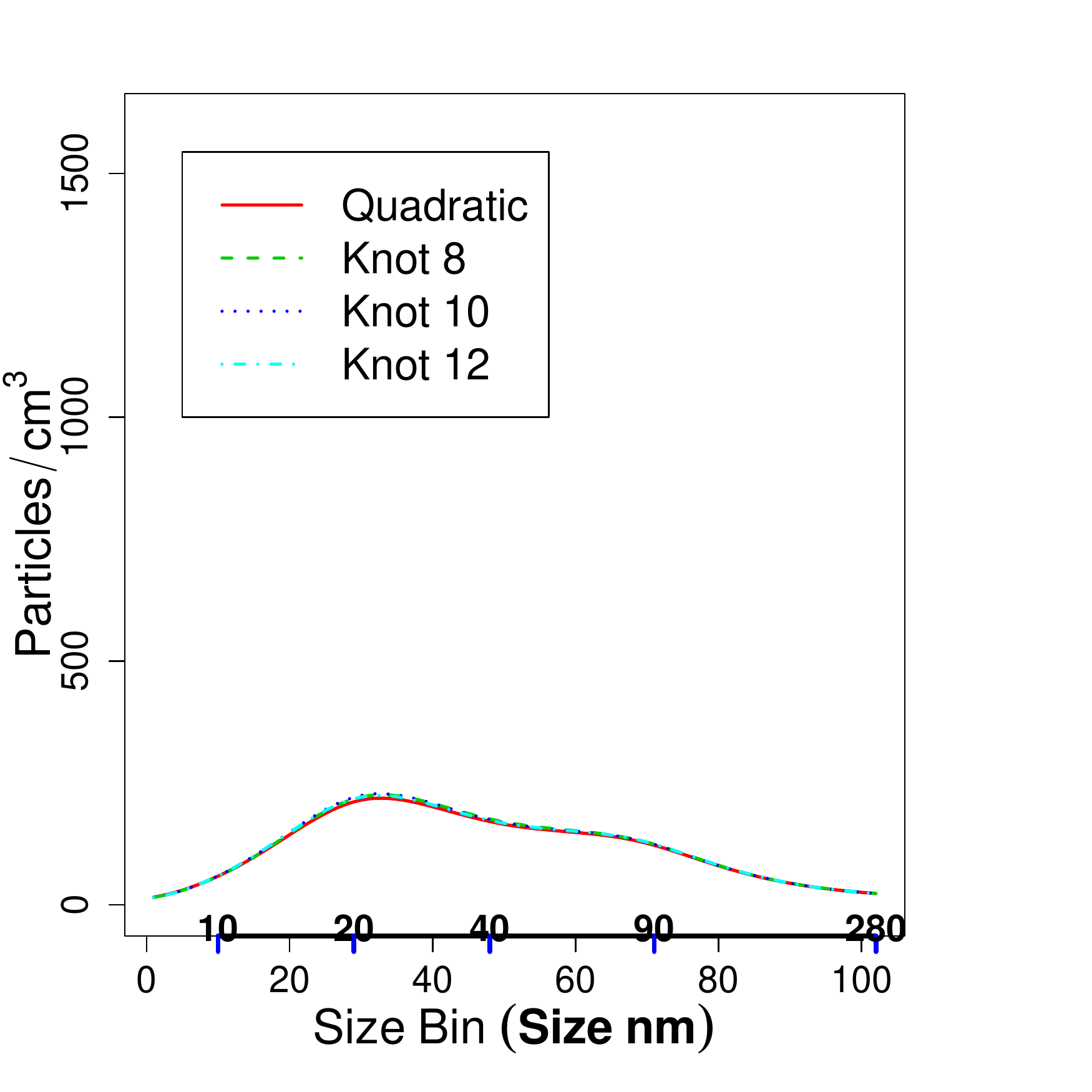}}
	\subfloat[5 minutes, Jump]{\includegraphics[width=5cm]{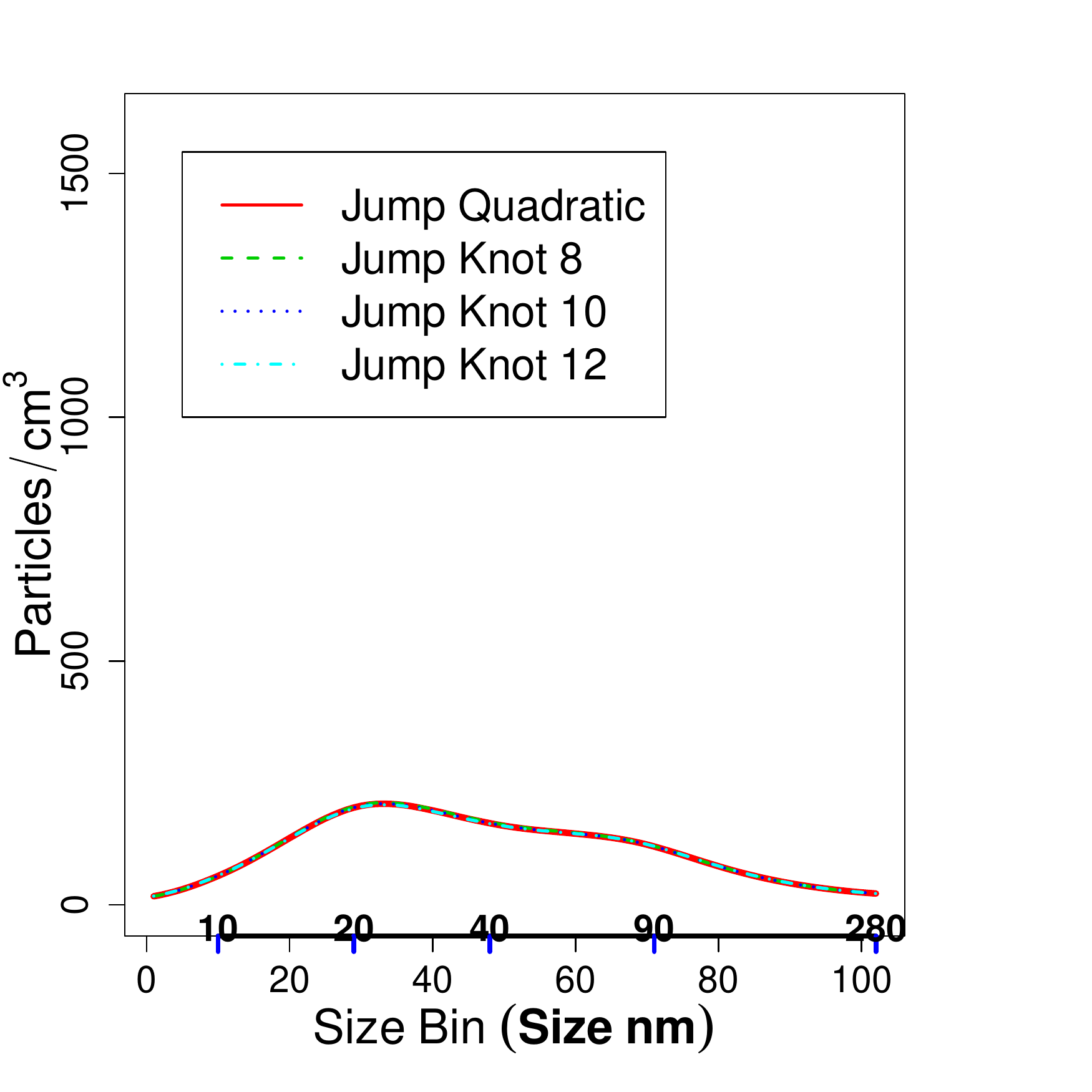}}
	\subfloat[5 minutes, Random Jump]{\includegraphics[width=5cm]{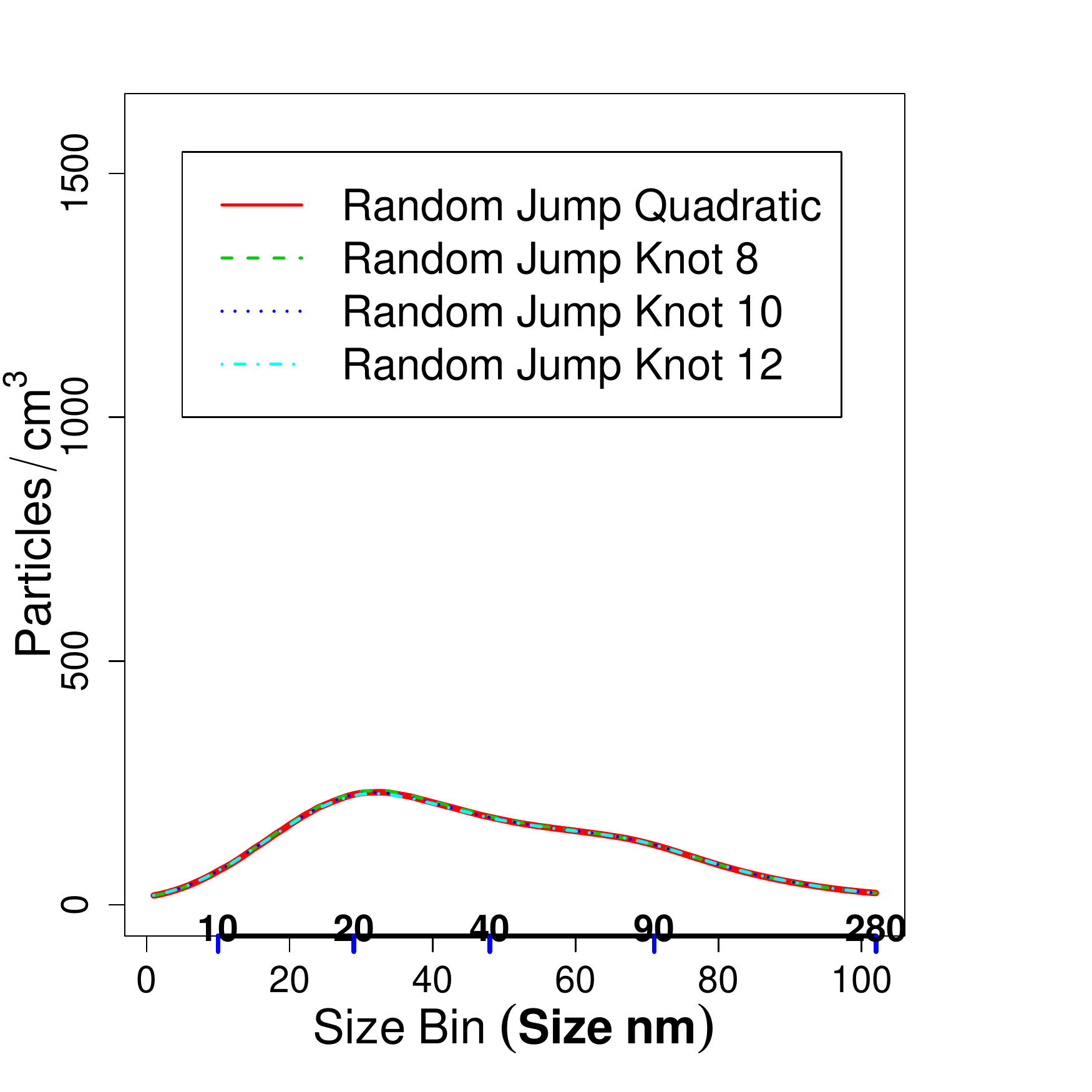}} \\
	\vspace{-2\baselineskip}
	\subfloat[15 minutes, Non-jump]{\includegraphics[width=5cm]{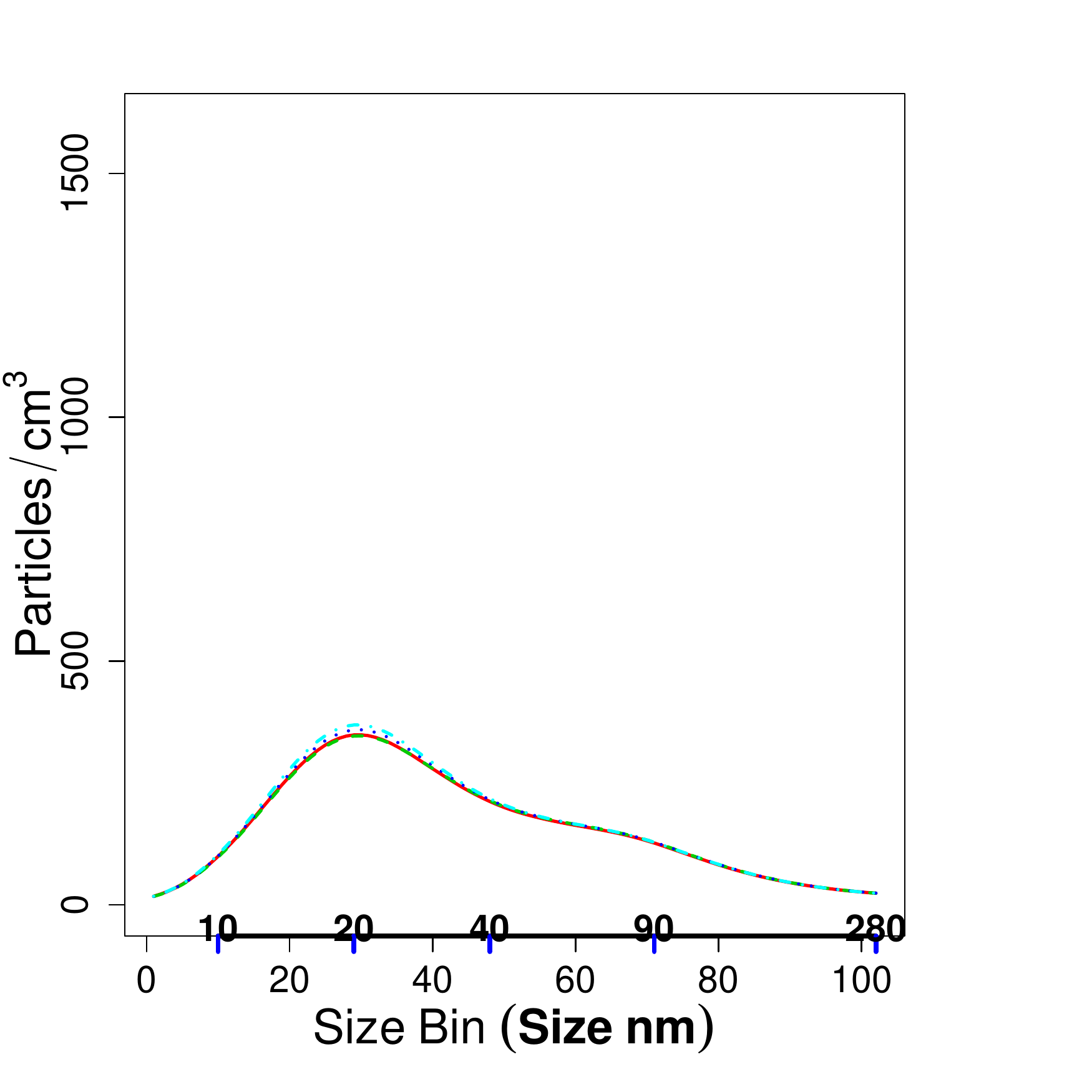}}
	\subfloat[15 minutes, Jump]{\includegraphics[width=5cm]{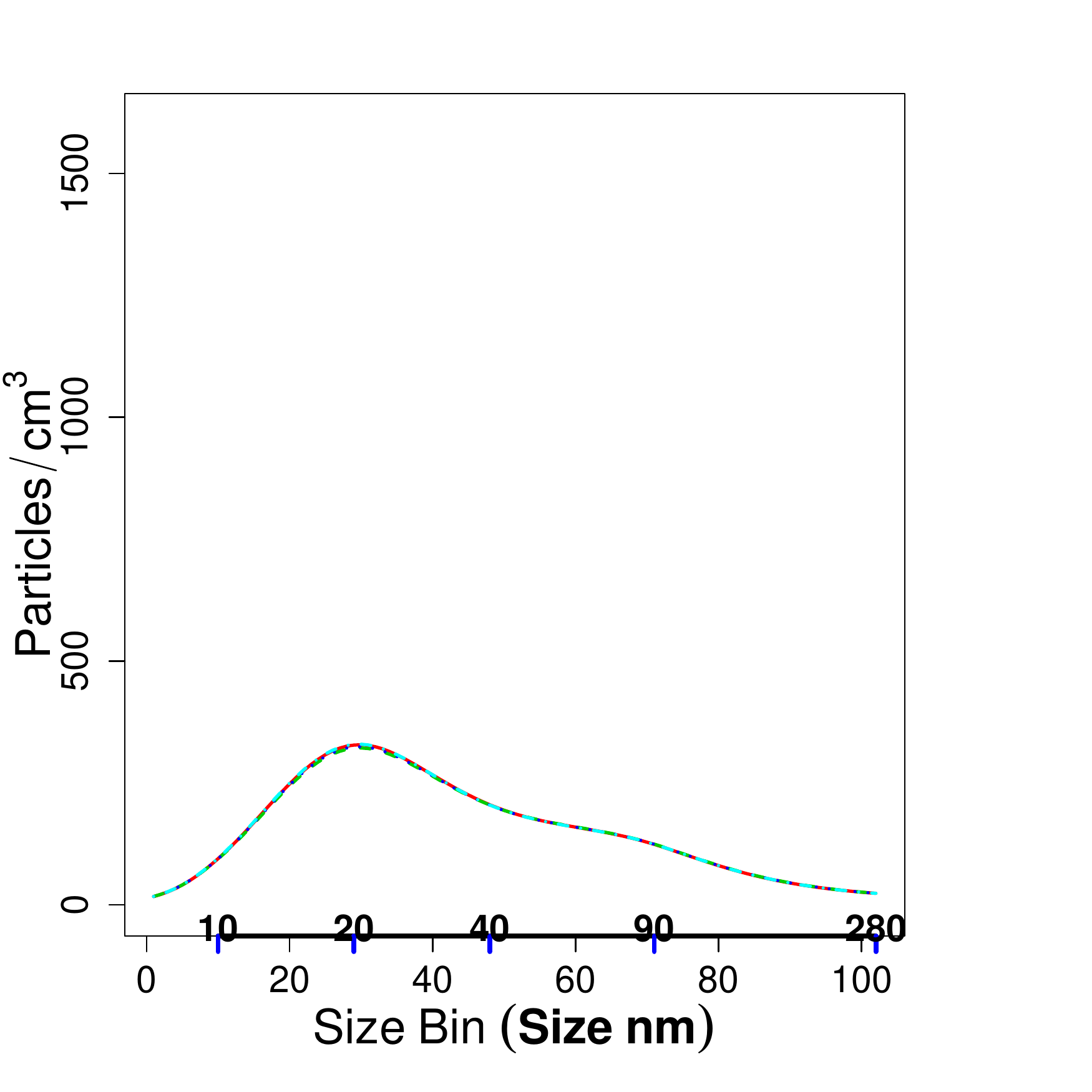}}
	\subfloat[15 minutes, Random Jump]{\includegraphics[width=5cm]{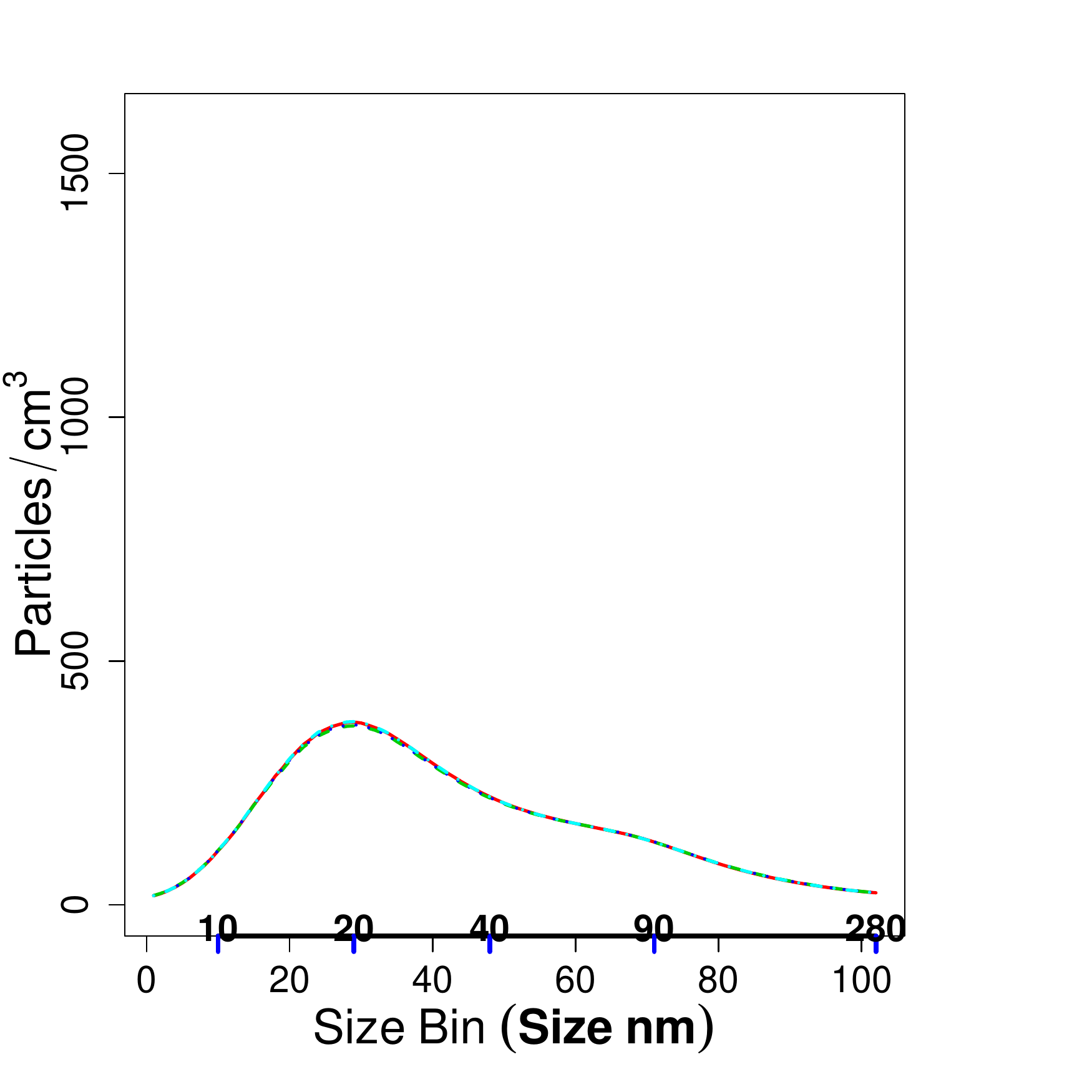}} \\
	\vspace{-2\baselineskip}
	\subfloat[20 minutes, Non-jump]{\includegraphics[width=5cm]{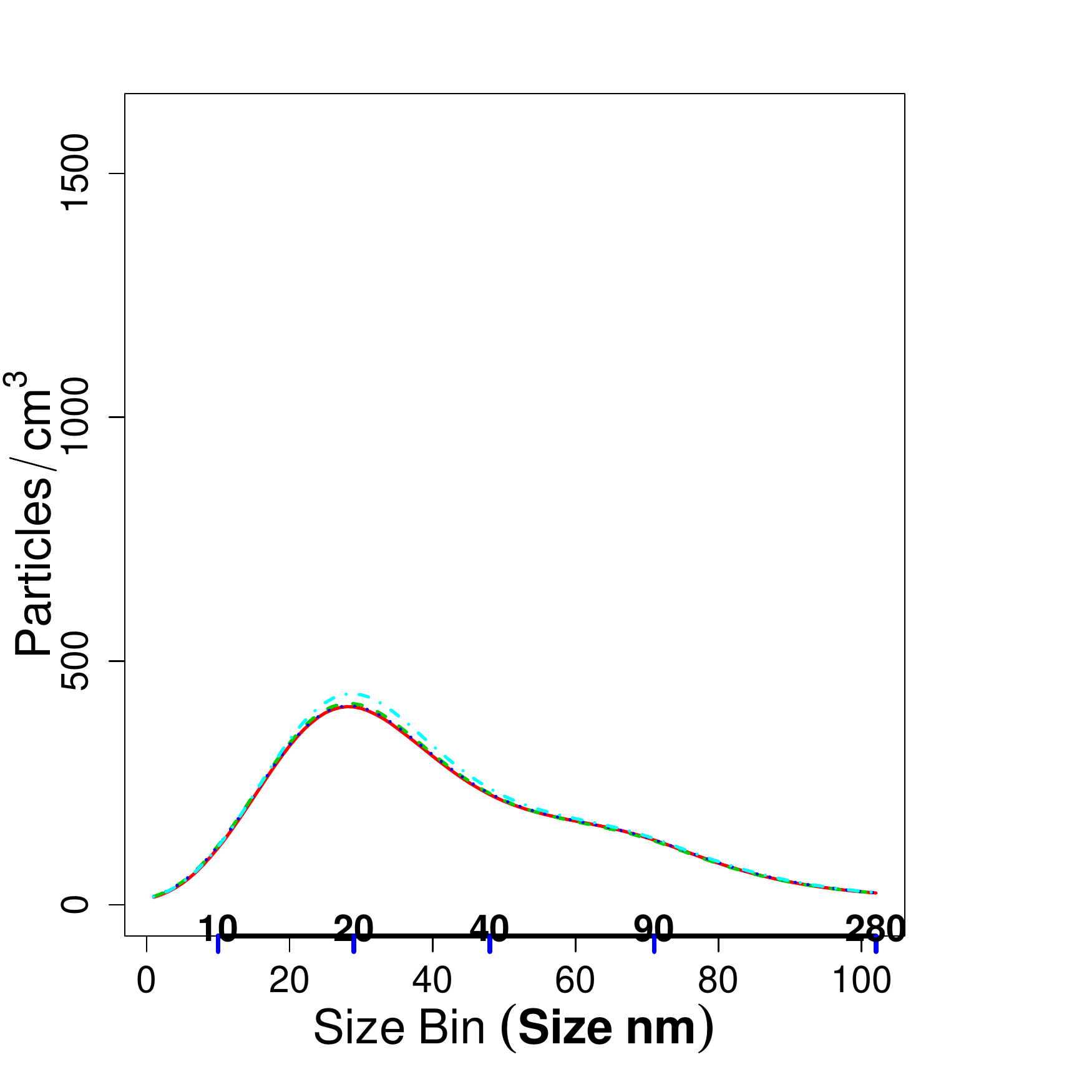}}
	\subfloat[20 minutes, Jump]{\includegraphics[width=5cm]{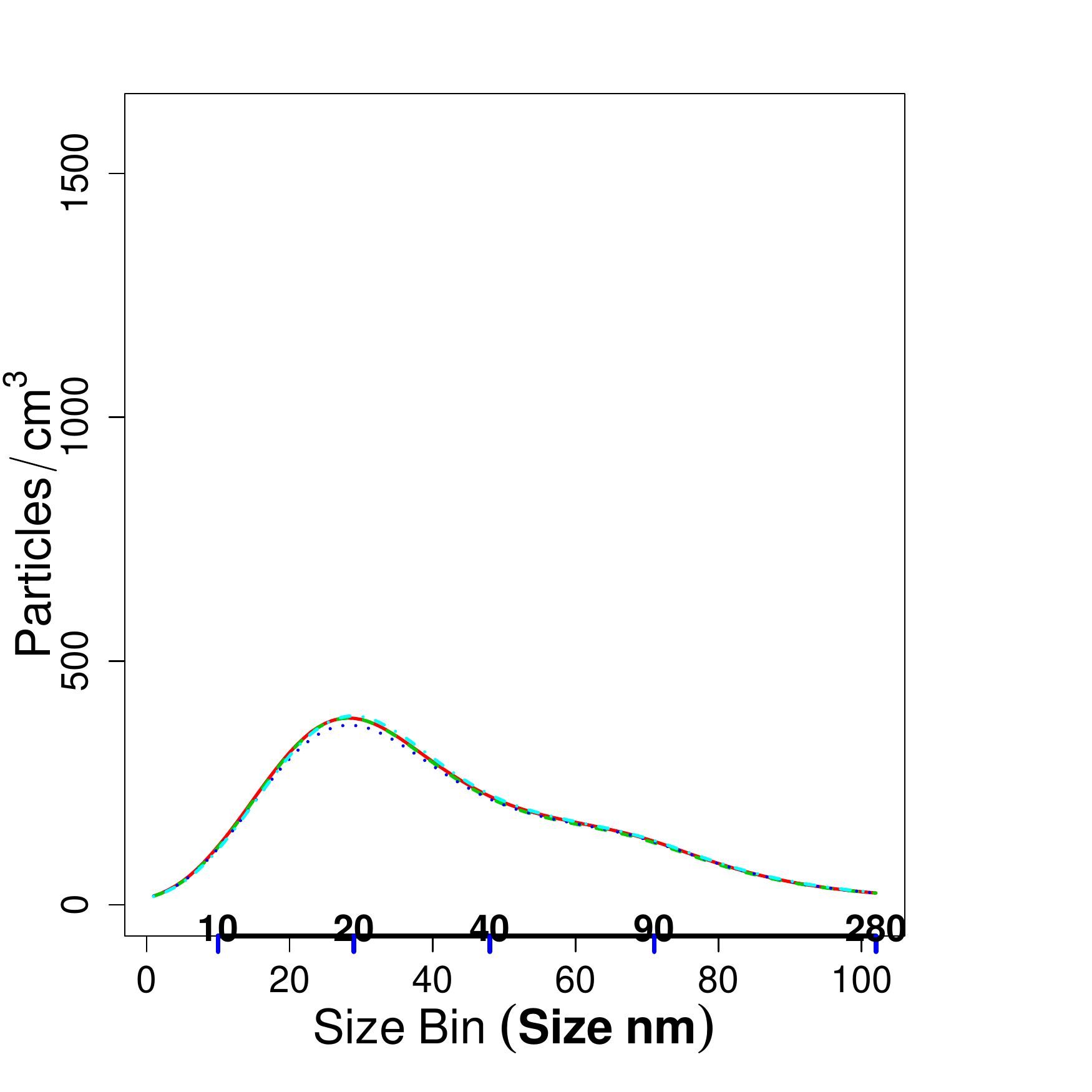}}
	\subfloat[20 minutes, Random Jump]{\includegraphics[width=5cm]{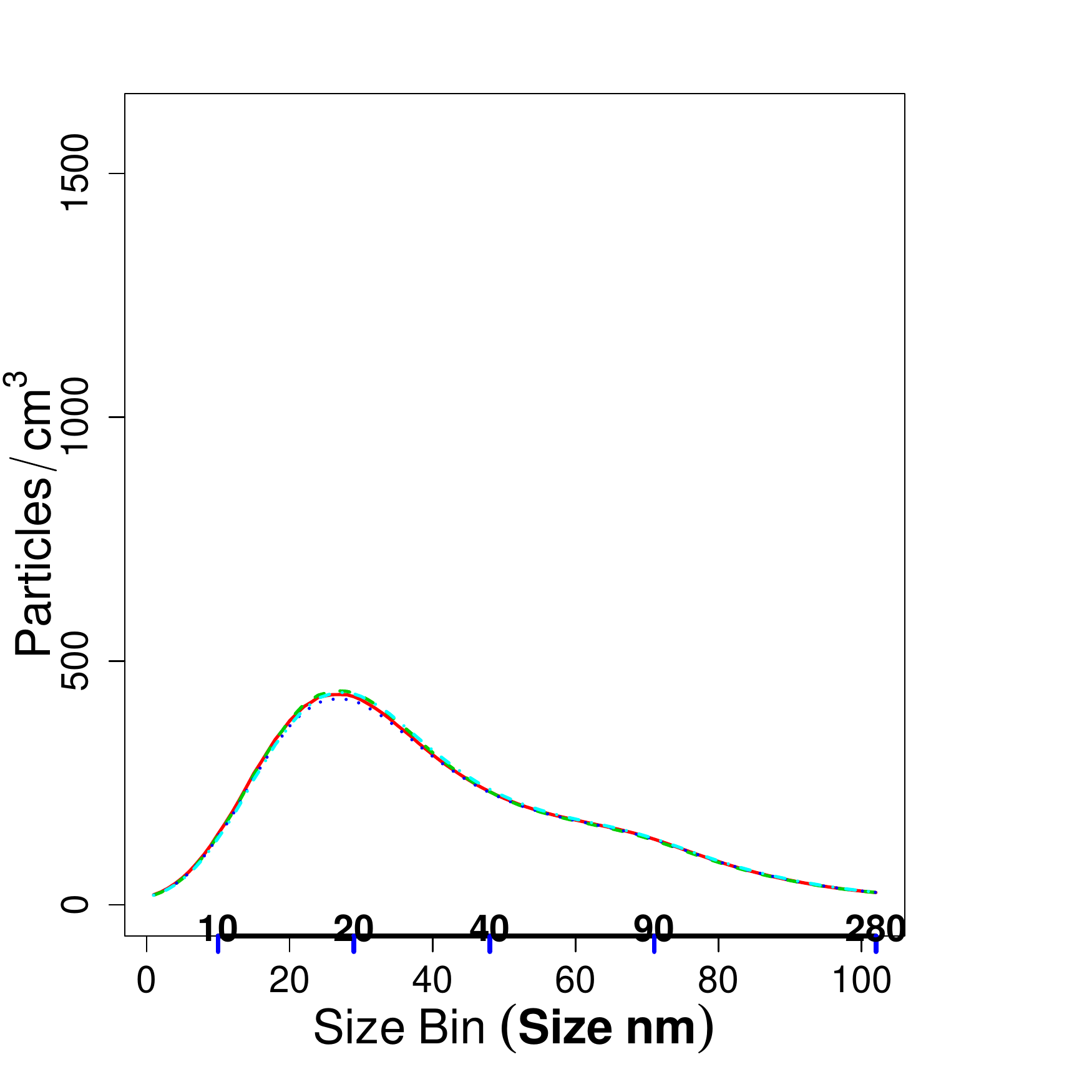}}
	 \caption{Mean predictive posterior counts for the windows-closed position for all models after the engine has been idling for $t=5$, row 1(a)(b)(c); $t=15$, row 2(d)(e)(f); and $t=20$, row 3(g)(h)(i). Non-jump models are in the left column (a)(c)(c), jump models are in the middle column (d)(e)(f), and random jump models are in the right column (g)(h)(i). Windows-closed models are very similar between all model types. Credible intervals are not plotted, but intervals overlap for all models.}
	\label{fig:windowsclosedallcurves}
	\end{figure}

Mean predictive posterior counts for the windows-open position were presented in the main manuscript. Mean predictive posterior counts for the windows-closed position for all models after the engine has been idling are presented in Figure~\ref{fig:windowsclosedallcurves} for $t=5$, row 1(a)(b)(c); $t=15$, row 2(d)(e)(f); and $t=20$, row 3(g)(h)(i). Non-jump models are in the left column (a)(c)(c), jump models are in the middle column (d)(e)(f), and random jump models are in the right column (g)(h)(i). Unlike for the windows-open position, windows-closed results are very similar both between all model types, including models with and without jumps and random jumps. Credible intervals are not plotted to ease visual interpretation, but there is overlap in these intervals for all models.

\subsection{Selected residual checks}

\begin{figure}[!ht]
		\centering
		\subfloat[No random effects]{\includegraphics[page=3,width=5cm]{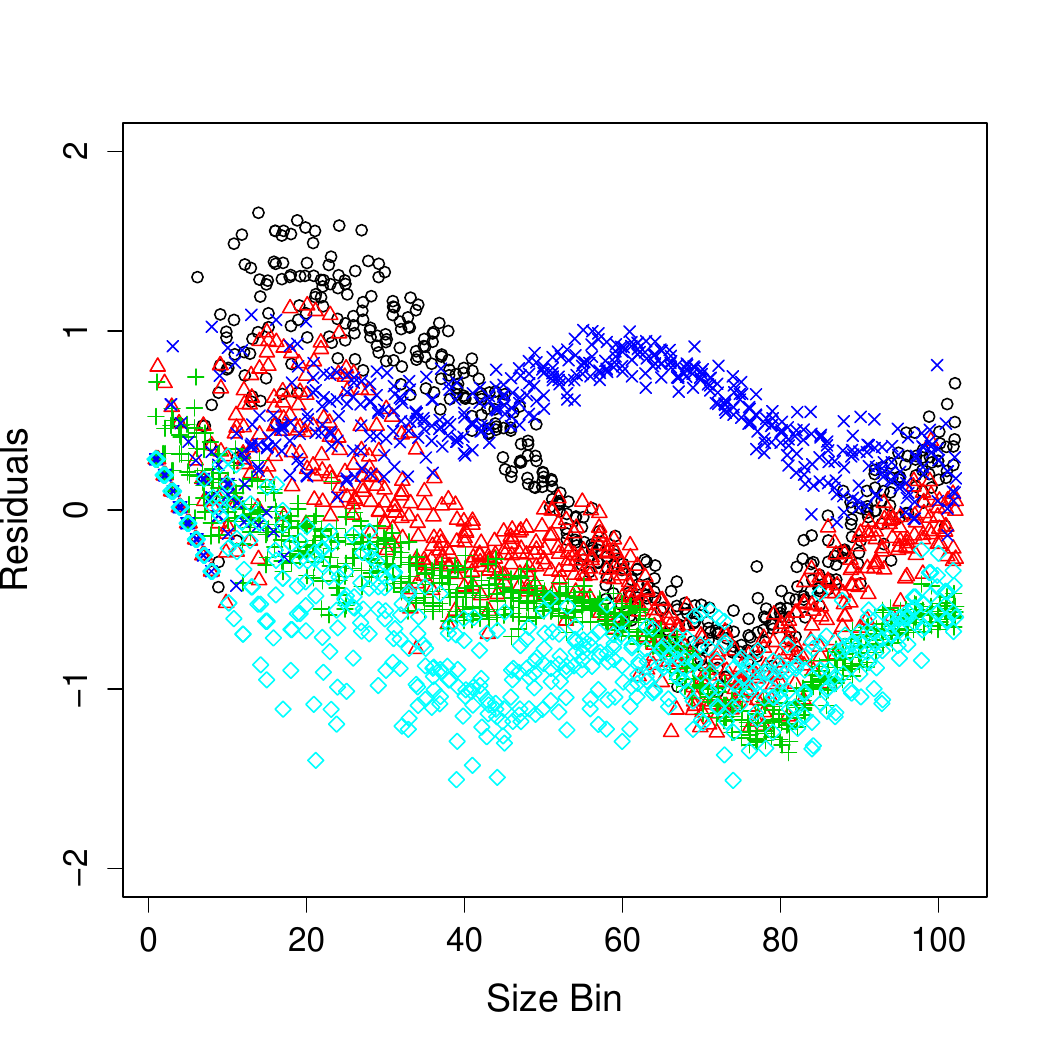}}
		\subfloat[Engine-off random effects]{\includegraphics[page=3,width=5cm]{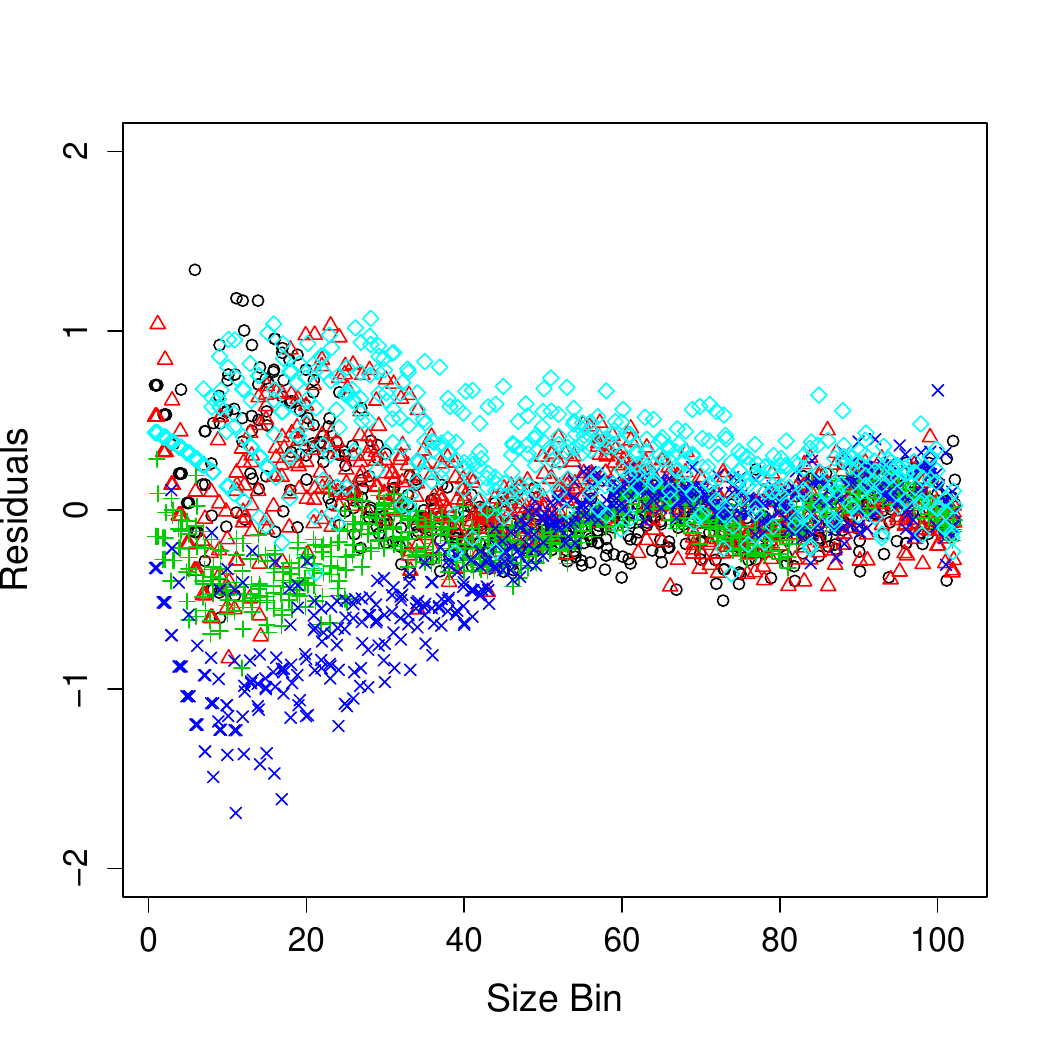}} 
		\subfloat[Engine-off and jump effects]{\includegraphics[page=3,width=5cm]{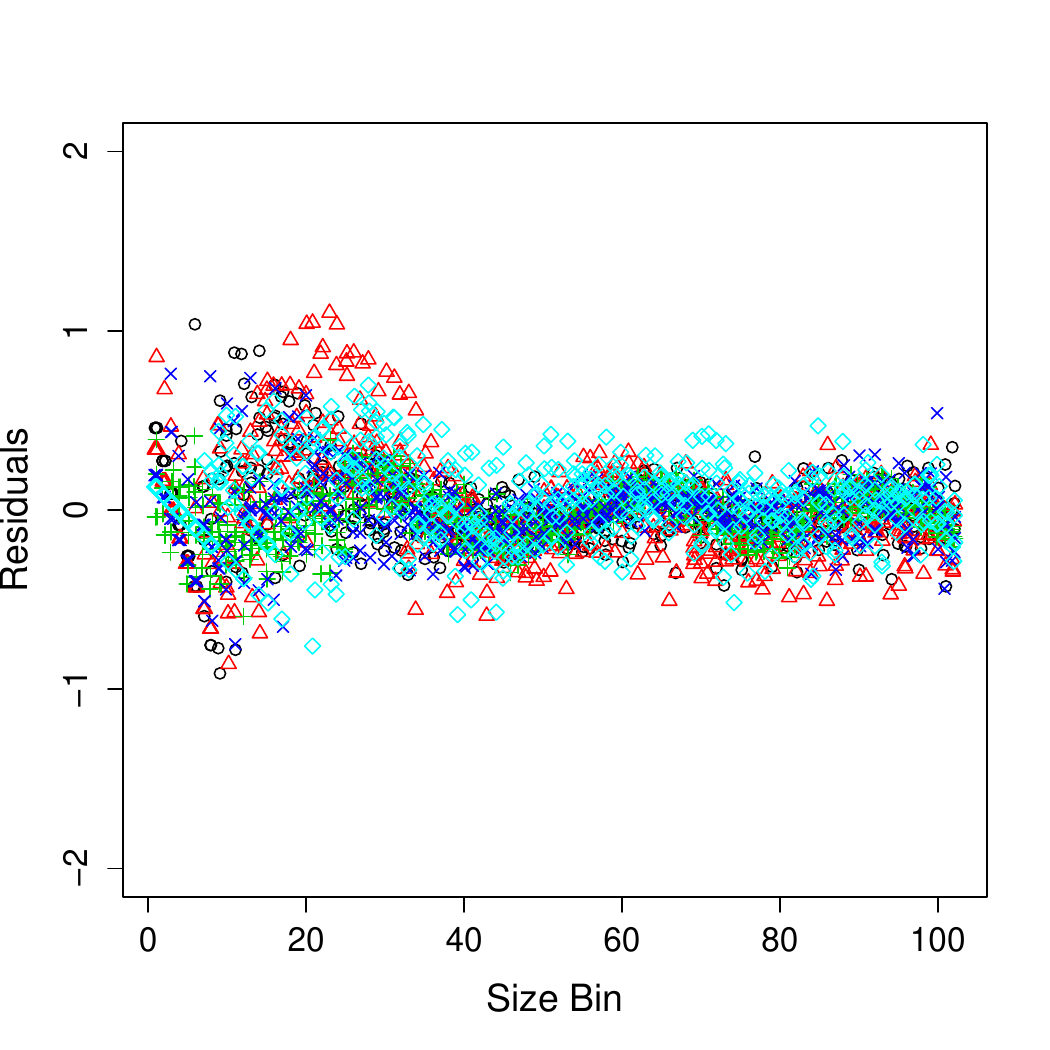}} \\
		\subfloat[No random effects]{\includegraphics[page=2,width=5cm]{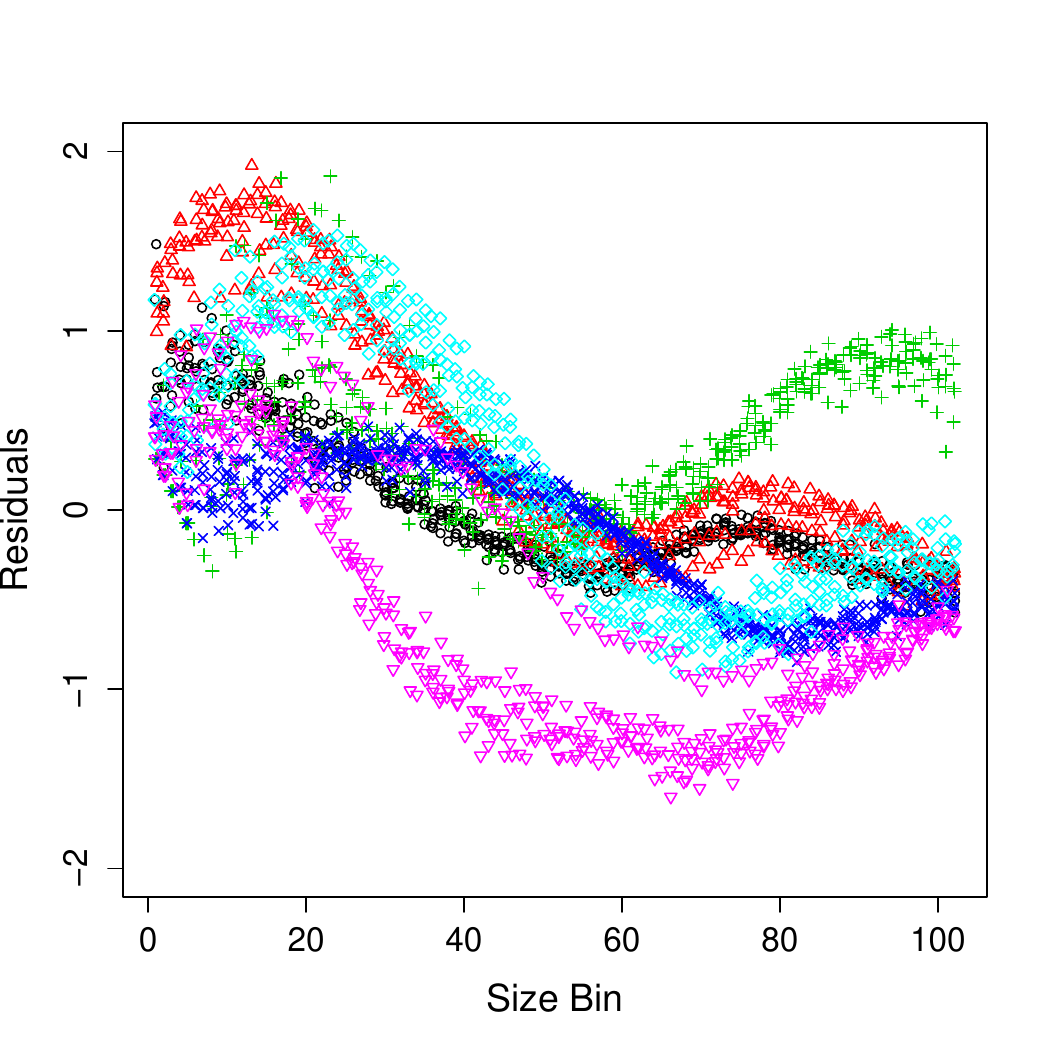}}
		\subfloat[Engine-off random effects]{\includegraphics[page=2,width=5cm]{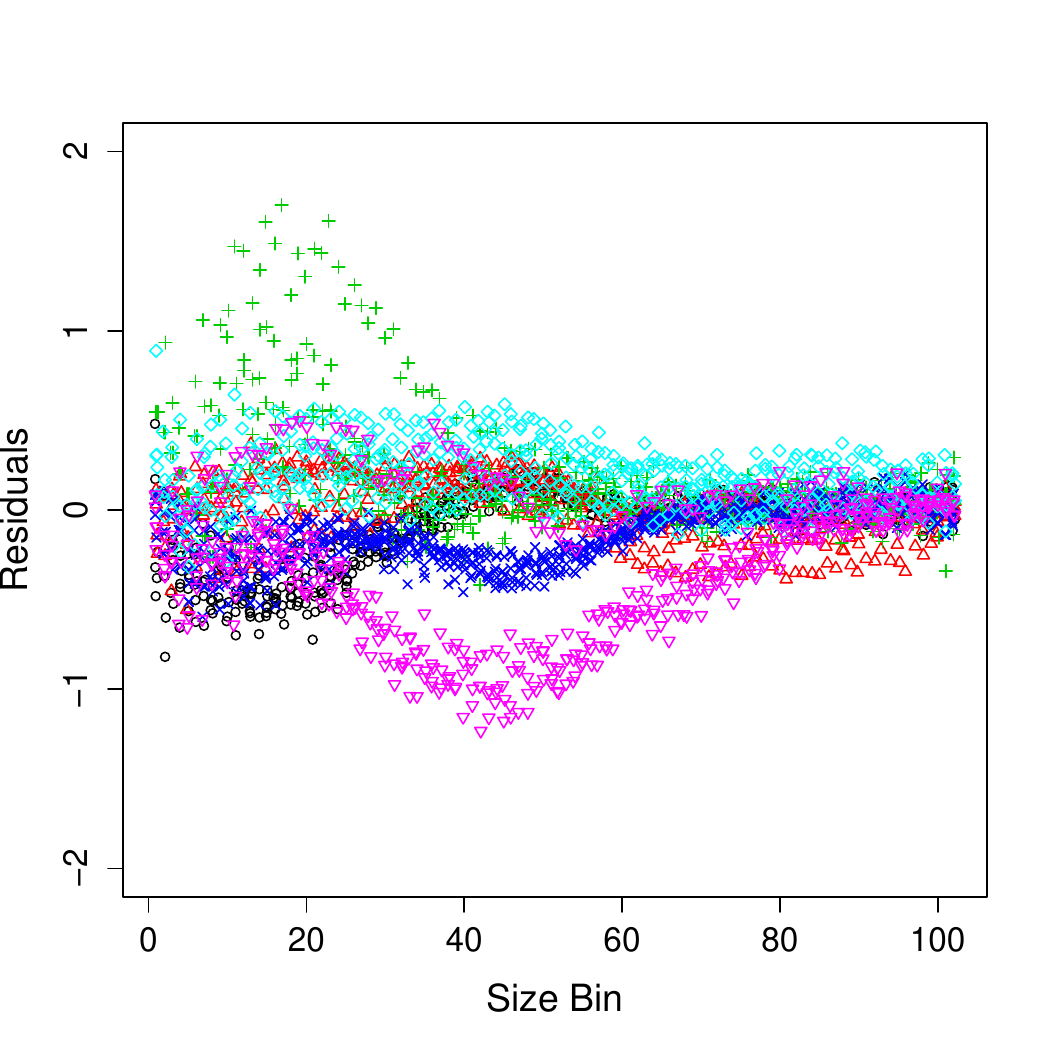}} 
				\subfloat[Engine-off and jump effects]{\includegraphics[page=2,width=5cm]{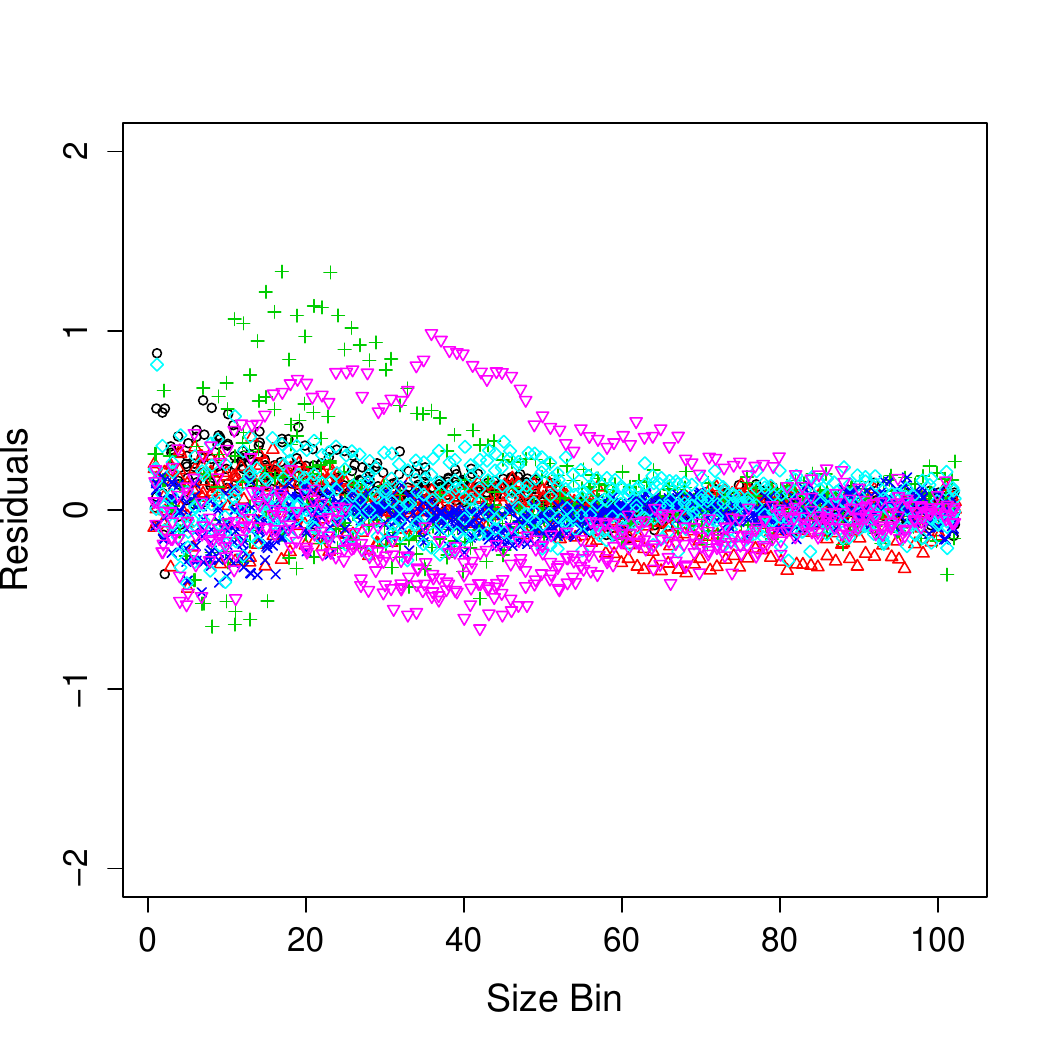}} \\
		\caption{Improvement in fit following the addition of engine-off random effects, and then both engine-off and jump random effects to a fixed effect only model. Figures (a) and (d) plot a subset of residuals colored by run and show residuals were extremely correlated by run before any random curves were included.  Figures~(b) and (e) plot the same subset of residuals after engine-off random curves by run are added to the model. Residuals by run are more mixed on the plot, indicating correlation by run is removed to a large extent, though it is still present. Figures~(c) and (f) plot residuals after the random jump is added.  Residual mixing is further increased. The variance of residuals is decreasing from Figures~(a)(d) to (b)(e) to (c)(f) with the further addition of random curves due to the removal of variation. }
		\label{fig:resid_rand}
		\end{figure}

Figure \ref{fig:resid_rand} shows improvement in fit following the addition of engine-off random effects, and then both engine-off and jump random effects to a fixed effect only model. Figures \ref{fig:resid_rand}(a) and (d) plot a subset of residuals colored by run and show residuals were extremely correlated by run before any random curves were included.  Figures \ref{fig:resid_rand}(b) and (e) plot the same subset of residuals after engine-off random curves by run are added to the model. Residuals by run are more mixed on the plot, indicating correlation by run is removed to a large extent, though it is still present. Figures \ref{fig:resid_rand}(c) and (f) plot residuals after the random jump is added.  Residual mixing is further increased. The variance of residuals is decreasing from Figures \ref{fig:resid_rand}(a)(d) to (b)(e) to (c)(f) with the further addition of random curves due to the removal of variation.	

\begin{figure}[!ht]
		 \centering
		\subfloat[]{\includegraphics[page=4,width=8cm]{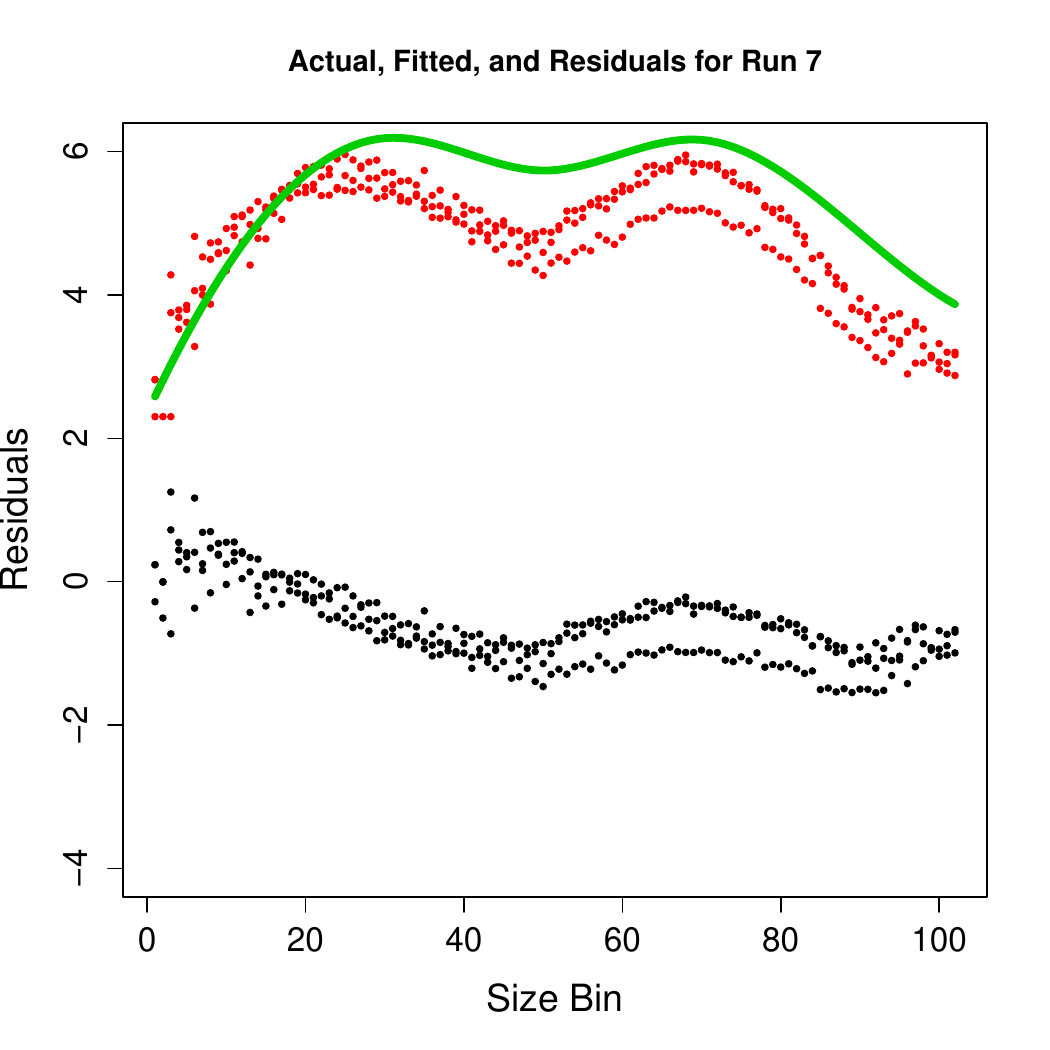}} 
		\subfloat[]{\includegraphics[page=4,width=8cm]{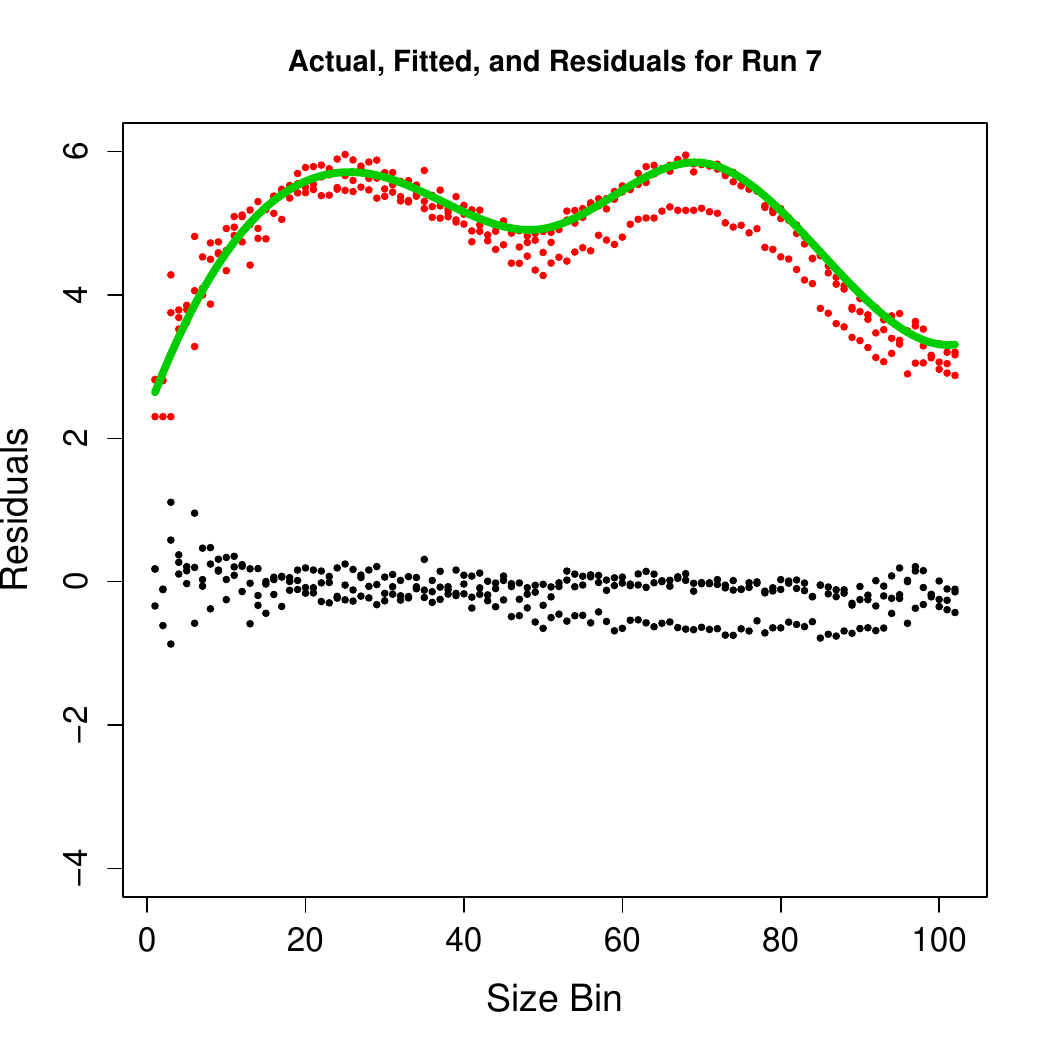}} \\
		\subfloat[]{\includegraphics[page=20,width=8cm]{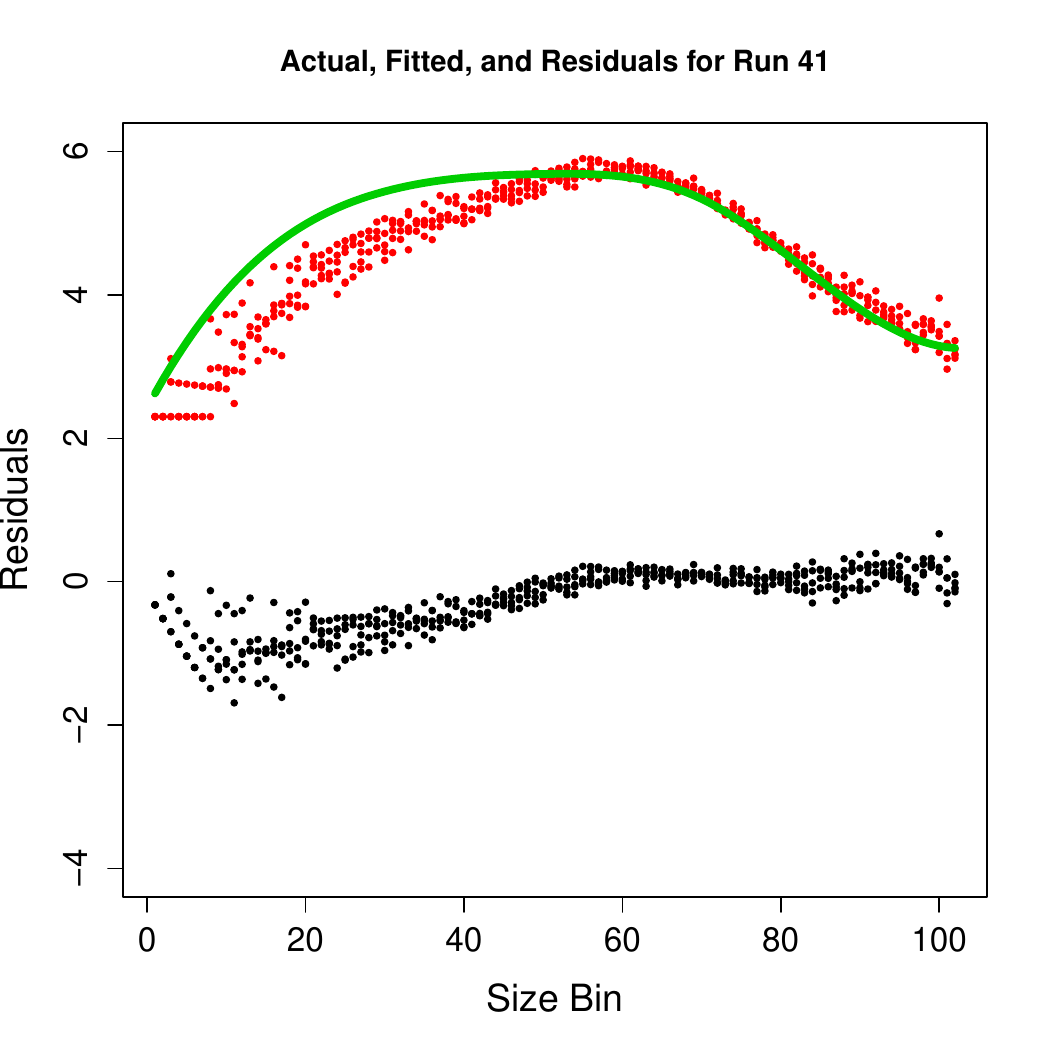}} 
		\subfloat[]{\includegraphics[page=20,width=8cm]{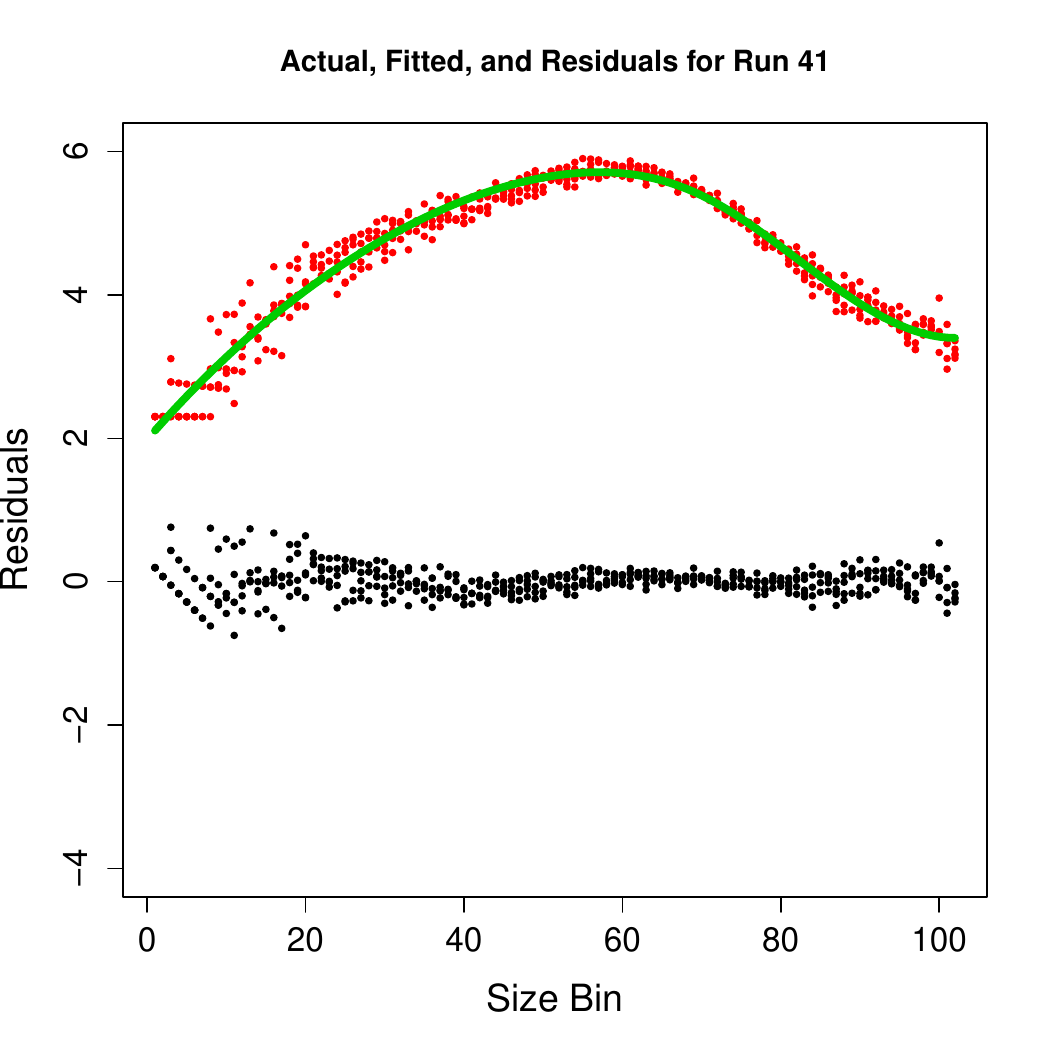}} \\

		\caption{Posterior mean baseline UFP size distributions $\E(\vect{\alpha^T B(s)}+\vect{\gamma_i^T B(s)} \mid \vect{Y})$, observed log counts $y_{ist}$, and residuals $\bar{e}_{ist}$ for two sample runs from when the engine is off. Posterior means are plotted with green curves,  observed log counts are plotted with red dots, and residuals are plotted in black.  The baseline UFP size distributions on the left, Figures (a) and (c), are fit with the jump quadratic model, while those on the right, Figures (b) and (d) are fit with the quadratic random jump model.  The improvement in fit from Figure (a) to (b) and from Figure (c) to (d) shows the value of adding the engine-on random jump.  }
		  \label{fig:randomeffectfit_ex}
		\end{figure}

	Figure \ref{fig:randomeffectfit_ex} plots posterior mean baseline UFP size distributions $\E(\vect{\alpha^T B(s)}+\vect{\gamma_i^T B(s)} \mid \vect{Y})$, observed log counts $y_{ist}$, and residuals $\bar{e}_{ist}$ for two sample runs from when the engine is off. Posterior means are plotted with green curves,  observed log counts are plotted with red dots, and residuals are plotted in black.  The baseline UFP size distributions on the left, Figures \ref{fig:randomeffectfit_ex}(a) and (c), are fit with the jump quadratic model, while those on the right, Figures \ref{fig:randomeffectfit_ex}(b) and (d) are quadratic random jump.  The improvement in fit from Figure \ref{fig:randomeffectfit_ex}(a) to (b) and from Figure \ref{fig:randomeffectfit_ex}(c) to (d) shows the value of adding the engine-on random jump. 
	
\begin{figure}[!ht]
		\centering
		\subfloat[Size 10, no random jump]{\includegraphics[width=5cm,page=1]{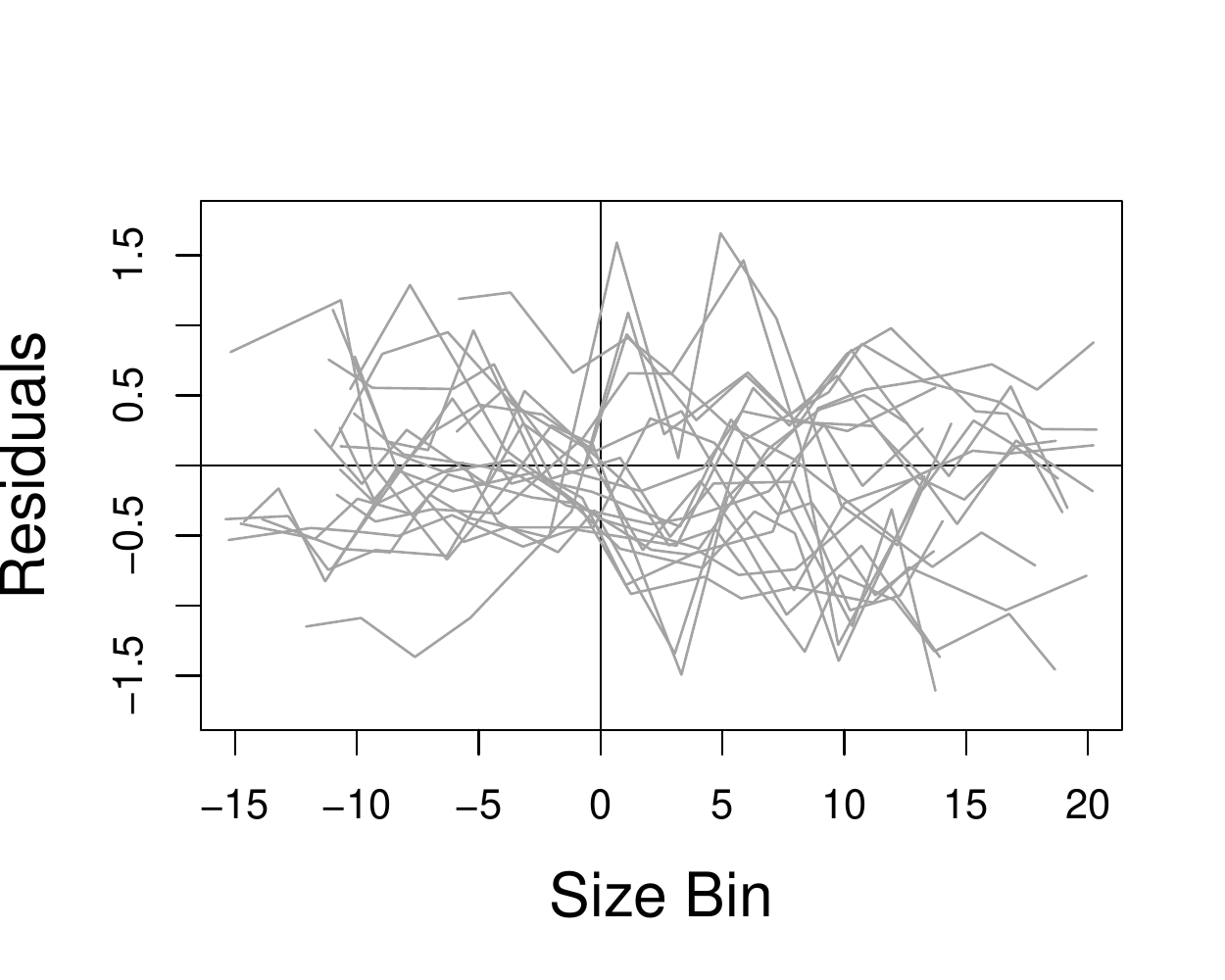}}
		\subfloat[Size 10, random jump]{\includegraphics[width=5cm,page=1]{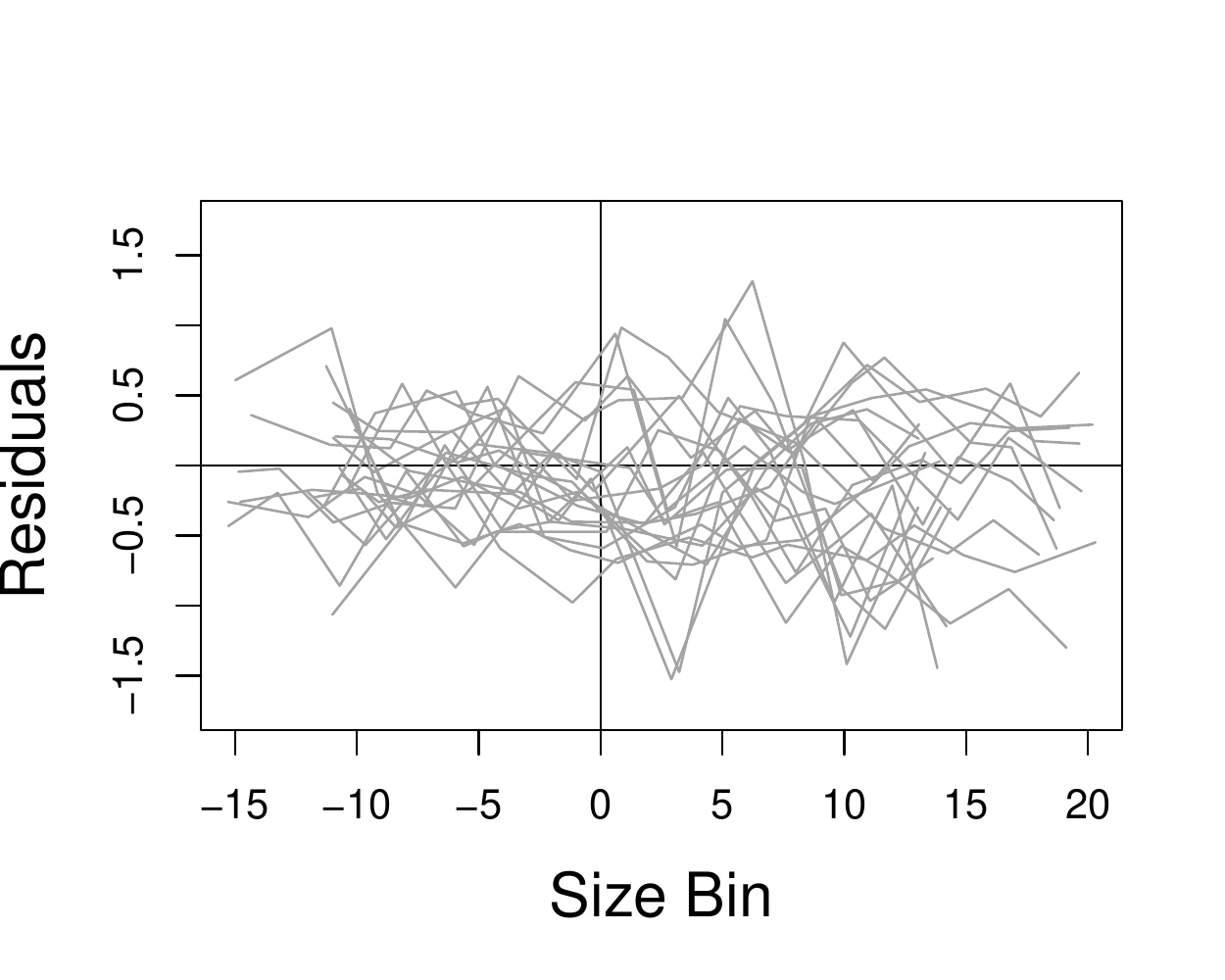}} \\
		\vspace{-2.5\baselineskip}
		\subfloat[Size 20, no random jump]{\includegraphics[width=5cm,page=2]{profile_plots_residuals_by_size_jump.pdf}}
		\subfloat[Size 20, random jump]{\includegraphics[width=5cm,page=2]{profile_plots_residuals_by_size_randomjump.pdf}} \\
		\vspace{-2.5\baselineskip}
		\subfloat[Size 30, no random jump]{\includegraphics[width=5cm,page=3]{profile_plots_residuals_by_size_jump.pdf}}
		\subfloat[Size 30, random jump]{\includegraphics[width=5cm,page=3]{profile_plots_residuals_by_size_randomjump.pdf}} \\
		\vspace{-2.5\baselineskip}
		\subfloat[Size 40, no random jump]{\includegraphics[width=5cm,page=4]{profile_plots_residuals_by_size_jump.pdf}}
		\subfloat[Size 40, random jump]{\includegraphics[width=5cm,page=4]{profile_plots_residuals_by_size_randomjump.pdf}}

		\caption{Residual profile plots over time for size bins 10, 20, 30, and 40 for the jump quadratic (left column) and random jump quadratic models (right column). The vertical black line indicates engine-on. Adding the random jump narrows residual range for engine-off for the entire range of sizes. Residual range for engine-on is also narrowed to a lesser extent.}
		\label{fig:resid_profile1}
		\end{figure}

\begin{figure}[!ht]
		\centering
		\subfloat[Size 50, no random jump]{\includegraphics[width=5cm,page=5]{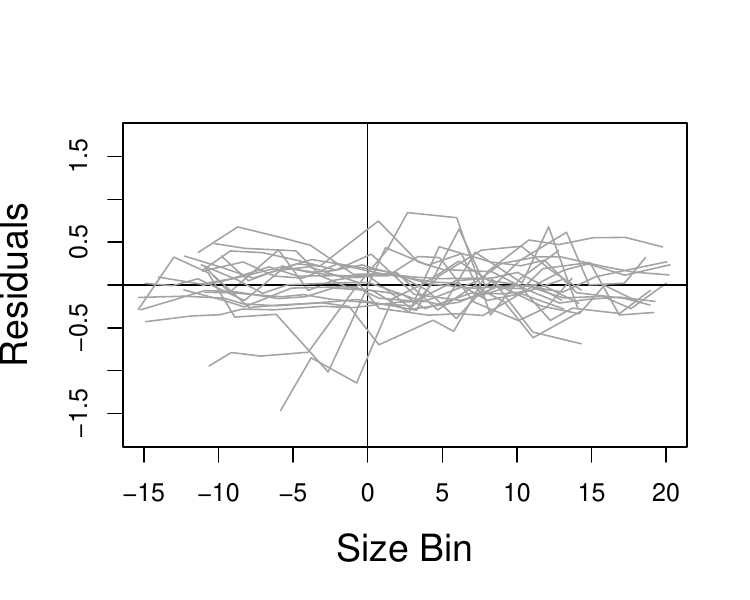}}
		\subfloat[Size 50, random jump]{\includegraphics[width=5cm,page=5]{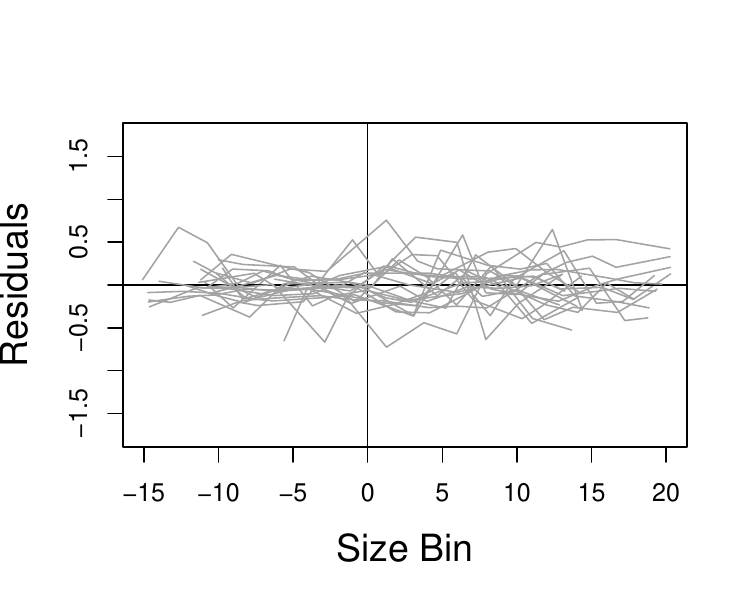}} \\
		\vspace{-2.5\baselineskip}
		\subfloat[Size 60, no random jump]{\includegraphics[width=5cm,page=6]{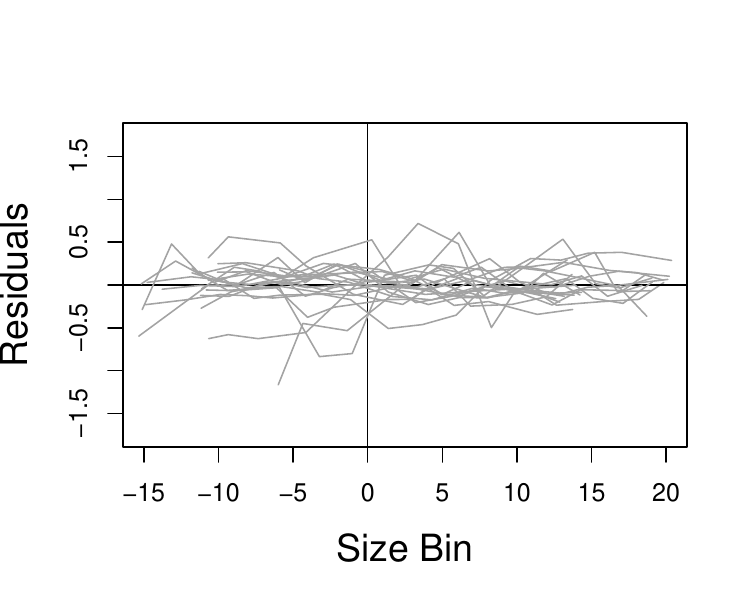}}
		\subfloat[Size 60, random jump]{\includegraphics[width=5cm,page=6]{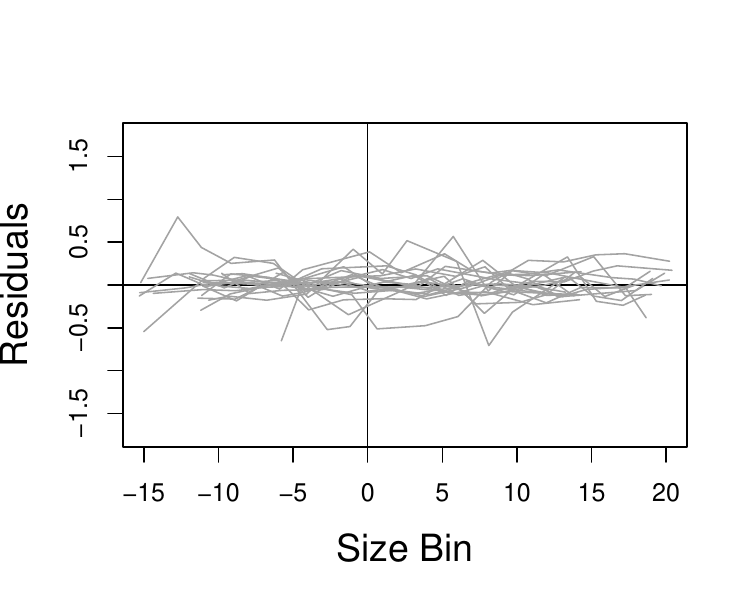}} \\
		\vspace{-2.5\baselineskip}
		\subfloat[Size 70, no random jump]{\includegraphics[width=5cm,page=7]{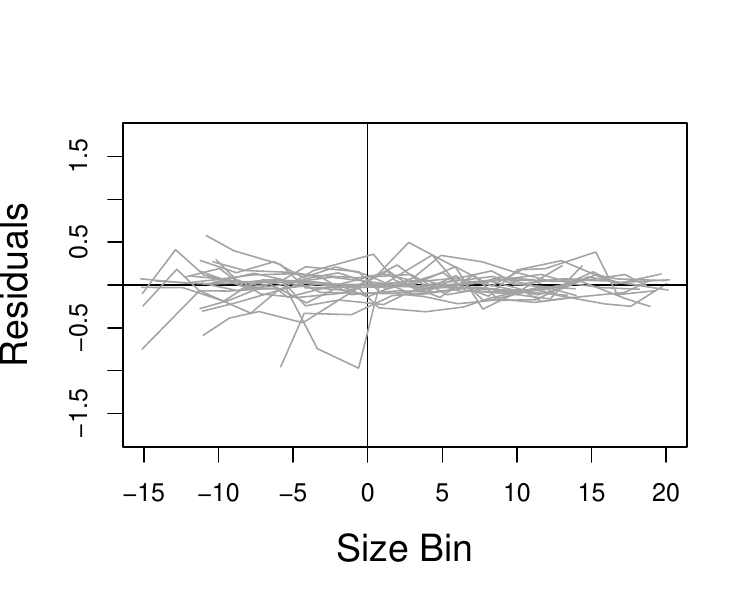}}
		\subfloat[Size 70, random jump]{\includegraphics[width=5cm,page=7]{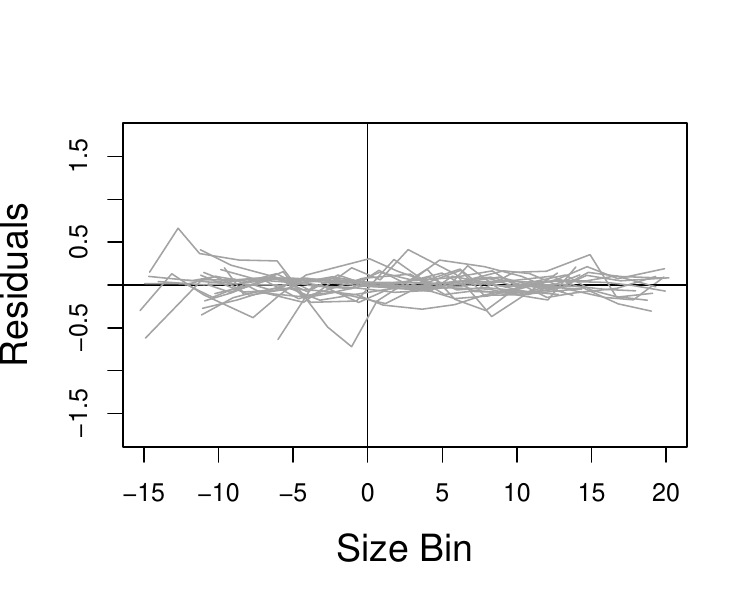}} \\
		\vspace{-2.5\baselineskip}
		\subfloat[Size 80, no random jump]{\includegraphics[width=5cm,page=8]{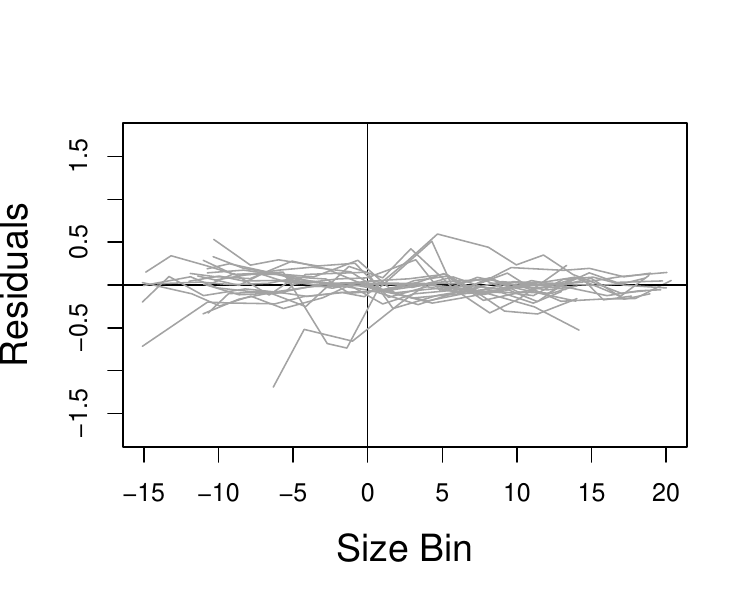}}
		\subfloat[Size 80, random jump]{\includegraphics[width=5cm,page=8]{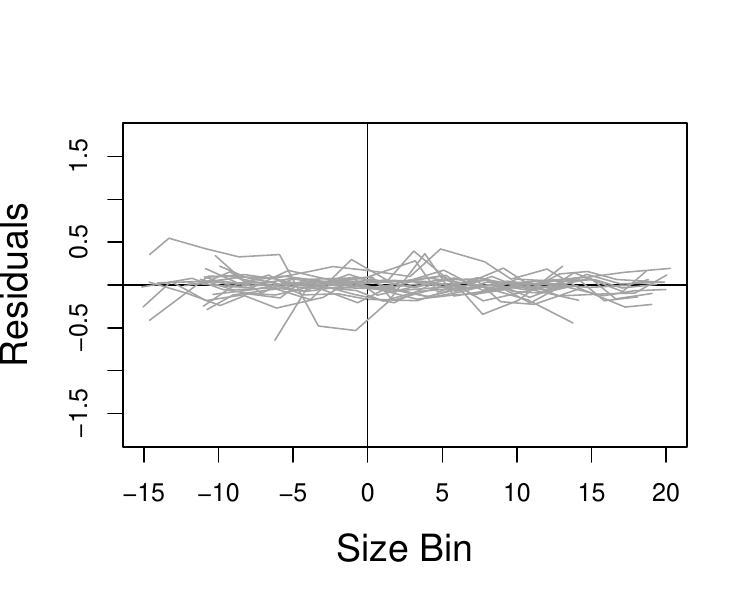}}

		\caption{Residual profile plots over time for size bins 50, 60, 70, and 80 for the jump quadratic (left column) and random jump quadratic models (right column). The vertical black line indicates engine-on. Adding the random jump narrows residual range for engine-off for the entire range of sizes. Residual range for engine-on is also narrowed to a lesser extent.}
		\label{fig:resid_profile2}
		\end{figure}

\begin{figure}[!ht]
		\centering
		\subfloat[Size 90, no random jump]{\includegraphics[width=5cm,page=9]{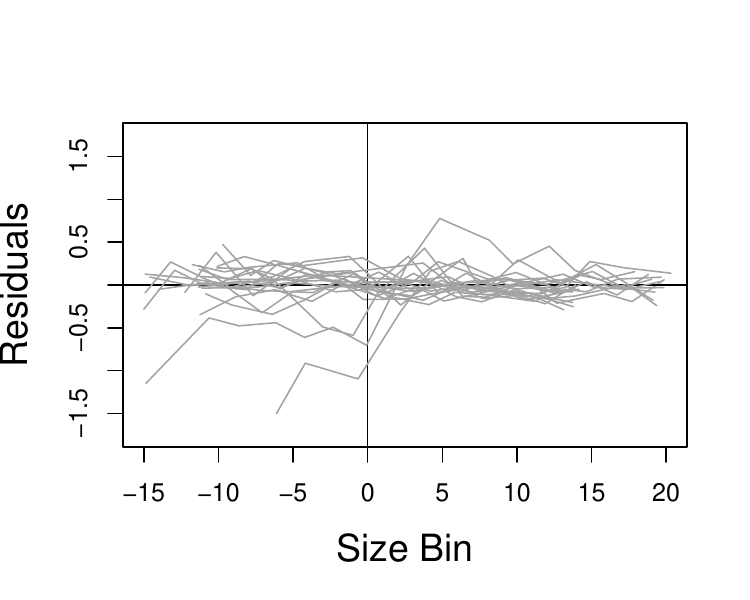}}
		\subfloat[Size 90, random jump]{\includegraphics[width=5cm,page=9]{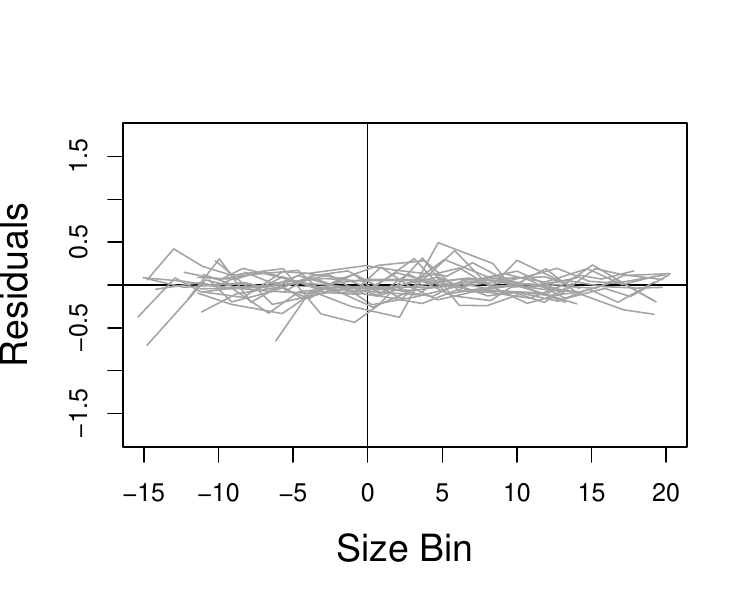}} \\
		\vspace{-2.5\baselineskip}
		\subfloat[Size 100, no random jump]{\includegraphics[width=5cm,page=10]{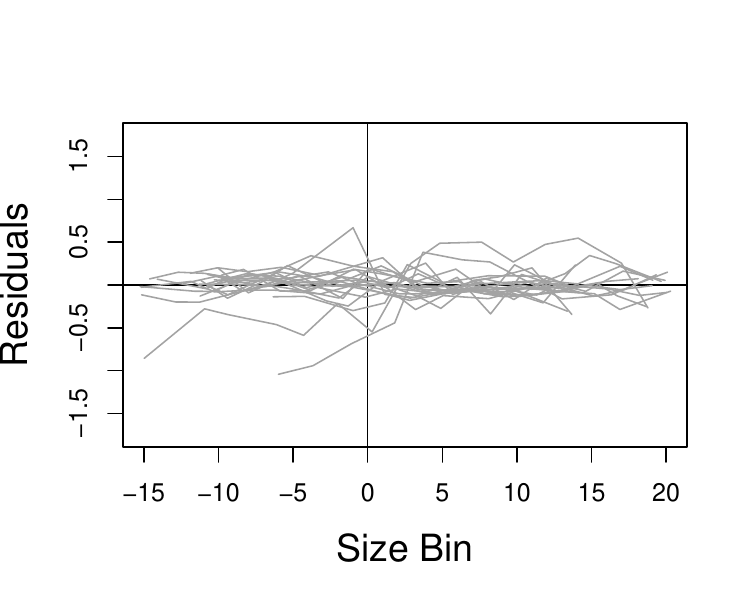}}
		\subfloat[Size 100, random jump]{\includegraphics[width=5cm,page=10]{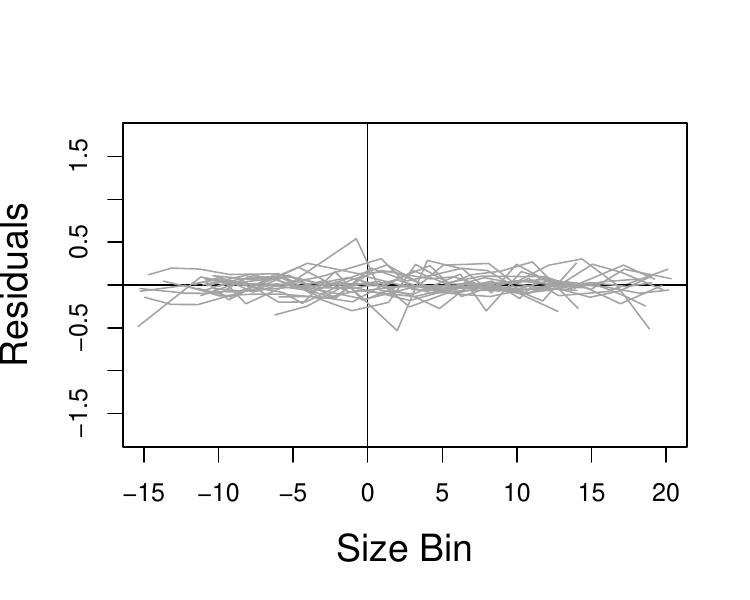}} \\

		\caption{Residual profile plots over time for size bins 90 and 100 for the jump quadratic (left column) and random jump quadratic models (right column). The vertical black line indicates engine-on. Adding the random jump narrows residual range for engine-off for the entire range of sizes. Residual range for engine-on is also narrowed to a lesser extent.}
		\label{fig:resid_profile3}
		\end{figure}

	Figures \ref{fig:resid_profile1}, \ref{fig:resid_profile2}, and \ref{fig:resid_profile3} show residual profile plots over time for select size bins for the jump quadratic (left column) and random jump quadratic (right column) models. The portion of the plot to the right of the vertical black line plots engine-on. Adding the random jump narrows residual range for engine-off for the entire range of sizes. Residual range for engine-on is also narrowed to a lesser extent.

\end{document}